\documentclass[colorlinks=true, linkcolor=blue, citecolor=blue, urlcolor=blue]{aa}  
\usepackage{orcidlink}
\usepackage{txfonts}
\usepackage{graphicx}	
\usepackage{amsmath}	
\usepackage{amssymb}	
\usepackage{xcolor}     
\usepackage{soul}       
\usepackage{array}
\usepackage{makecell}
\usepackage{enumitem}
\usepackage{arydshln}
\usepackage{pdflscape}
\usepackage{multirow}
\usepackage{etoolbox}

\NewDocumentCommand\michele{O{Rewritten}m}{\textsf{\color{red}[Michele: #1]\(\rightarrow\)[``\emph{#2}'']}}

\def\ksmpc{${\, \mathrm{km}\, \mathrm{s}^{-1}\, \mathrm{Mpc}^{-1}}$}

\def\j14{J1433+6007}

\newcommand{\sref}[1]{Section~\ref{#1}}

\makeatletter
\newcommand{\mycustombox}[1]{%
  \fbox{%
    \begin{minipage}{\linewidth-2\fboxsep-2\fboxrule}
      \vspace{0.15cm} 
      
      \raggedleft 
      \textbf{Accepted to} \textit{Physical Review D}\\
      Draft version: \today
      
      \vspace{0.15cm}
    \end{minipage}%
  }%
}

\patchcmd{\aa@maketitle}{\fbox}{\mycustombox}{}{}
\makeatother

\makeatletter
\renewcommand*\aa@textidlineempty{\aa@headings}
\renewcommand*\aa@headfont{\small\mathversion{normal}}
\makeatother

\begin{document} 

\title{TDCOSMO XXV: Measuring $H_0$ to 6.5\% precision with quasar strong lensing and maximally flexible mass-sheet dynamics}
\author{William~Sheu\orcidlink{0000-0003-1889-0227}\inst{\ref{af:ucla}}\fnmsep\thanks{Corresponding author: wsheu@astro.ucla.edu}
    \and Tommaso~Treu\orcidlink{0000-0002-8460-0390}\inst{\ref{af:ucla}}
    \and Martin~Millon\orcidlink{0000-0001-7051-497X}\inst{\ref{af:eth}, \ref{af:geneve}}
    \and Frédéric~Dux\orcidlink{0000-0003-3358-4834}\inst{\ref{af:eso}, \ref{af:epfl}}
    \and Devon~Williams\orcidlink{0000-0002-8386-0051}\inst{\ref{af:ucla}}
    \and Shawn~Knabel\orcidlink{0000-0001-5110-6241}\inst{\ref{af:ucla}}
    \and Simon~Birrer\orcidlink{0000-0003-3195-5507}\inst{\ref{af:stony}}
    \and Pritom~Mozumdar\orcidlink{0000-0002-8593-7243}\inst{\ref{af:ucla}}
    \and Giacomo~Queirolo\orcidlink{0009-0009-8222-2865}\inst{\ref{af:montpellier}}
    \and Anowar~J.~Shajib\orcidlink{0000-0002-5558-888X}\inst{\ref{af:uchicago}, \ref{af:kavli_chicago}, \ref{af:bangladesh}}
    \and Michele~Cappellari\orcidlink{0000-0002-1283-8420}\inst{\ref{af:oxford}}
    \and Kenneth~C.~Wong\orcidlink{0000-0002-8459-7793}\inst{\ref{af:tokyo}}
    \and Ildar~M.~Asfandiyarov\orcidlink{0009-0000-9605-9979}\inst{\ref{af:ubai}}
    \and Otabek~A.~Burkhonov\orcidlink{0000-0003-1169-6763}\inst{\ref{af:ubai}}
    \and Frédéric~Courbin\orcidlink{0000-0003-0758-6510}\inst{\ref{af:iccub}, \ref{af:icrea}, \ref{af:ieec}}
    \and Shuhrat~A.~Ehgamberdiev\orcidlink{0000-0001-9730-3769}\inst{\ref{af:ubai}, \ref{af:ssu}}
    \and Sofía~Rojas-Ruiz\orcidlink{0000-0003-2349-9310}\inst{\ref{af:ucla}}
    \and Asadulla~M.~Shaymanov\orcidlink{0009-0002-7815-5945}\inst{\ref{af:ubai}}
    \and Talat~A.~Akhunov\orcidlink{0000-0001-5115-6310}\inst{\ref{af:nuu}, \ref{af:ubai}}
    }
\institute{Department of Physics and Astronomy, University of California, Los Angeles, CA 90095, USA \label{af:ucla}
\and Institute for Particle Physics and Astrophysics, ETH Zurich, Wolfgang-Pauli-Strasse 27, CH-8093 Zurich, Switzerland \label{af:eth}
\and D\'epartement de Physique Th\'eorique, Universit\'e de Gen\`eve, 24 quai Ernest-Ansermet, CH-1211 Gen\`eve 4, Switzerland \label{af:geneve}
\and European Southern Observatory, Alonso de Córdova 3107, Vitacura, Santiago, Chile \label{af:eso}
\and Institute of Physics, Laboratory of Astrophysics, Ecole Polytechnique Fédérale de Lausanne (EPFL), Observatoire de Sauverny, 1290 Versoix, Switzerland \label{af:epfl}
\and Department of Physics and Astronomy, Stony Brook University, Stony Brook, NY 11794, USA \label{af:stony}
\and Laboratoire Univers et Particules de Montpellier (LUPM), CNRS \& Université de Montpellier (UMR 5299), Parvis Alexander Grothendieck, F-34095 Montpellier Cedex 05, France \label{af:montpellier}
\and Department of Astronomy \& Astrophysics, University of Chicago, Chicago, IL 60637, USA \label{af:uchicago}
\and Kavli Institute for Cosmological Physics, University of Chicago, Chicago, IL 60637, USA \label{af:kavli_chicago}
\and Center for Astronomy, Space Science and Astrophysics, Independent University, Bangladesh, Dhaka 1229, Bangladesh \label{af:bangladesh}
\and Sub-Department of Astrophysics, Department of Physics, University of Oxford, Denys Wilkinson Building, Keble Road, Oxford, OX1 3RH, UK \label{af:oxford}
\and Research Center for the Early Universe, Graduate School of Science, The University of Tokyo, 7-3-1 Hongo, Bunkyo-ku, Tokyo 113-0033, Japan \label{af:tokyo}
\and Ulugh Beg Astronomical Institute, 33 Astronomicheskaya St., Tashkent 100052, Uzbekistan \label{af:ubai}
\and ICC-UB Institut de Ciències del Cosmos, Universitat de Barcelona, Martí Franquès, 1, 08028 Barcelona, Spain \label{af:iccub}
\and Institut Català de Recerca i Estudis Avançats (ICREA), Pg. Lluís Companys 23, 08010 Barcelona, Spain \label{af:icrea}
\and Institut dEstudis Espacials de Catalunya (IEEC), Edifici RDIT, Campus UPC, Castelldefels,
08860 Barcelona, Spain \label{af:ieec}
\and Samarkand State University, 15, University boulevard, 140104, Samarkand, Uzbekistan \label{af:ssu}
\and National University of Uzbekistan, Department of Astronomy and Astrophysics, 100174 Tashkent, Uzbekistan \label{af:nuu}
} 
\date{\hspace*{\fill} \today \hspace*{\fill}}
\abstract{
We present a blind time-delay cosmography measurement of the {Hubble-Lema\^{\i}tre} constant $H_0$ based on the quadruply imaged quasar SDSSJ1433+6007.  Our analysis combines deep \textit{Hubble Space Telescope} imaging, extended time-delay monitoring from the Wendelstein and Maidanak Observatories, and spatially resolved stellar kinematics from the Keck Cosmic Web Imager and Reionization Mapper.  We build a robust lens model to reconstruct the mass distribution and high-signal-to-noise kinematic maps to break the mass-sheet degeneracy (MSD), explicitly accounting for the lens galaxy's oblateness, rotation, and anisotropy.  Furthermore, we constrain the external convergence ($\kappa_{\rm ext}$) by characterizing the line-of-sight environment using wide-field photometry from the Dark Energy Spectroscopic Instrument (DESI) Legacy Survey data release 10.  We incorporate these constraints into our joint lensing and dynamical model, running multiple iterations to estimate random and systematic uncertainties.  Accounting for maximal flexibility of the mass-sheet transformation, and assuming a flat $\Lambda$CDM cosmology and an $\Omega_{\rm m, 0}$ prior from DESI data release 2, we infer $H_0 = 73.2^{+4.8}_{-4.7}$ \ksmpc\ (a $6.5\%$ precision), and an internal mass-sheet parameter $\lambda_{\rm int}=1.12^{+0.05}_{-0.06}$.  Notably, $\lambda_{\rm int}$ is $2\sigma$ away from unity for this system, highlighting the importance of treating it as a free parameter. Our $H_0$ measurement is consistent with the result from our 2025 milestone paper, and it will be included in our next hierarchical analysis to improve the overall precision.  Moving forward, the comprehensive pipeline demonstrated herein establishes a robust framework that can be readily applied to future strongly lensed systems to further refine cosmological constraints.
}
\keywords{Gravitational lensing: strong -- Cosmology: cosmological parameters -- Cosmology: distance scale}

\titlerunning{SDSSJ1433+6007 ``soup-to-nuts'' $H_{0}$ analysis}
\authorrunning{W. Sheu et al.}
\maketitle
\nolinenumbers

%
\section{Introduction} \label{sec:introduction}
The standard cosmological model, $\Lambda$CDM, provides a remarkably consistent description of the Universe, successfully linking the temperature fluctuations of the Cosmic Microwave Background (CMB) to the formation of large-scale structure. However, the precision of modern cosmology has revealed a discrepancy in this picture: the ``$H_0$ Tension.'' This refers to the statistically significant discrepancy between the value of the {Hubble-Lema\^{\i}tre} constant ($H_0$) predicted by early-universe physics and the value measured directly in the local (``late'') universe \citep{COSMOVERSE25}.

On the one side, the Planck collaboration infers $H_0 = 67.4 \pm 0.5$ km s$^{-1}$ Mpc$^{-1}$ by fitting the $\Lambda$CDM model to CMB power spectra \citep{Planck2020}. This measurement is incredibly precise but is model-dependent; it relies on the assumption that standard physics (e.g., standard recombination, fixed number of relativistic species) holds true from the epoch of recombination to the present day.  Other such early-universe methods include using baryon acoustic oscillations \citep[BAO;][]{desidr2} or Big Bang nucleosynthesis \citep[BBN;][]{Krolewski2025}, both of which yield measurements comparable to the \textit{Planck} measurement.

On the other side, the local distance ladder, anchored by Cepheid variables and Type Ia supernovae (SNIa), provides a direct, geometric measurement of the expansion rate. The SH0ES collaboration, utilizing \textit{Hubble Space Telescope} (\textit{HST}) photometry, reports a value of $H_0 = 73.04 \pm 1.04$ km s$^{-1}$ Mpc$^{-1}$ \citep{Riess2022}. This value is approximately $5\sigma$ higher than the \textit{Planck} prediction. Other local probes, such as those using the Tip of the Red Giant Branch (TRGB) to calibrate SNIa, have yielded values that sit between these two extremes \citep{Freedman2021, Freedman_2025}. Yet another set of measurements based on megamasers and surface brightness fluctuations are in agreement with the SH0ES value \citep{Pesce2020,Jensen2025}. An up-to-date review of recent measurements is given by \citet{COSMOVERSE25}.

The persistence of this tension, despite exhaustive searches for systematic errors in both the CMB and distance ladder analyses, suggests the exciting possibility of new physics \citep{valentino2021}. Theoretical solutions range from early dark energy and additional relativistic species to modifications of General Relativity on cosmological scales \citep{Poulin_2019, Knox_2020}. To distinguish between residual systematics and genuine physical anomalies, independent probes of $H_0$—completely disjoint from the systematics of the CMB and the calibration of the Cepheid distance ladder—are critically needed.

Strong gravitational lensing of time-variable sources serves as exactly such an independent arbiter. As first proposed by \citet{Refsdal1964}, the arrival time difference between multiple images of a lensed quasar depends on the ``time-delay distance,'' $D_{\Delta t}$, which is inversely proportional to $H_0$. This method is geometric, independent of the sound horizon scale used in CMB and baryon acoustic oscillations measurements, and does not require the rung-by-rung calibration of the distance ladder. Consequently, time-delay cosmography has matured into a competitive cosmological probe, providing constraints on $H_0$ with precision comparable to the traditional methods \citep[e.g.,][]{Suyu2010, TreuMarshall2016, Suyu2017, Wong2020, treu2023, birrer2024, tdcosmo2025}.  This is most commonly achieved by identifying and studying the time delays of strongly lensed quasar systems \citep[e.g.,][]{lemon2018, lemon2022, sheu2024}, or more rarely, strongly lensed supernovae \citep[e.g.,][]{quimby2014, Kelly2015, sheu2023, Kelly2023, pascale2025}.

In this paper, we showcase a comprehensive ``soup-to-nuts'' analysis to determine $H_0$ using a quadruply lensed quasar system SDSS\j14 (hereafter \j14), developing a pipeline for future lensed quasar studies built upon previous work by members of our team \citep{Shajib_2023,Paic2026}.  This encompasses a re-measurement of the excess time delays between quasar images (\sref{sec:td_measurements}), inspection of the nearby field imaging and line-of-sight measurement (\sref{sec:external_shear}), reconstruction of the \textit{HST} imaging with lens modeling (\sref{sec:lens_model}), reduction and analysis of the spatially-resolved kinematic information (\sref{sec:IFU_analysis}) that is used to break the mass sheet degeneracy \citep{Falco1985,Birrer_2020,B+T21}, and finally the construction of the full dynamical model of the lens galaxy to constrain cosmology (\sref{sec:dynamical_model}).  Before the analysis, we briefly review the lensing formalism used throughout the manuscript (\sref{sec:lensing_formalism}), then discuss the previous literature regarding this system (\sref{sec:j14}).  In \sref{sec:results}, we display and discuss our final cosmological results. Finally, in \sref{sec:conclusion}, we summarize our results and offer concluding remarks.  

In keeping with our collaboration protocol and to avoid potential experimenter biases, we blind our cosmological parameters throughout our analysis; only after we completed our analysis did we unblind these factors and reveal our results in \sref{sec:results}, without modification.  To enforce blindness, we abstained from printing and showing unblinded values prior to the set unblinding date\footnote{Unblinding took place during a collaboration-wide teleconference held on April 7th, 2026}, only displaying blinded values (by multiplying the sample by a random number) to probe relative uncertainties.  The parameters we blinded were $H_0$, $D_{\Delta t}$, and $D_{\rm d}$.  


\section{Lensing Formalism} \label{sec:lensing_formalism}
We adopt the standard formalism of strong gravitational lensing.  The mapping from the source plane position $\boldsymbol{\beta}$ to the image plane position $\boldsymbol{\theta}$ is governed by the lens equation:
\begin{equation}
    \boldsymbol{\beta} = \boldsymbol{\theta} - \boldsymbol{\alpha}(\boldsymbol{\theta}),
\end{equation}
where $\boldsymbol{\alpha}(\boldsymbol{\theta}) = \nabla \psi(\boldsymbol{\theta})$ is the deflection angle and $\psi(\boldsymbol{\theta})$ is the deflection potential.  The deflection potential $\psi(\boldsymbol{\theta})$ is fundamentally related to the dimensionless surface mass density, or convergence, by:
\begin{equation}
     \kappa (\boldsymbol{\theta}) \equiv \frac{1}{2} \nabla^{2} \psi (\boldsymbol{\theta}).
\end{equation}
The excess time delay of an image at position $\boldsymbol{\theta}$ relative to the case of no lensing is given by:
\begin{equation} \label{eq:tds}
    \Delta t(\boldsymbol{\theta}) = \frac{D_{\Delta t}}{c} \left[ \frac{1}{2} (\boldsymbol{\theta} - \boldsymbol{\beta})^2 - \psi(\boldsymbol{\theta}) \right],
\end{equation}
which consists of geometric and gravitational time-delay components within the square brackets.  The time-delay distance $D_{\Delta t}$ combines the angular diameter distances to the lens ($D_{\rm d}$), to the source ($D_{\rm s}$), and between them ($D_{\rm ds}$): 
\begin{equation}
    D_{\Delta t} \equiv (1 + z_{\rm l}) \frac{D_{\rm d} D_{\rm s}}{D_{\rm ds}}.
\end{equation}
In observations of a source with variable luminosity, we can only measure the excess time delay between images (positioned at $\boldsymbol{\theta_{A}}$ and $\boldsymbol{\theta_{B}}$): 
\begin{equation} \label{eq:tds_excess}
    \Delta t_{AB} = \Delta t(\boldsymbol{\theta_{B}}) - \Delta t(\boldsymbol{\theta_{A}}) \propto D_{\Delta t}
\end{equation}
Because each angular diameter distance in $D_{\Delta t}$ is inversely proportional to $H_0$, it follows that $D_{\Delta t} \propto H_0^{-1}$; therefore, measuring excess time delays alongside a robust model of $\kappa(\boldsymbol{\theta})$ yields $H_0$.  Imaging information is used to model the convergence profile of the lensing galaxy by reconstructing source structures within the lensed images.  

\subsection{The mass-sheet degeneracy}
\label{subsec:msd}

While time-delay cosmography is conceptually robust, it is subject to specific systematic uncertainties.  By far the most important of these is the mass-sheet degeneracy \citep[MSD;][]{Falco1985, S+S13}.  
The MSD arises from a transformation of the lens mass profile wherein the surface mass density $\kappa(\boldsymbol{\theta})$ can be scaled by a factor $\lambda$ and supplemented by a constant mass sheet $(1-\lambda)$, such that:
\begin{equation}
    \kappa_{\lambda}(\boldsymbol{\theta}) = \lambda \kappa(\boldsymbol{\theta}) + (1-\lambda).
\end{equation}
This mass-sheet transform (MST), in conjunction with a linear scaling of the source plane position by $\boldsymbol{\beta}_{\lambda} = \lambda \boldsymbol{\beta}$, leaves all imaging observables identical.  However, it rescales the predicted time delays and time-delay distance by $1/\lambda$, and thus the inferred $H_0$ is scaled by a factor $\lambda$.  In other words, lens modeling alone is completely degenerate with this theoretical and often unphysical mass-sheet parameter $\lambda$.  However, stellar kinematic information, such as the velocity dispersion and the root-mean-squared velocity, is not subject to the MSD, making it an invaluable tool in breaking the degeneracy.

As shown in \citet{Shajib_2023}, the true galaxy convergence profile $\kappa_{\rm gal}$, while accounting for the MSD, can be described as:
\begin{equation} \label{eq:kappa_gal}
    \kappa_{\text{gal}}(\boldsymbol{\theta}) \approx (1 - \kappa_{\text{ext}}) [\lambda_{\text{int}}\kappa_{\text{model}}(\boldsymbol{\theta}) + (1 - \lambda_{\text{int}})\kappa_{\rm S}(\boldsymbol{\theta})].
\end{equation}
Here, $\kappa_{\text{ext}}$ represents the external convergence contribution from line-of-sight mass structures, $\lambda_{\rm int}$ is the internal mass-sheet transform parameter that rescales the convergence profile $\kappa_{\text{model}}$ recovered from lens modeling, and $\kappa_{\rm{S}}$ describes a predetermined ``variable'' mass-sheet profile.  This profile behaves like a true mass sheet within the regime constrained by strong lens modeling but rapidly approaches zero at larger radii.  An appropriate functional form we adopt in this paper is given by:
\begin{equation} \label{eq:vmst}
    \kappa_{\text{s}}(\theta) = \frac{\theta_{\rm S}^2}{\theta^2 + \theta_{\rm S}^2},
\end{equation}
where the scale radius $\theta_{\rm S}$ should be large enough such that it is indistinguishable from an exact mass sheet with the images used to constrain the lens model \citep{blum2020, Birrer_2020, Shajib_2023}.  

A useful parameterization of the mass-sheet variables is:
\begin{equation} \label{eq:lambda_total}
    \lambda_{\rm total}\equiv \lambda_{\rm int}(1 - \kappa_{\rm ext}),
\end{equation}
which represents the linear transformation of the lens model convergence due to the total mass-sheet contribution (see Equation~\ref{eq:kappa_gal}).    
The external convergence can be constrained using a ground-based photometric survey of the galaxies in the lens galaxy’s environment or along the line of sight \citep{Rusu2017, Wells_2023}, while the internal mass-sheet component can be constrained by kinematic information, as previously stated.
Only by restricting these two contributions to the overall manifestation of a mass-sheet can one achieve an accurate measurement of $H_0$.  In this paper, we will account for the MSD in our dynamical modeling of \j14, using spatially resolved kinematic measurements.  

\section{SDSSJ1433+6007} \label{sec:j14}
\begin{figure}
 \includegraphics[width=1\linewidth]{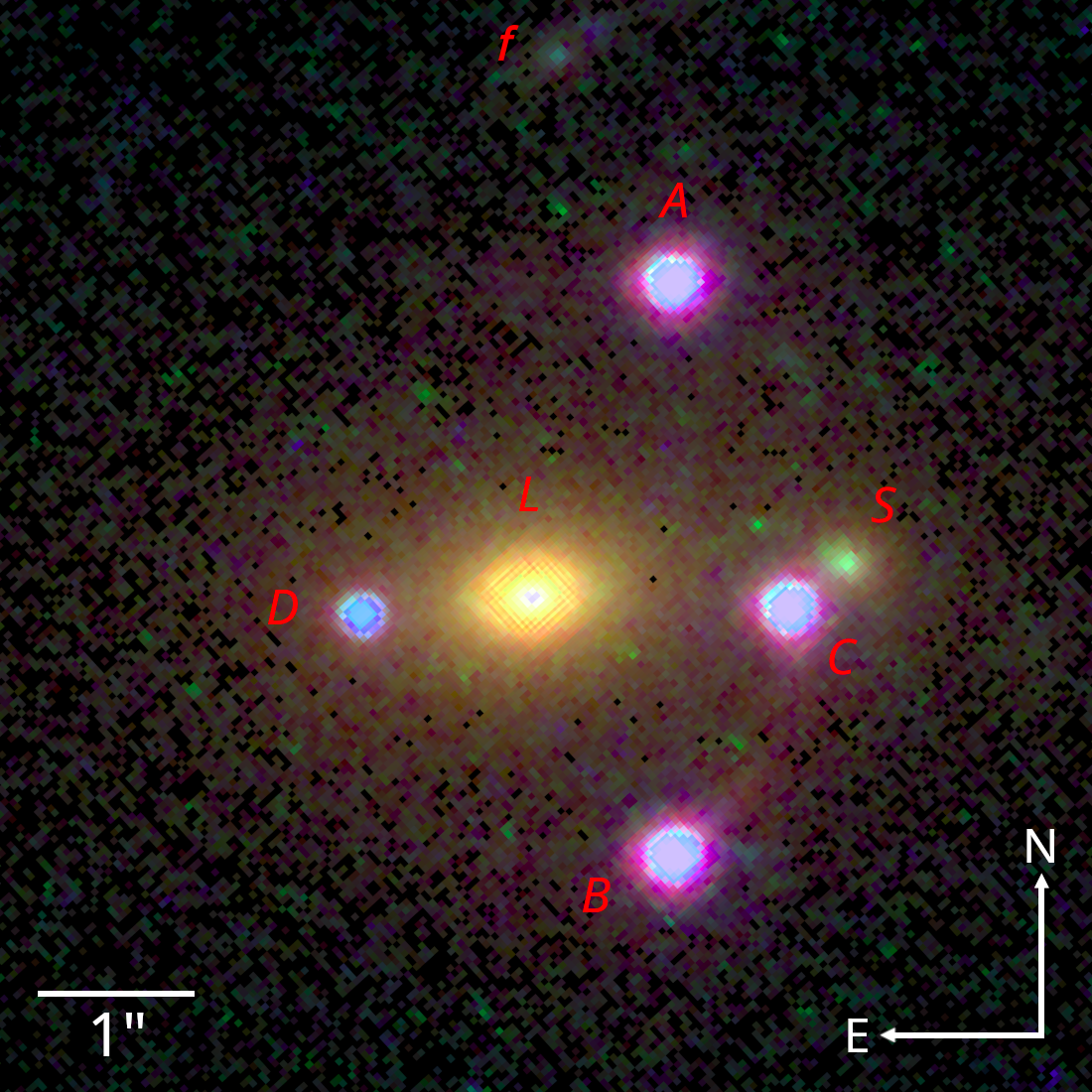}
 \caption{\textit{HST} imaging of \j14, with the lens galaxy ($L$), satellite galaxy ($S$), and the lensed quasar images ($A$, $B$, $C$, and $D$) labeled.  The faint background galaxy ($f$) is discussed in \sref{subsec:identify_perturbers}.  The colored image is generated using the \textit{HST} F160W, F814W, and F475X bands for the red, green, and blue channels, respectively.}
 \label{fig:hst_rgb}
\end{figure}

The quadruply imaged quasar system J1433+6007 was initially discovered by \citet{Agnello_2017} through an outlier-selection algorithm searching the Sloan Digital Sky Survey \citep[SDSS;][]{Abazajian_2009} DR12 catalog.  Since its discovery, observations have been made with the \textit{HST} Wide Field Camera 3 through programs 15320 (PI: T. Treu; in the F160W, F814W, and F475X filters) and 15177 (PI: A. Nierenberg; in the F105W and F140W filters).  See Figure~\ref{fig:hst_rgb} for a color \textit{HST} image of \j14 and our labeling scheme.  The system displays a classic quadruply-imaged cusp lensing orientation, despite the presence of a satellite galaxy $S$ near image $C$.  The source galaxy hosting the lensed quasar can be seen in a faint arc between images $B$ and $C$, which is visible in all \textit{HST} F160W, F814W, and F475X bands.  We find that the source galaxy light profile is very prominent in the F160W band after accounting for and subtracting out the lens galaxy light (see Figure~\ref{fig:f160w_model}).  \j14 has a lens redshift of $z_{\rm l} = 0.407$ and a source redshift of $z_{\rm s} = 2.737$, with an aperture-integrated velocity dispersion of $261\pm{6}\text{ (statistical)}\pm 7 \text{ (systematic)}$ km s$^{-1}$ measured with the Echellette Spectrograph and Imager (ESI) at the W. M. Keck Observatory \citep{Mozumdar_2023}.  From the kinematic information in \sref{sec:IFU_analysis}, we find that the satellite galaxy shares the same redshift as the main perturber.  \citet[][hereafter Q25]{q25} present over three years' worth of monitoring observations from the Wendelstein Observatory, resulting in a marginalized time-delay measurement uncertainty of $3.9\%$.  

\subsection{Improvements from literature}
\label{subsec:literature_improvements}

Using their time-delay measurements and the \textit{HST} imaging, Q25 performed a time-delay cosmography analysis on \j14, recovering a measurement of $H_0 = 71.7^{+3.9}_{-3.6}$ \ksmpc.  In this paper, we independently revisit this measurement, equipped with more data, the detailed analysis required to leverage it, and a maximally flexible model to account for MSD systematics.  Q25 did not utilize kinematic information nor line-of-sight information in their measurement of $H_0$, relying solely on their lensing model and time-delay information.  As previously discussed in \sref{subsec:msd}, this may result in a value degenerate with the MSD, from both internal and external sources.  In contrast, we use high-resolution spatially-resolved kinematics from the Keck Cosmic Web Imager \citep[KCWI;][]{KCWI2018} instrument to build a dynamical model which jointly fits the kinematic and lensing/imaging data.
Additionally, we use wide-field imaging of the environment around \j14 (up to $120''$) to probe line-of-sight substructures, and to identify possible perturbers that could contribute a significant flexion to the lens model.  We also offer an improved lens model by accounting for a more complex source light structure in the IR range.  Furthermore, we introduce an improved measurement of the time delays using additional observations from the Maidanak telescope \citep[see][]{Burkhonov_etal}.  Finally, by combining these advancements with an MSD-robust dynamical model, we can more accurately constrain $H_0$.

\section{Time-delay measurements} \label{sec:td_measurements}
\begin{figure*}
\begin{center}
 \includegraphics[width=1\linewidth]{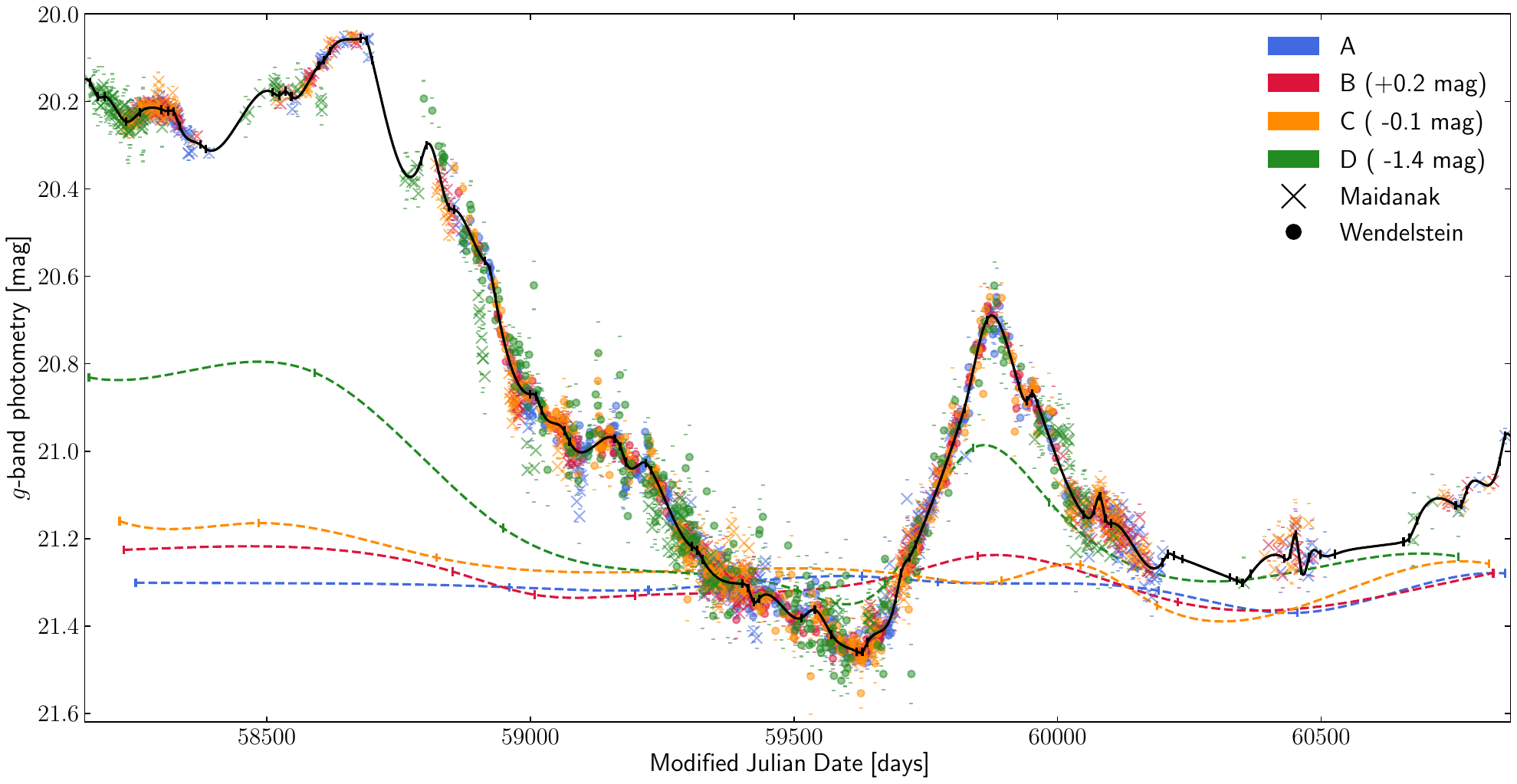}
 \caption{The full light curve for the quasar images of \j14.  The lensed images' photometries are stacked after accounting for time delay and magnification differences (given in the top-right legend).  We label measurements from the Maidanak Observatory with a cross ($\times$), and measurements from the Wendelstein Observatory with a dot ($\boldsymbol{\cdot}$).  As stated in the main text and shown in Figure~\ref{fig:td_posteriors}, we do not directly combine the different observatories' light curves in our analysis.  For this figure, we add a flux (additive) and magnitude (multiplicative) shift to match their photometries.  The light curve (in black) is then fit to the combined data, with respectively-colored microlensing trends shown as dashed line for each lensed image.  The purpose of this figure is to illustrate the overall coverage across both sets of observatory observations, highlighting the additional span provided by the new observations from the Maidanak Observatory.}
 \label{fig:td_full}
 \end{center}
\end{figure*}

Q25 presented high-cadence light curves for the lensed quasar images in \j14, derived from three years of monitoring at the Wendelstein Observatory (February 2020 to June 2023, with a mean cadence of roughly four nights). We augment that dataset with observations from the Maidanak Observatory \citep{Ehgamberdiev_2018} spanning May 2019 to June 2025. While the Maidanak light curve is more sparse—with observations approximately every seven nights—and offers only a marginal improvement to the statistical constraint, it provides crucial confirmation that microlensing was not significant during the shorter Wendelstein baseline.  This allows us to make further informed constraints on the microlensing effects of the spline light curve modeling in Q25.  See Figure~\ref{fig:td_full} for the light curve coverage provided by the Maidanak and Wendelstein Observatories.

We adopt the methodology presented by \citet{Dux_2025}, which is based on high-cadence $r$-band monitoring campaigns conducted with the 2.6-m ESO VLT Survey Telescope and the MPG 2.2-m telescope. 
Photometric calibration and data preprocessing are performed using the \textsc{lightcurver}\footnote{\url{https://duxfrederic.github.io/lightcurver/}}  pipeline \citep{Dux_2024}.  We then use \textsc{STARRED} \citep{starred0, starred1} to forward-model the quasar point sources and deblend the extended lensing galaxy emission.
This method processes all epochs simultaneously by fitting a high-resolution pixelated model of the background and point sources convolved with epoch-specific point source function (PSF) kernels derived from field stars. 
Time delays are estimated using the \textsc{PyCS3}\footnote{\url{https://cosmograil.gitlab.io/PyCS3/}} toolbox by simultaneously aligning the light curves with free-knot splines to represent intrinsic quasar variability, while polynomials or splines are employed to model extrinsic microlensing variations \citep{pycs30, pycs31, Millon_2020}.  Final uncertainties and covariance matrices are determined by applying this procedure to mock light curves containing injected red noise, thereby accounting for the degeneracies between intrinsic and extrinsic variability. 
Due to differences in the filters used at the two observatories, we do not directly combine the light curves; instead, we treat them as independent datasets and combine their time-delay constraints at the likelihood level. 

\begin{figure}
 \includegraphics[width=1\linewidth]{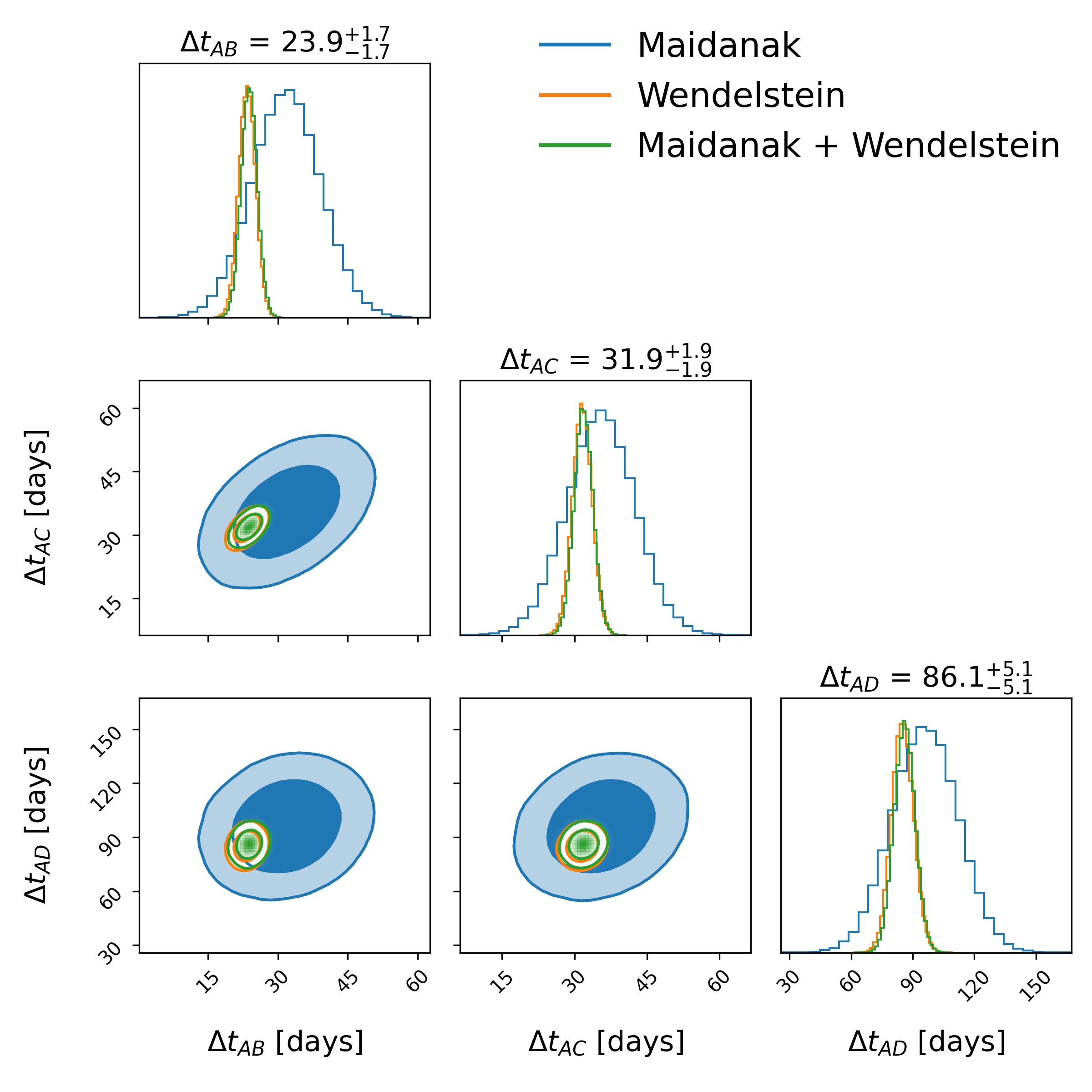}
 \caption{The time-delay measurement (relative to quasar image $A$) posteriors from the Maidanak Observatory (this work; blue), Wendelstein Observatory (Q25 with more informed microlensing assumptions; orange), and the combined likelihood distribution (green).  Both independent probes are consistent with each other, and the resulting combined posterior yields an uncertainty improved by an average factor of 1.5 compared to previous results.}
 \label{fig:td_posteriors}
\end{figure}

In Figure~\ref{fig:td_posteriors}, we present the Maidanak, Wendelstein, and combined observatories' posteriors for the time delays relative to quasar image $A$.  The total time-delay precision relative to image $A$ is $11.1\%$, $3.9\%$, and $3.6\%$, respectively.
Assuming purely independent Gaussian errors, this yields a nominal combined precision of $2.6\%$ (compared to $3.8\%$ in Q25), which improves a key component of the $H_0$ uncertainty budget (Equation~\ref{eq:tds}).  The true joint precision depends on the covariance between the relative delays, which we utilize in our lens model (see \sref{sec:lens_model}).

\section{Line-of-sight measurements} \label{sec:external_shear}
\begin{figure*}
\begin{center}
 \includegraphics[width=1\linewidth]{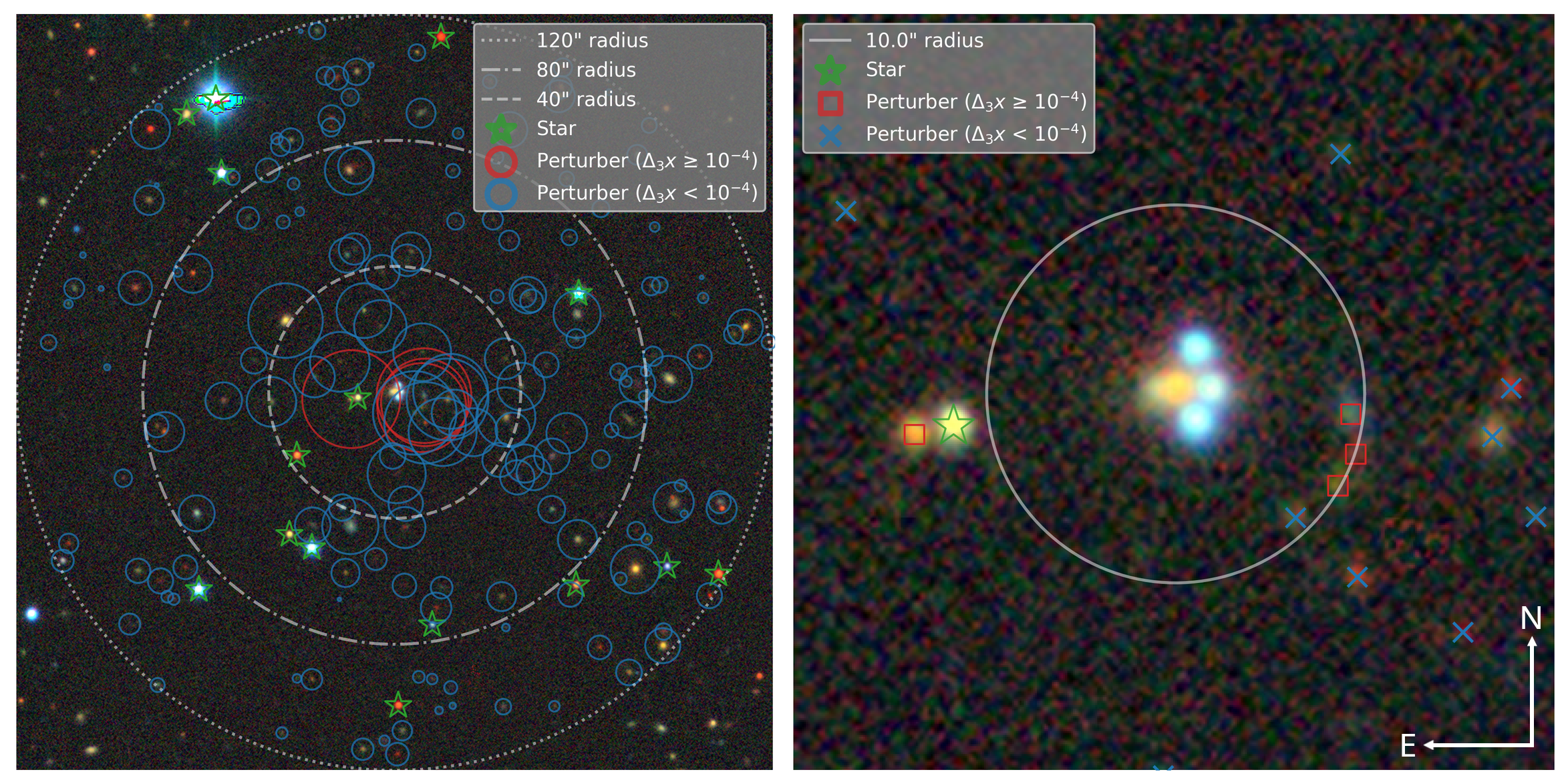}
 \caption{Ground-based imaging of the field surrounding \j14, from the DESI Legacy Surveys DR10.  The color image is generated using the stacked Mosaic-3 $z$, 90Prime $r$, and 90Prime $g$ band imaging for the red, green, and blue channels, respectively.  Left: the $120''$ field around \j14, with all objects detected by \textsc{Tractor} labeled.  Stars are labeled with green stars, and galaxies are labeled with circles, with the area of the circle being proportional to the $\log \Delta_{3}x$.  Galaxies with a flexion shift $\ge 10^{-4}$ are colored red.  Right: the $40''$ field, following the same color scheme as the left image.  However, the galaxies with a flexion shift $\ge 10^{-4}$ are labeled with a square, whereas the galaxies with a flexion shift $< 10^{-4}$ are labeled with a cross symbol ($\times$).  The four perturbers with flexion shifts $\ge 10^{-4}$ (captured in both fields) are accounted for in our lens model.}
 \label{fig:los_combined}
\end{center}
\end{figure*}

The external convergence $\kappa_{\rm ext}$, when not accounted for, acts as an external mass sheet to bias our cosmological values.  Therefore, we constrain $\kappa_{\rm ext}$ by investigating nearby galaxies along the line-of-sight (LOS) to approximate their contribution to the MST.  We use the Dark Energy Spectroscopic Instrument (DESI) Legacy Imaging Surveys data release 10 to probe structures up to $120''$ away from \j14 in the $g$, $r$, and $z$-bands \citep{Dey_2019}.  The $g$ and $r$ bands were observed with the 90Prime instrument \citep[on the Bok telescope, as part of the Baryon Oscillation Spectroscopic Survey;][]{Dawson_2012} over four exposures each filter, and the $z$ band was observed with Mosaic-3 instrument \citep[on the Mayall telescope, as part of the Mayall z-band Legacy Survey;][]{Silva2016} over three exposures.  From the stacked imaging, the $g$, $r$, and $z$ bands' extended sources limiting magnitudes ($5\sigma$) are 24.23, 23.62, and 22.89, respectively \citep{Dey_2019}.  These authors' source extraction algorithm \textsc{Tractor} photometrically catalogs these objects of interest, identifying 177 objects within the $120''$ radius (and outside a $3''$ radius) of \j14 \citep[excluding stars identified by \textit{Gaia};][]{gaia2018}.  All the objects have photometric redshift estimates from \citet{zhou2023}.

\subsection{Identifying perturbers} \label{subsec:identify_perturbers}


From the stellar mass ($M_{*}$) versus velocity dispersion ($\sigma_{\rm v}$) relation in \citet{zahid2016}, $K$-correcting the photometry according to their respective photometric redshifts, and assuming a constant $M_{*}/L$ across the field (calibrated using the lens galaxy of \j14), we estimate the velocity dispersion and uncertainty for each galaxy.  If we additionally assume a spherical isothermal sphere (SIS) lens profile, the Einstein radius can be approximated through: 
\begin{equation}
    \theta_{\rm E} \approx 4\pi\frac{\sigma^2_{\rm v}}{c^2}\frac{D_{\rm ps}}{D_{\rm s}},
\end{equation}
where $\sigma_{\rm v}$ is the velocity dispersion, $D_{\rm s}$ is the angular diameter distance from the observer to the source, and $D_{\rm ps}$ is the angular diameter distance between the perturber and the source.  We estimate these distance values using the photometric redshift distribution of the object and a fiducial $\Lambda$CDM cosmology of $H_0 = 70$~km~s$^{-1}$~Mpc$^{-1}$ and $\Omega_{\rm m, 0}=0.3$.  Our choice of a reasonable cosmology is negligible compared to the uncertainties of the photometric redshifts, when calculating their effective Einstein radius.
The second-order lensing distortion (i.e., the flexion shift $\Delta_{3}x$) can then be calculated through:
\begin{equation} \label{eq:flexion}
    \Delta_{3}x = f(\xi)\frac{\theta^2_{\rm E, main}\theta^2_{\rm E, pert}}{\delta^3},
\end{equation}
where $\theta_{\rm E, main}$ is the Einstein radius of the primary lens, $\theta_{\rm E, pert}$ is the estimated Einstein radius of the perturber, $\delta$ is the 2D angular separation between the two objects \citep{McCully_2017}. The distance ratio $\xi$ is defined as:
\begin{equation}
    \xi \equiv \frac{D_{\rm dp}D_{\rm s}}{D_{\rm p}D_{\rm ds}}, 
\end{equation}
where $D_{\rm dp}$ and $D_{\rm p}$ are the angular diameter distances between the lens and perturber, and the observer and perturber, respectively.  
Finally, $f(\xi)$ in Equation~\ref{eq:flexion} is a piecewise function defined as:
\begin{equation}
    f(\xi) \equiv \begin{cases}
        1 & \text{if } D_{\rm p} < D_{\rm d}, \\
        (1-\xi)^2  & \text{otherwise.} 
\end{cases} \end{equation}

From this analysis, we find that four objects exceed a conservative flexion shift threshold of $\Delta_{3}x \ge 10^{-4}$ \citep{McCully_2017}, and are therefore included in the lens model (see Figure~\ref{fig:los_combined}).  Their broad Einstein radius uncertainties from this procedure are also incorporated into the lens model likelihood function.  Although several assumptions are made in these measurements, this procedure allows for the identification of any possible perturbers that may cause slight Fermat potential shifts in the data, which the lens model then more accurately constrains.  Because three of them are in close spatial proximity to one another (contained within a $2''$ radius aperture, $10''$ from the main perturber) and possess similar photometric redshifts ($z_{\rm s} = 0.79$, $0.83$, and $0.71$), their combined effect on the lens model would be highly degenerate.  Therefore, we choose to aggregate these three objects before explicitly adding them into the lens model.  Their combined center is calculated as an average weighted by their respective flexion shifts, and their combined Einstein radius prior is estimated by aggregating the total flexion shift contribution of the three perturbers.

Aside from these perturbers, there is also a faint galaxy $3\farcs6$ North of the main galaxy (labeled as $f$ in Figure~\ref{fig:hst_rgb}).  While visible in the \textit{HST} imaging ($m=25.0, 24.1, 23.7$ in the F475X, F814W, and F160W bands, respectively), it is undetected in the DESI Legacy imaging.  While this galaxy is relatively close to the lens, we do not expect this galaxy to significantly affect our Fermat potential measurements.  The color, magnitude, and morphology are consistent with that of a background galaxy, which would contribute little to no flexion to our model.  However, even if it were at the lens redshift, we expect it to have a much lower $M_{*}/L$ compared to the lens and other nearby galaxies simply due to its size and dimness.  Therefore, we do not model any potential contributions from this object and mask out its light profile in the lens model.

\subsection{Measuring external convergence} \label{subsec:kappa}

\begin{figure}
\begin{center}
 \includegraphics[width=1\linewidth]{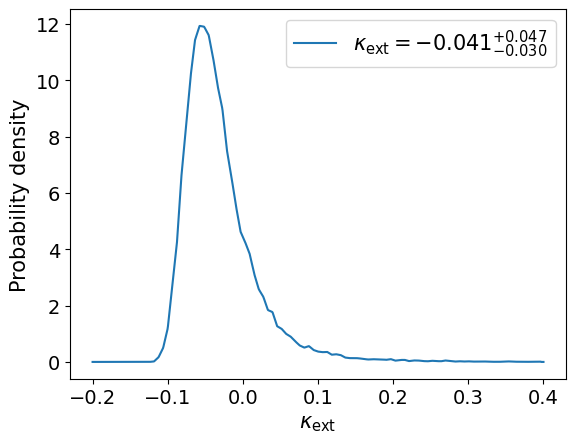}
 \caption{The probability density function of the $\kappa_{\rm ext}$ posterior.  This was obtained by using the 120$''$ LOS field imaging to measure a weighted count of galaxies of the nearby region, and generating different realizations from the cosmological $N$-body simulation Millennium simulations to probe the external convergence posterior.}
 \label{fig:kappa_pdf}
\end{center}
\end{figure}

Using the catalog of galaxies identified by the \textsc{Tractor} algorithm within the $120''$ radius field, we follow the pipeline by \cite{Wells_2023} to probe the LOS substructures.  Succinctly, the algorithm performs a count of galaxies within the catalog (weighted by criteria such as the redshifts, inverse distance, and potential), and utilizes the Millennium Simulation \citep{Springel2005} to generate fields similar to the weighted counts observed.  In the pipeline and catalog, we impose a 23-magnitude cutoff to improve completeness, as it is well below the detection magnitudes for the 90Prime filters.  From these simulations, we can infer the total convergence contribution of these LOS structures as $\kappa_{\rm ext}$, where for \j14, we calculate a posterior of $\kappa_{\rm ext} = -0.041^{+0.047}_{-0.030}$ (see Figure~\ref{fig:kappa_pdf} for the full posterior).  We find that \j14 lies in a field that is slightly underdense compared to the average cosmic density ($\kappa_{\rm ext} = 0$), though our measurement is still consistent within $1\sigma$ of being underdense, average, and overdense.  

\section{Lens model} \label{sec:lens_model}
\begin{figure*}
\begin{center}
 \includegraphics[width=1\linewidth]{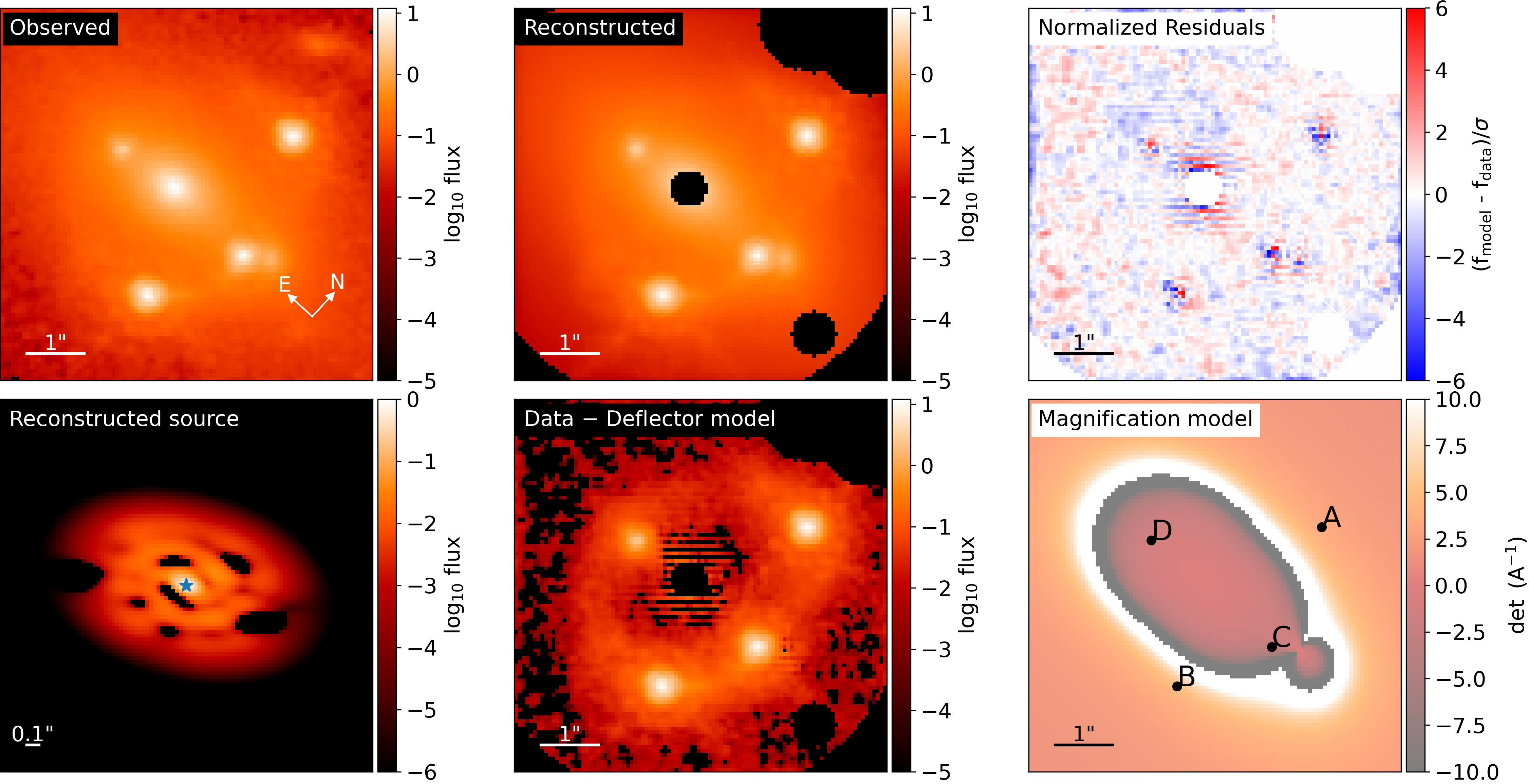}
 \caption{Representative plots illustrating the lens model of \j14, and its reconstruction of the \textit{HST} F160W filter data.  The top left panel shows the \textit{HST} F160W observed image.  The top middle panel shows the reconstructed image from our best-fit model.  The top right panel shows the normalized residuals that are minimized in fitting for the model.  The bottom left panel shows the reconstructed source of the best-fit model, with a star denoting the location of the point source.  The bottom middle panel shows the \textit{HST} observed image subtracted by the best-fit light profiles of the lens and satellite galaxies, illustrating the lensed arcs.  The bottom right panel shows the magnification map of the lensing profile of the best-fit model, with the lensed point source locations labeled.  We measure a $\chi^2_{\nu=8813}$ of 0.76 for the best-fit model to the F160W data.}
 \label{fig:f160w_model}
\end{center}
\end{figure*}

Using \textit{HST} imaging in the F160W, F814W, and F475X filters (PID: 15320, PI: T. Treu), we construct a precise model of the lens galaxy 2D potential via flux based reconstruction.  While there are also F105W and F140W band exposures publicly available for this system (PID: 15177, PI: A. Nierenberg), we choose to focus only on the three aforementioned bands because they have significantly higher exposure times (1500 to 2000 seconds, compared to $<500$ seconds), and any additional features revealed in these IR bands are already emphasized in the F160W band.  The source galaxy light profile, which is important in constraining the lens model, is prominent within the F160W band (after accounting for the lens light profile; see Figure~\ref{fig:f160w_model}), whereas it is faintly visible in the F814W and F475X bands.  The F160W band is then Nyquist subsampled to a pixel size of $0\farcs065$ (from a native pixel size of $0\farcs13$), as the four well-dithered exposures allowed for this higher resolution.  The F814W and F475X bands were left at a pixel size of $0\farcs04$, as these bands do not contribute nearly as much lensing information compared to the F160W band.  We use the software \textsc{Lenstronomy}\footnote{\url{https://lenstronomy.readthedocs.io/en/latest}} \citep{birrer2018, lenstronomyII} to construct our lens model. It has been shown to be robust in the Time-Delay Lens Modeling Challenge \citep[TDLMC;][]{Ding2021}, where two independent teams used \textsc{Lenstronomy} to recover lens model parameters within statistical consistency (in Rung 2).  \textsc{Lenstronomy} has since been instrumental for its reliable lens modeling capabilities within the TDCOSMO collaboration, and has been extensively tested both internally and externally \citep[e.g.,][]{Shajib_2022, Schmidt_2022, Ertl_2023, Sheuc_2024}.

\subsection{Model specifications} 
We construct a model that jointly fits for all three \textit{HST} bands, using \textsc{Lenstronomy} to constrain the same deflection model while varying certain light profile parameters between bands.  All flux normalization amplitudes are linearly fit after the nonlinear parameters (e.g., deflection profile parameters, light profile parameters related to position and shape) are sampled.  For the deflection profile, we use an elliptical power-law \citep[EPL;][]{Tessore_2015} profile for each of the main deflector and the satellite, an external shear component, and SIS profiles for the perturbers with flexion shifts $\ge 10^{-4}$.  We also include broad priors on the satellite and perturbers' Einstein radii estimated from their luminosities ($K$-corrected from the \textit{HST} F160W and the 90Prime photometry, respectively; see \sref{subsec:identify_perturbers}).  As we are allowing for a flexible mass sheet in our dynamical analysis, we only account for a power-law profile in our lens model.  The systematics introduced by a composite lens profile are already accounted for by our implementation of a maximally conservative mass sheet \citep[e.g.,][]{tdcosmo2025, Paic2026}.  The lens galaxy and satellite are positioned at the same redshift (as confirmed by our integral field spectroscopy data in \sref{sec:IFU_analysis}) of $z_{\rm l} = 0.407$.  

As discussed in \sref{subsec:identify_perturbers}, we use two SIS profiles to account for the nearby perturbers that have exceeded the flexion shift threshold of $\ge 10^{-4}$.  The aggregated perturber is set at the average photometric redshift (weighted by their inverse variance) of $z_{\rm p}=0.76$, and the isolated perturber is set at its measured photometric redshift of $z_{\rm p}=0.48$.  We incorporate the uncertainty of their photo-$z$ measurements into the Einstein radius prior uncertainty, which dominates the error budget compared to the photometry uncertainties.  

To model the lens-plane light of the main deflector, we employ two elliptical Sérsic profiles \citep{sersic} in each band: an exponential profile (Sérsic index of 1) and a de Vaucouleurs \citep{vaucouleurs1948} profile (Sérsic index of 4). Because the deflector's emission in the F814W and F475X bands is significantly fainter and less complex than in the F160W band and since the F160W center is masked due to correlated noise (discussed below), we jointly constrain the lens light centroids across all bands and fix them to the center of the main deflector's EPL profile. Furthermore, we tie the axis ratios of the two Sérsic components together for the F814W and F475X bands, as their simpler light distributions do not warrant an additional degree of freedom. Although the F160W band provides the strongest lensing constraints, we adopt the F475X light model for our dynamical analysis; its wavelength range ($\sim4000$ to $6000$ \AA) closely matches our spectral coverage (see \sref{subsec:spect_reduction}). For the satellite galaxy, we fit a single Sérsic profile per band, fixing its center and ellipticity to its corresponding EPL profile. Finally, the lensed quasar images are modeled using a point spread function (PSF). For each band, this PSF is initially constructed by stacking nearby stars in the \textit{HST} field and then iteratively reconstructed at the pixel level.
Our choice to Nyquist subsample the F160W band allows us to achieve higher resolution within the arcs and reduce large-scale aliasing artifacts.  However, this in turn introduces correlated noise within the brightest regions of the image.  Therefore, we opt to add masks near the center of the lens galaxy in the F160W model.  We note that the F160W band is most informative to the lens model because of its ability to resolve the source arcs (as the arcs are inherently brighter in the observed IR band); a robust model is achieved by combining this with the astrometry-constraining power of the optical F814W and F475X bands; this is achieved by iteratively aligning the astrometry between all three filters.  

\begin{table*} 
\begingroup
\renewcommand{\arraystretch}{1.25}
\begin{center}
 \caption{Description and posteriors of the parameters sampled by the lens model, which are passed into the dynamical model in \sref{sec:dynamical_model}.  The lens model is constrained by the \textit{HST} F160W, F814W, and F475X imaging data ($\mathcal{D}_{\rm I}$), as well as with the time-delay information ($\mathcal{D}_{\rm T}$).  The $D_{\Delta t}$ parameter in our lens model was blinded during our analysis.}   
 \label{tab:lens_model}
 \begin{tabular}{cclc}
  \hline
Parameter & & Description & Lens model posterior $p(\omega|\mathcal{D}_{\rm I}, \mathcal{D}_{\rm T})$ 
  \\
  \hline
$\theta_{\rm E}$ & [$''$] & Einstein radius & $1.538^{+0.005}_{-0.007}$ \\
$\gamma$ &  & Logarithmic slope of the power-law convergence profile & $1.954^{+0.015}_{-0.015}$ \\
$q_{\rm m}$ &  & Minor-to-major axis ratio of the power-law convergence profile & $0.754^{+0.006}_{-0.006}$ \\
$R_{\rm eff}$ & [$''$] & Effective radius of the \textit{HST} F475X light profile & $3.03^{+0.17}_{-0.13}$ \\
$q_{\rm l}$ &  & Minor-to-major axis ratio of the \textit{HST} F475X light profile & $0.70^{+0.01}_{-0.01}$ \\
${D}_{\Delta t}$ & [Mpc] & Time-delay distance  & $2317^{+111}_{-111}$ \\ 

\hline
\end{tabular}
\end{center}
\endgroup
\end{table*}

We use an exponential profile to describe the source light for each band.  Because the F160W band holds the most information regarding the source galaxy light, we also include a component modeled using a shapelet basis set for its source galaxy light profile \citep{shapelets1, shapelets2}.  This allows us to reconstruct and incorporate sub-features within the source galaxy profile, providing a more accurate and constraining fit.  We use shapelet orders of $n_{\rm max}=9$, 10, and 11 (see \sref{subsec:systemaics}), which we find are sufficiently complex to capture the details within the source.  The center of the shapelets basis is fixed to the center of the exponential profile, which is shared across all three of the source exponential profiles (across filters).  

We fit for the lensed quasar images individually on the image plane, and include Gaussian priors (on the source plane) to ensure that the lensed image positions, when traced back to the source plane, are close ($\sigma=0\farcs0003$) to the center of the exponential light profile.  We also allow a deviation from the predicted quasar image position on the image plane (with a prior of $\sigma=0\farcs001$) for each image to account for slight instrumental or systematic differences.

\subsection{Incorporating time delays}

To better inform our lens model, we include the time-delay measurements into the modeling process as an additional multivariate Gaussian prior into the likelihood function, derived from the time-delay covariance matrix (see Figure~\ref{fig:td_posteriors}).  However, as the time delays are expected to be directly proportional to $D_{\Delta t}$ in addition to the Fermat potential, we must also sample and constrain $D_{\Delta t}$ in our pipeline. The resulting posterior on $D_{\Delta t}$ from our lens model is then marginalized over in the dynamical model (\sref{sec:dynamical_model}).  

Our procedure for lens modeling consists of using a particle swarm optimization operation \citep{Kennedy1995} to locate a maximum of the lens likelihood function that is likely close to the global maximum, followed by a Markov chain Monte Carlo (MCMC) pipeline to obtain the posterior of the model parameters \citep{emcee}.  The chains are then run until convergence is reached, before extracting the final posteriors.

\subsection{Systematic tests} \label{subsec:systemaics}

We run the pipeline modeling with the three different shapelets orders of the F160W source light profile ($n_{\rm max}=9$, 10, and 11).
We also test two marginally different masks for the F160W image to probe additional systematics, varying the full aperture radius of the imaging data (from the lens galaxy center) from 60 pixels to 61 pixels ($\Delta r\approx0\farcs065$) while slightly increasing the masked region North of image $A$ to ensure that the total number of unmasked pixels is conserved.  These tests result in six different model configurations.
We find that the models are consistent with each other within the uncertainties. In order to account for these different choices in the error budget, the distribution used in our dynamical fit is a simple combination of the resulting chains of each model.  We generate our final posterior distributions by sampling each set of these systematic chains, using their Bayesian information criteria (BIC) to calculate their respective weights.

From our lens modeling, we pass the six parameter posteriors ($\omega \in \{\theta_{\rm E}, \gamma, q_{\rm m}, R_{\rm eff}, q_{\rm l}, \widetilde{D}_{\Delta t}\}$) into our dynamical modeling to describe the mass model, as well as the best-fit variables describing the F475X lens light profile shape.  In Table~\ref{tab:lens_model}, we present the lens model posterior results for these parameters.  The full lens model posteriors are provided in Appendix~\ref{app:A} (Tables~\ref{tab:full_lens_params},~\ref{tab:full_lens_light_params}, and~\ref{tab:full_source_params}).  The lens model and residuals for the best-fit model (with shapelets order of $n_{\rm max}=10$ in the F160W filter) are shown in Figure~\ref{fig:f160w_model}, resulting in a reduced chi-square of $\chi_{\nu=52695}^2 = 1.06$.  We also show the lens modeling results in the F814W and F475X filters in Appendix~\ref{app:A} (Figures~\ref{fig:f814w_model} and~\ref{fig:f475x_model}).  

\section{Integral field spectroscopy} \label{sec:IFU_analysis}
\begin{figure*}
\begin{center}
 \includegraphics[width=1\linewidth]{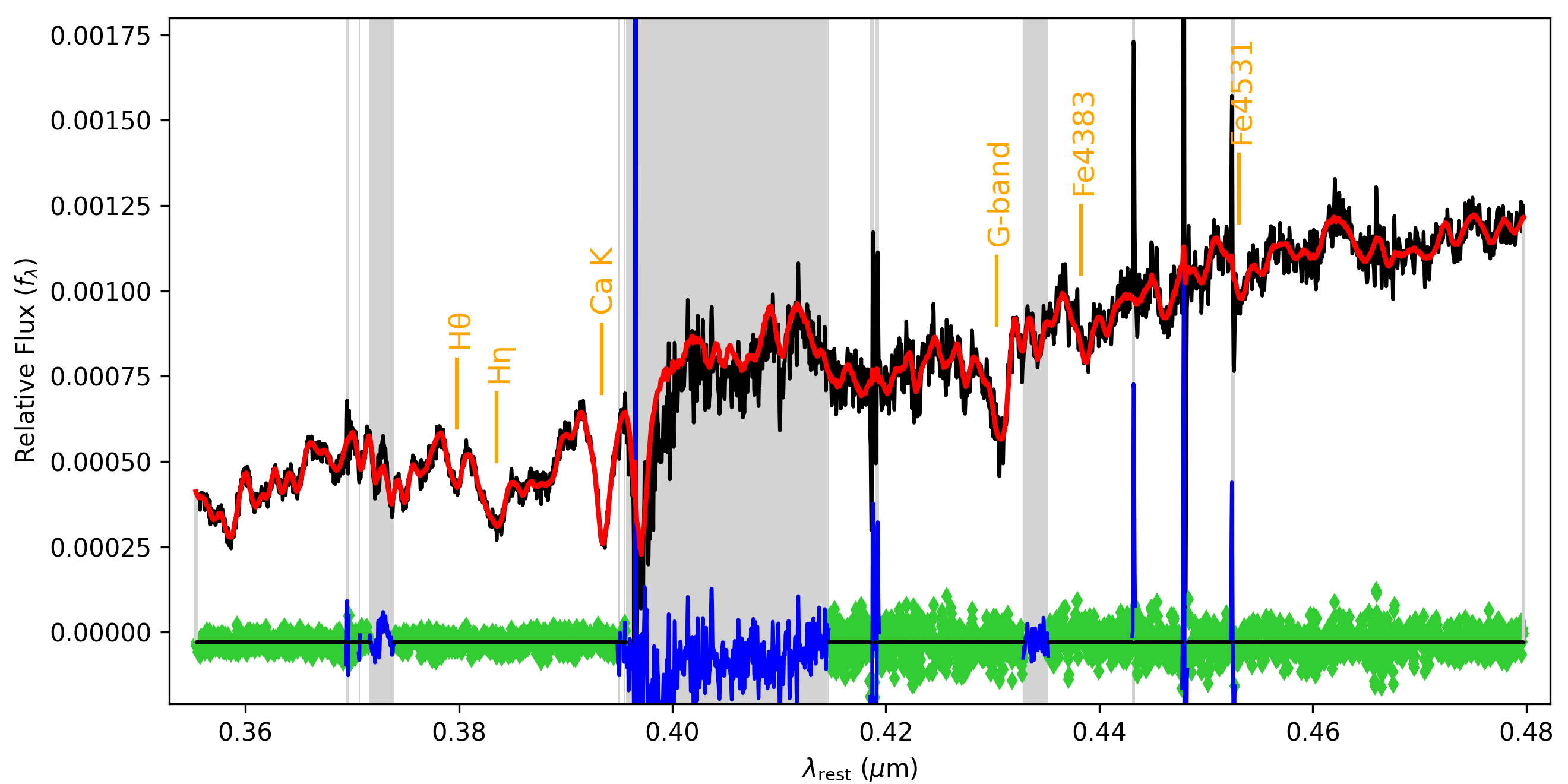}
 \caption{The $1''$ aperture spectrum of the lens galaxy of \j14.  The data is in black, the \textsc{pPXF} fit (using the \textsc{MILES} stellar library) is in red, and the residuals are in green.  The shaded region represents the wavelength ranges excluded from the fit, and the blue shows the data points that are being masked out.  Prominent absorption lines in the spectra have been annotated in orange.  The \textsc{pPXF} model fits the data remarkably well ($\chi^2_{\nu = 2755}=0.98$), and there are strong absorption lines to tightly constrain the velocity dispersion at $\sigma_{\rm v} = 260 \pm 5$ km s$^{-1}$ (statistical).}
 \label{fig:aperture_spec}
\end{center}
\end{figure*}

The KCWI instrument \citep{KCWI2018} on the W. M. Keck Observatory is a powerful integral field spectrograph located on the summit of Maunakea, Hawaii.  The main KCWI instrument can be concurrently processed by a blue channel (KCWI-blue) and red channel \citep[the Keck Cosmic Reionization Mapper or KCRM;][]{kcrm}.  Our configuration uses a $1\times 1$ binning, a dual amp readout, and the smallest integral field unit (IFU) slicer size to maximize its spatial resolution.  The KCWI-blue channel has a wavelength range from 3600 to 5600 \AA, while the KCRM channel has a range of 5600 to 8850 \AA.  The red channel has the added disadvantage of being more susceptible to cosmic ray contamination; therefore, KCRM observations are taken three at a time and median-coadded to compensate.  KCRM also suffers from a slightly lower signal-to-noise ratio (S/N) compared to the KCWI-blue channel.  However, despite this, the KCRM data well constrains prominent absorption lines outside of the KCWI-blue coverage
(see Figure~\ref{fig:aperture_spec}).  We choose to mask a region between the two channels (from $5569$ to $5834$ \AA\,observer frame) to avoid any distortions in the vicinity of the dichroic  Over two nights on May 7th 2024 and June 8th 2024, we observed \j14 for 8 hours on KCWI-blue, and 7.25 hours on KCRM excluding overhead times (PID: U077, PI: T. Treu). 

While the KCWI slicer (with $1\times 1$ binning) has a native, rectangular pixel size of $0\farcs35\times 0\farcs147$, we drizzle and subsample to a square pixel size of $0\farcs1334$ using our well-dithered observing pattern.  The KCWI-blue side has a spectral resolution of $R=3600$, while the KCRM side has a spectral resolution of $R>2000$.  We also truncate the full spectral range to 5000 to 6750 \AA\ in the observed frame (3554 to 4797 \AA\ rest frame) to avoid telluric lines and dichroic uncertainties at the edges of the instruments' wavelength ranges.  This 1750 \AA\ observed range captures plenty of prominent absorption lines used for constraining the lens galaxy kinematics.  In Figure~\ref{fig:aperture_spec}, we present the 1$''$ diameter aperture spectrum of the lens galaxy, along with the fitted model, masked region, and strong absorption lines present. 

\subsection{Spectral reduction and fitting} 
\label{subsec:spect_reduction}

\begin{figure*}
\begin{center}
 \includegraphics[width=1\linewidth]{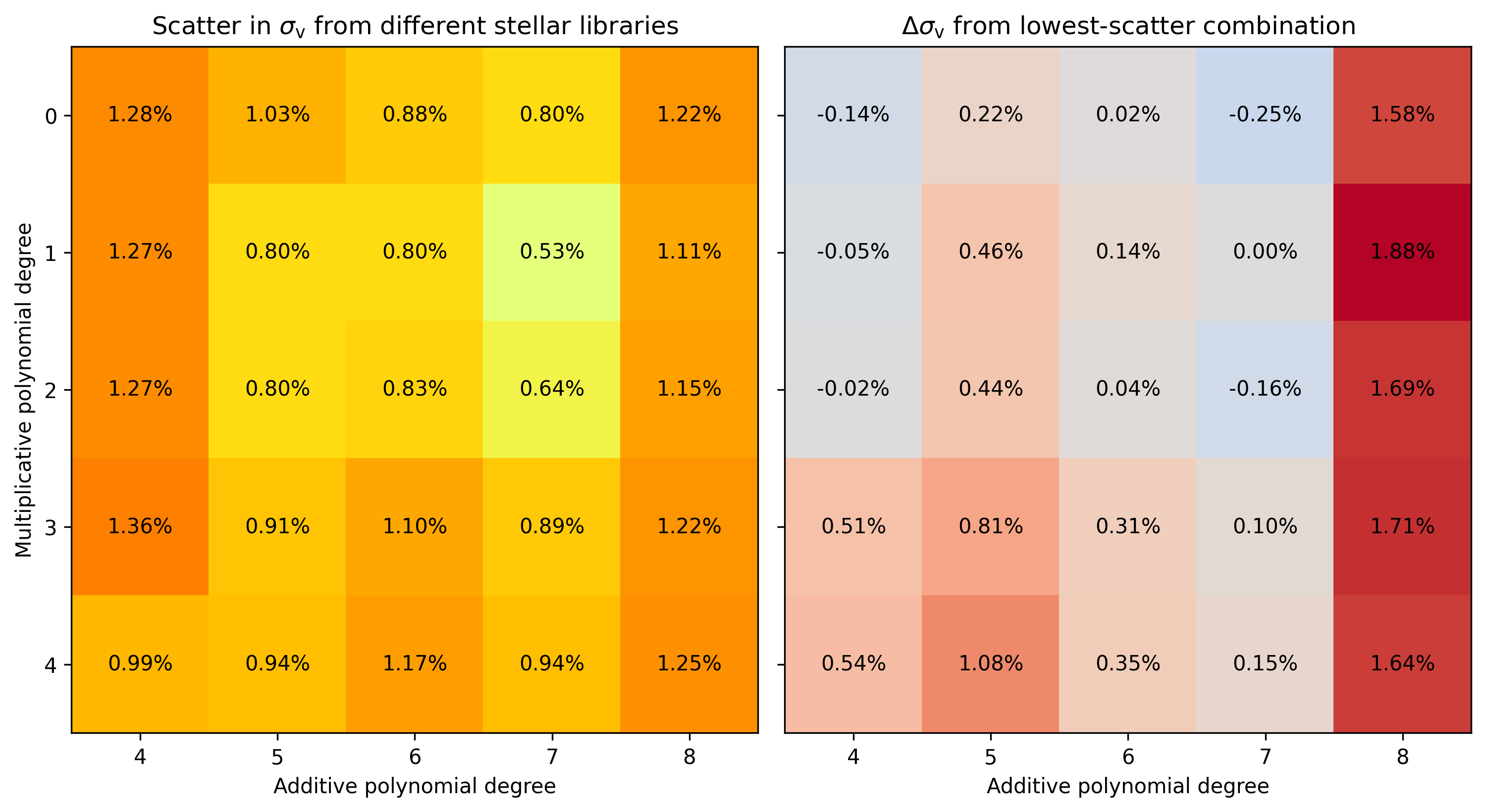}
 \caption{Comparison plots to gauge the effects of different additive and multiplicative polynomial orders on the measurement of $\sigma_{\rm v}$ across three stellar libraries.  Left: The effect of different polynomial orders on the BIC-weighted scatter of $\sigma_{\rm v}$ across stellar libraries.  We find that additive and multiplicative degrees of seven and one, respectively, best reduce the scatter between stellar libraries.  Hence, we use this configuration moving forward.  Right: The effect of different polynomial orders on the BIC-weighted mean of $\sigma_{\rm v}$ across stellar libraries, relative to the measurement at the smallest-scatter configuration.  We observe a clear overfitting regime at an additive degree of eight.  If we exclude this regime, $\Delta {\sigma}_{\rm v}$ between different polynomial degrees ranges from $-0.25\%$ to $+1.08\%$.}
 \label{fig:mva}
\end{center}
\end{figure*}

\begin{figure*}
\begin{center}
 \includegraphics[width=1\linewidth]{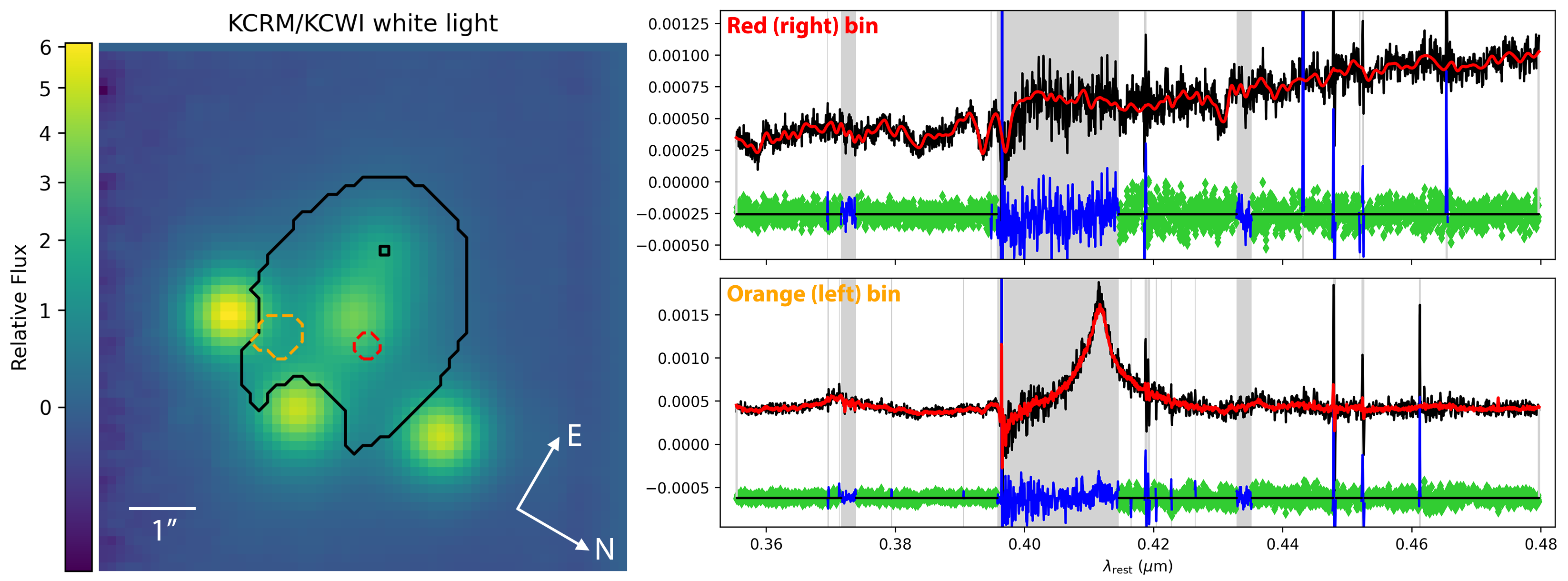}
 \caption{Plots describing the KCRM/KCWI integral-field data. Left: the white-light image integrated from 5000 to 6750 \AA\ (observer frame).  The solid black outline describes the full extent of our kinematic map, and the dashed red and orange outlines denote specific bins from our final kinematic map for which we show their spectra to the right.  Right: the 1D spectra integrated over the spaxels in their respective region on the left.  Both spectra plots follow the same layout as Figure~\ref{fig:aperture_spec}, using \textsc{pPXF} and the \textsc{MILES} stellar library.  The red bin (top right) shows the spectrum near the lens galaxy center, relatively isolated from the quasar light, while the orange bin (bottom right) shows the spectrum near quasar image $B$.  Despite the quasar light contamination in the orange bin, a decent fit to the lens galaxy kinematics is achieved.}
 \label{fig:kcwi_white_light}
\end{center}
\end{figure*}

In Figure~\ref{fig:aperture_spec}, we show the integrated spectra and fit of \j14 within an aperture of $1''$ diameter.  
We use the \textsc{pPXF}\footnote{\url{https://pypi.org/project/ppxf}} package  \citep{Cappellari2017, Cappellari2023} for all kinematic fitting of the spectra in this analysis; an example is illustrated in Figure~\ref{fig:aperture_spec}.  
Additionally, because some spaxels are in close proximity to the lensed quasar images, we use the brightest spaxels of quasar images $A$, $B$, and $C$ as additional stellar templates to fit for the quasar contribution to the spectra, following \citet{Shajib_2023}.  We mask out $>3\sigma$ outliers in the spectral fit, as well as the H$\gamma$, H$\delta$, and H$\kappa$ lines, to avoid contamination from nebular gas emission and thermal broadening. We allow H$\theta$ and H$\eta$ to remain in the fit, as they are useful stellar absorption lines and their nebular emission should be negligible.  From probing a $0\farcs5$ diameter aperture of the satellite galaxy using this procedure, we have also confirmed that it lies at the same redshift as the main perturber.

To analyze the data while properly accounting for both statistical and systematic uncertainties of the velocity dispersion $\sigma_{\rm v}$, we  follow the procedures laid out by \citet{Knabel2025}.  As the line-of-sight velocity $v_{\rm LOS}$ measurement is expected to be fairly simple and robust (i.e., the centers of absorption features), we do not probe the systematic uncertainties for $v_{\rm LOS}$.  \citealp{Knabel2025} has shown that the main systematic uncertainty contribution comes from differences in stellar libraries.  Therefore, we experiment with different stellar libraries (\textsc{Indo-US}, \citealt{indous04}; \textsc{X-Shooter}, \citealt{xshooter22}; and \textsc{MILES}, \citealt{miles11}), in addition to different additive polynomial orders (from four to eight) and multiplicative polynomial orders (from zero to four) in our \textsc{pPXF} fits.  
We note that each of these stellar libraries has been ``cleaned'' of irregular spectra which would contribute to nonphysical features in the fit.
Firstly, we vary different combinations of additive and multiplicative polynomial orders, and measure the BIC-weighted variance of $\sigma_{\rm v}$ across the three stellar libraries on the $1''$ aperture spectra.  We find that an additive polynomial order of seven and a multiplicative polynomial order of one minimize the scatter across libraries (see Figure~\ref{fig:mva}).  The overall effect on the BIC-weighted mean value of $\sigma_{\rm v}$ (with respect to an additive polynomial order of seven and a multiplicative polynomial order of one) ranges from $-0.25\%$ to $+1.88\%$, with the higher end attributed to an over-fitted regime at an additive polynomial order of eight.  If we exclude the plausible over-fitted regime of an additive polynomial order of eight, this range shrinks to $-0.25\%$ to $+1.08\%$, a percent-level accuracy.

The analysis of the datacube involves binning the spaxels into Voronoi bins such that their total S/N are comparable between bins.  This is done using the \textsc{Python} package \textsc{Vorbin}\footnote{\url{https://pypi.org/project/vorbin}} \citep{Cappellari2003}.  However, to assess the systematic variance and covariance between bins, we firstly only use six bins each with an S/N~$\approx 30$~\AA$^{-1}$.  This way, the statistical contribution to the uncertainties becomes negligible, and the systematic contribution can be accurately probed.  Using the additive and multiplicative polynomial orders that minimize $\sigma$ scatter (seven and one, respectively), we measure the $\Delta$BIC between the different stellar libraries by bootstrapping these six bins and calculate the libraries' associated weights $w$, as described by \citealp{Knabel2025}. 
For the \textsc{Indo-US}, \textsc{X-Shooter}, and \textsc{MILES} stellar libraries, we measure $\Delta$BIC values of $3.4 \pm 16.9$, $18.9 \pm 25.5$, and $0.0 \pm 16.6$, and weights of $0.39$, $0.16$, and $0.45$, respectively.  Using $w$ to weight each library's measurements of $\sigma$, we calculate the covariance matrix between the six Voronoi bins and find an average off-diagonal square root covariance of $0.88\%$ and an average diagonal standard deviation of $1.20\%$.

Using the white-light image of the datacube (see Figure~\ref{fig:kcwi_white_light}), we fit for the KCWI/KCRM PSF, which needs to be accounted for in our dynamical model \citep[e.g., see][for the potential uncertainties associated with errors on the PSF]{FKM2026}.  From the white-light image, we forward-model a simulated image using an exponential and a de Vaucouleurs profile to model the lens, and four point sources for the quasar images.  We do not subtract the source galaxy flux because, based on the \textit{HST} F475X band (which shares a very similar wavelength coverage to the KCWI/KCRM region we are probing), the source flux is negligible (see Figure~\ref{fig:f475x_model}).  We also mask out any flux contribution from the satellite galaxy, and fit for the PSF with two concentric 2D Gaussian distributions.  This configuration should well-replicate a Moffat PSF profile \citep{Moffat1969}, capturing the core and wings of the light as distinct Gaussian profiles.  

\subsection{Kinematic maps} \label{subsec:kinematic_maps}

\begin{figure*}
\begin{center}
 \includegraphics[width=1\linewidth]{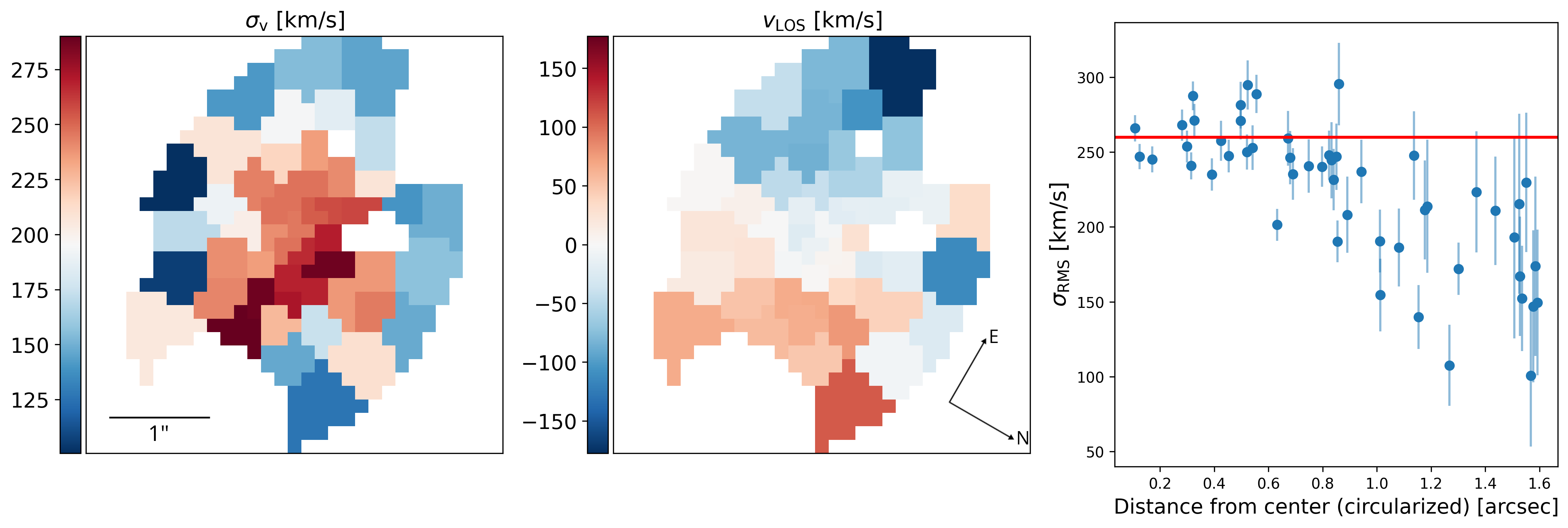}
 \caption{Kinematic map for \j14 using KCRM/KCWI.  Left and middle: the velocity dispersion and the line-of-sight velocity maps, respectively.  There is clear rotation displayed in the $v_{\rm LOS}$ map which aligns well with the light profile major axis ($\Delta$PA$=15.7\pm 8.5$ degrees), indicating an oblate symmetry.  Right: the root-mean-squared velocity versus the distance between the flux-weighted bin center and the center of the lens, circularized to account for its light profile ellipticity.  The red horizontal line denotes the $1''$ aperture velocity dispersion measurement of 260~km~s$^{-1}$.  The kinematic map for $\sigma_{\rm RMS}$ is utilized in our dynamical model described in \sref{sec:dynamical_model}.} 
 \label{fig:kinematic_maps_2}
\end{center}
\end{figure*}

In our final kinematic maps, we bin to an S/N $\approx 10$~\AA$^{-1}$, up to approximately the elliptical Einstein radius, resulting in 55 Voronoi bins.  We also exclude spaxels near quasar images $B$, $C$, and $D$ to avoid cases where the quasar light and associated Poisson noise overwhelm the lens galaxy signal.  Additionally, we also exclude spaxels near the satellite galaxy, which may bias kinematic measurements.  See Figure~\ref{fig:kcwi_white_light} for the full binned region, as well as a set of representative binned-spectra fits.  We also exclude four outlier bins with $\sigma_{\rm RMS} > 300$~km~s$^{-1}$ measurements, which we ascribe to a combination of quasar light contamination, a low signal-to-noise ratio in the deflector, and spurious noise spikes coinciding with critical absorption lines.  Figure~\ref{fig:kinematic_maps_2} illustrates our final kinematic maps for $\sigma_{\rm v}$, $v_{\rm LOS}$, and $\sigma_{\rm RMS}$, where $\sigma_{\rm RMS} \equiv \sqrt{\sigma_{\rm v}^2 + v_{\rm LOS}^2}$.  

We add in quadrature the systematic uncertainties and covariance we measured previously to the statistical uncertainties, measured from \textsc{pPXF}, for each bin.  Most remarkable is the $v_{\rm LOS}$ map, which shows that the \j14 lens galaxy is a rotating elliptical galaxy.  By measuring the specific angular momentum of $\lambda_{\rm R} = 0.149$ \citep[defined in][]{Emsellem_2007}, we find that this system is very near the nominal border distinguishing fast from slow rotating galaxies in \citet{Cappellari2016}.  However, given that \j14 is a massive elliptical galaxy lens (suggesting a density-supported rather than rotationally-supported structure), strong evidence indicates that this is likely a slow-rotating system with a large inclination angle \citep{knabelslac}.  From measuring the kinematic position angle using the \textsc{Python} package \textsc{PaFit}\footnote{\url{https://pypi.org/project/pafit}}, we find that it aligns well with the light position angle ($\Delta{\rm PA} = 15.7 \pm 8.5$ degrees) and hence has an oblate symmetry, with a $99.97\%$ certainty \citep{Krajnovic2006}.  The $\sigma_{\rm RMS}$ map, which tracks the second moment of the line-of-sight velocity, is the constraint used in our dynamical model.

\section{Dynamical model} \label{sec:dynamical_model}
\begin{table*} 
\begin{center}
 \caption{Description and priors of the parameters sampled by our dynamical model.}   
 \label{tab:pop_names}
 \begin{tabular}{ccll}
  \hline
  Parameter & & Description & Dynamical model prior $p(\pi)$ 
  \\
  \hline
$\theta_{\rm E}$ & [$''$] & Einstein radius of the convergence profile & $\sim \mathcal{U}(0, 10)$ \\
$\gamma$ &  & Logarithmic slope of the convergence profile & $\sim \mathcal{U}(0, 5)$ \\
$q_{\rm m}$ & & Minor-to-major axis ratio of the convergence profile & $\sim \mathcal{U}(0.05, 1)$ \\
$R_{\rm eff}$ & [$''$] & Effective radius of the \textit{HST} F475X light profile & $\sim \mathcal{U}(0, 20)$ \\
$q_{\rm l}$ & & Minor-to-major axis ratio of the \textit{HST} F475X light profile & \multirow{1.6}{*}{$q_{\rm l, int}(q_{\rm l}, i) \sim \mathcal{N} (0.74, 0.08)$} \\
$i$ & [$^\circ$] & Inclination angle & \\
$\kappa_{\rm ext}$ & & External convergence & $\sim \mathcal{U}(-0.2, 0.4)$ \\
$\theta_{\rm S}$ & [$''$] & Scale radius of the variable mass-sheet & $\sim \mathcal{U}(7.72, 15.43)$ \\
{$\beta_{\rm ani, cyl}$} & & {Cylindrically-aligned, constant anisotropy parameter} & $\sigma_z/\sigma_R \sim \mathcal{U}(0, 1)$ \\
{$\beta_{\rm ani, sph}$} & & {Spherically-aligned, constant anisotropy parameter} & $\sigma_r/\sigma_\theta \sim \mathcal{N}(1, 0.07)$ \\
$\log (M_{\rm BH})$ & [$\log M_{\odot}$] & Logarithmic central black hole mass & $\sim \mathcal{N}(8.78, 0.35)$ \\
${\lambda}_{\rm int}$ & & Internal mass-sheet parameter & $ \sim \mathcal{U}(0.88, 1.21)$ \\ 
${\Omega}_{\rm m, 0}$ & & Matter density parameter & DESI-DR2 or Pantheon+ \\
${H}_0$ & [\ksmpc] & {Hubble-Lema\^{\i}tre} constant & $\sim \mathcal{U}(0, 150)$ \\

\hline
\end{tabular}
\end{center}
\end{table*}

Using the \textsc{Python} package \textsc{Jampy}\footnote{\url{https://pypi.org/project/jampy}}, we construct a dynamical model for \j14, constrained by both the lensing model and kinematic maps from the previous sections \citep{Cappellari2008, Cappellari2020}.  Given a 2D convergence profile, a 2D light profile, inclination angle, anisotropy parameterization, and angular diameter distance to the lens, our model generates kinematic measurement predictions at a given position relative to the lens galaxy.  We assume a constant anisotropy for our analysis and hence sample the radius-independent anisotropy parameter; see Table~\ref{tab:pop_names} for definitions.  The likelihood function we maximize is:
\begin{equation}
    p(\pi|\mathcal{D}) = \frac{p(\pi)}{p(\mathcal{D})} p(\mathcal{D}|\pi),
\end{equation}
where $\mathcal{D}$ represents the super-set of data used for our analysis (kinematic data $\mathcal{D}_{\rm K}$, \textit{HST} imaging $\mathcal{D}_{\rm I}$, time-delay measurements $\mathcal{D}_{\rm T}$, and LOS data $\mathcal{D}_{\rm E}$), and $\pi$ represents the parameters that we fit for in our dynamical model.  These parameters are listed and described in Table~\ref{tab:pop_names}.  By expanding $\mathcal{D}$ to components, we arrive at:
\begin{equation}
    p(\pi|\mathcal{D}) \propto \frac{p(\pi)}{p(\mathcal{D})}p(\mathcal{D}_{\rm K} | \pi) p(\mathcal{D}_{\rm E} | \pi) p(\mathcal{D}_{\rm I}, \mathcal{D}_{\rm T} | \pi).
\end{equation}
Note that while the datasets should be independently measured, they are still marginally dependent due to observing the same system.  Therefore, the right-hand side (RHS) is proportional, not equal, to $p(\pi|\mathcal{D})$.  This does not affect our log-likelihood maximization, only resulting in an additional additive constant.  Further simplifying the RHS via Bayes' theorem results in:
\begin{equation}
    p(\pi|\mathcal{D}) \propto p(\pi)^j p(\pi | \mathcal{D}_{\rm K}) p(\pi | \mathcal{D}_{\rm E}) p(\pi | \mathcal{D}_{\rm I}, \mathcal{D}_{\rm T}),
\end{equation}
where $p(\pi)$ represents the prior on $\pi$ shown in Table~\ref{tab:pop_names}, and $p(\pi | \mathcal{D}_{\rm X})$ is the dynamical model posterior based on its fit to the respective dataset(s).  The exponent $j$ is to ensure that the prior is not applied multiple times.  As we only use these priors in the dynamical modeling likelihood (i.e., uniform priors are assumed for relevant parameters in $\pi$ in the posteriors from the lens model and LOS analysis), $j=1$.

$\mathcal{D}_{\rm E}$ constrains only $\kappa_{\rm ext}$ in our dynamical model, while $\mathcal{D}_{\rm I}$ and $\mathcal{D}_{\rm T}$ (i.e., the lens model) constrain the parameter set $\omega$.  All parameters in $\omega$, with the exception of $D_{\Delta t}$, are in $\pi$ ($\pi \cap \omega = \omega - \{D_{\Delta t}\}$), and thus these posteriors can be probed directly in our likelihood function.  The constraint on $D_{\Delta t}$ can be translated to a constraint on the fitted cosmology ($\Omega_{\rm m, 0}$ and $H_0$).  Therefore, we can further transform the RHS as:
\begin{equation}
    p(\pi|\mathcal{D}) \propto p(\pi) p(\pi | \mathcal{D}_{\rm K}) p(\kappa_{\rm ext} | \mathcal{D}_{\rm E}) p(\pi \cap \omega, \Omega_{\rm m, 0}, H_0 | \mathcal{D}_{\rm I}, \mathcal{D}_{\rm T}),
\end{equation}
and finally:
\begin{multline}
    p(\pi|\mathcal{D}) \propto p(\pi) p(\pi | \mathcal{D}_{\rm K}) p(\kappa_{\rm ext} | \mathcal{D}_{\rm E}) p(\pi \cap \omega| \mathcal{D}_{\rm I}, \mathcal{D}_{\rm T}) \\p(D_{\Delta t}| \mathcal{D}_{\rm I}, \mathcal{D}_{\rm T}) p(\Omega_{\rm m, 0}, H_0 | D_{\Delta t}).
\end{multline}
$p(\pi | \mathcal{D}_{\rm K})$ represents the likelihood constraints provided by fitting our dynamical model to the kinematic data, $p(\kappa_{\rm ext} | \mathcal{D}_{\rm E})$ represents the posterior on $\kappa_{\rm ext}$ from the LOS measurements (see \sref{subsec:kappa}), $p( \pi \cap \omega| \mathcal{D}_{\rm I}, \mathcal{D}_{\rm T})$ represents the posterior from our lens model using both the imaging and time-delay information (see \sref{sec:lens_model}), and finally $p({\Omega}_{\rm{m}, 0}, {H}_0 | {D}_{\Delta t})$ is the likelihood of a cosmology parameterization for a given ${D}_{\Delta t}$ value.  


When building our dynamical model within the \textsc{Jampy} framework, we must provide a 2D surface profile of the tracer stellar population, and a 2D surface potential profile of the mass profile (alongside other secondary parameters such as $i$, the kinematic PSF, the anisotropy configuration, black hole mass, etc).  The 2D surface profile distribution is taken from our F475X light profile in the lens model, where we fix the best-fitting profile and vary its expansion/contraction by the total effective radius $R_{\rm eff}$ posterior (see \sref{sec:lens_model} for more details on the light model configuration).  The 2D surface potential profile is parameterized as Equation~\ref{eq:kappa_gal}, accounting for a power-law convergence, an external convergence, and a variable mass-sheet contribution.  To convert the unit-less convergence profile into physical mass units ($\Sigma(\boldsymbol{\theta})$):
\begin{equation}
    \Sigma_{\rm gal}(\boldsymbol{\theta}) = \frac{c^2}{4\pi G} \frac{D_{\Delta t}}{D_{\rm d}^2(1 + z_{\rm l})} \kappa_{\rm gal}(\boldsymbol{\theta}),
\end{equation}
where ${D}_{\Delta t}$ and $D_{\rm d}$ are derived from the sampled cosmology.  Internally, \textsc{Jampy} converts the provided 2D light and density profiles into 3D profiles, which cannot be analytically done with an arbitrary parameterization.  Therefore, we first decompose these 2D profiles through multi-Gaussian expansion (using \textsc{MgeFit}\footnote{\url{https://pypi.org/project/mgefit}}; \citealp{Cappellari2002}), since each individual 2D Gaussian component can be analytically converted into a 3D component.  For both mass and light, we fit 20 Gaussian components to the 2D surface profile, logarithmically sampled up to $50R_{\rm eff}$.

Throughout our fitting process, we implement an iterative process to prune Voronoi bins that result in overwhelming $\chi^2$ contributions $>3\sigma$ (or a $0.28\%$ statistical outlier) to the overall likelihood function.  This is to ensure that our dynamical model is not being overfitted by these outlying bins, which could be caused by isolated velocity features in the lens galaxy.  MCMC chains are well converged before extracting the model posteriors.  

We take many precautions to prevent any possible source of bias in our dynamical model.  Using toy models, \citet{FKM2026} demonstrated where these systematics may arise in the process.  They showed that it is necessary to properly model the PSF kinematic data, which we accurately account for in \sref{subsec:spect_reduction}.  Based on real data and simulations, constant anisotropy is a better description than Osipkov-Merritt \citep{VM2026, FKM2026}, which is what we adopt in this analysis.  Recent Schwarzschild modeling \citep{schwarzschild79} indicates that orbital distributions are nearly isotropic \citep{Cappellari2026}. This evidence effectively addresses the concerns raised by \citet{FKM2026}, which were predicated on more extreme orbital scenarios.  We agree with \citet{FKM2026} that inconsistencies may occur from a radial color gradient within the tracer population profile, which is why we build the tracer profile using observations from a similar wavelength range as the kinematic measurements.  Modern analysis techniques, high-fidelity data, and refined stellar libraries (see \sref{sec:IFU_analysis} and \citealp{Knabel2025}) keep our kinematic systematics low, including spatial bin covariance.  This methodology has been thoroughly tested (\citealp{Knabel2025}; \citealp{tdcosmo2025}), superseding the overly simplified scenarios presented in \citet{FKM2026}.

\subsection{Intrinsic axis ratio prior} \label{subsec:inc}
The inclination angle ($i$), or the orientation of the lens galaxy relative to our LOS, can be defined in the following equation:
\begin{equation} \label{eq:inc}
    q_{\rm l, int}(q_{\rm l}, i) \equiv \sqrt{\frac{q_{\rm l}^2-\cos^2i}{\sin^2i}} ,
\end{equation}
where $q_{\rm l}$ is the observed minor-to-major axis ratio of the 2D light profile, and $q_{\rm l, int}$ is the intrinsic axis ratio of the 3D galaxy.  We use the posterior of $q_{\rm l, int}(q_{\rm l}, i) \sim \mathcal{N} (0.74, 0.08)$ from \citet[first row of Table~1]{Li_2018} as a joint prior on our sampled values $q_{\rm l}$ and $i$, which was derived using a sample of slow rotator early-type galaxies in SDSS-IV DR14 MaNGA.  While formally $i \in [0^\circ, 90^\circ]$, we allow for $i \in [0^\circ, 180^\circ]$ as this is analytically identical and allows for the Markov chains to better capture the expected posterior.  

\subsection{Scale radius prior}
We sample scale radii $\theta_{\rm S}$ (which defines our variable mass sheet in Equation~\ref{eq:vmst}) such that the variable mass sheet is large enough to be agnostic of the lensing information.  Through simulations, \citet{Birrer_2020} find that at $\theta_{\rm S} \ge 5\theta_{\rm E}$, the variable mass sheet follows very closely to a true mass sheet, which we use as lower bound for $\theta_{\rm S}$.  We also set an upper bound of $\theta_{\rm S} \le 10\theta_{\rm E}$, as unreasonably high values $\theta_{\rm S}$ result in an artificially tighter, physically-motivated constraints on $\lambda_{\rm int}$ (see \sref{subsec:lambda_prior}).  Since the error of the Einstein radius from the lens model is negligible compared to the range spanned by these bounds, we essentially sample $\theta_{\rm S} \sim \mathcal{U}(7\farcs72, 15\farcs43)$.

\subsection{Anisotropy parameter prior} \label{subsec:ani_prior}
We use constant anisotropy, which has been found to be a good description of both observed \citep{Cappellari2026} and simulated galaxies \citep{VM2026}, and is more accurate than the Osipkov-Merritt model \citep{Ospikov,Merritt1985} used a few years ago by our collaboration.  Because \j14 is a rotating elliptical galaxy (see Figure~\ref{fig:kinematic_maps_2}), a cylindrically-aligned velocity ellipsoid is more appropriate as a solution for the Jeans equation in our dynamical model \citep{Cappellari2008}.  However, as an additional systematics check, we fit for two different models: with a cylindrically-aligned or spherically-aligned velocity ellipsoid solution. These different orientations also necessitate a slightly different definition of the constant anisotropy parameter $\beta_{\rm ani}$. For the cylindrically-aligned case $\beta_{\rm ani}$ is defined as:
\begin{equation} \label{eq:beta_cyl}
    \beta_{\rm ani,~cyl} \equiv 1 - (\sigma_{z}/\sigma_{R})^2, 
\end{equation}
and for the spherically-aligned case:
\begin{equation} \label{eq:beta_sph}
    \beta_{\rm ani, ~sph} \equiv 1 - (\sigma_{\theta}/\sigma_{r})^2.
\end{equation}
$\sigma_{z}$ and $\sigma_{R}$ are the LOS and radial velocity dispersions in a cylindrical coordinate system, and $\sigma_{\theta}$ and $\sigma_{r}$ are the tangential and radial velocity dispersions in a spherical coordinate system. We include a prior of $\sigma_z/\sigma_R \sim \mathcal{U}(0, 1)$ for the cylindrically-aligned model, based on physically and empirically motivated bounds \citep{SAURON}.

For the spherically-aligned model, we establish a Gaussian prior on the velocity dispersion ratio of $\sigma_r/\sigma_\theta \sim \mathcal{N}(1, 0.07)$. This prior was derived empirically by analyzing a sample of 13 representative early-type galaxies from \citet[Figure~10]{Cappellari2026}. For each galaxy, the ratio $\sigma_r/\sigma_{\rm tang}$ (which is equivalent to $\sigma_r/\sigma_\theta$ in our spherical model) was averaged across the radial range $R_{\rm e}/30 < r < R_{\rm e}$. This specific radial annulus was chosen because it exhibits relatively low scatter, meaning the residual radial variation is primarily driven by data or numerical noise rather than intrinsic physical features. Crucially, this selection intentionally excludes the very center of the galaxy, where the dynamics can be significantly influenced by the supermassive black hole and would necessitate a logistic JAM model rather than the constant anisotropy assumed here.

Averaging the values across these 13 galaxies yields a mean of $0.98$ and a standard deviation of $0.07$. Given the remarkable proximity of the mean to isotropy, we adopt a rounded mean of 1.0. We do not derive separate priors for fast versus slow rotators, as the resulting sub-samples would be too small for reliable statistics and the anisotropy values are comparable between the two populations.

\subsection{Central black hole mass prior}
In our dynamical model, we also include a prior on the central black hole mass $M_{\rm BH}$, as it has been found that it may significantly affect the kinematics near the center of the elliptical galaxy \citep[e.g.,][]{han2025, tdcosmo2025}.  As such, we include a prior on $\log(M_{\rm BH})$, calculated using the well-studied linear scaling relation between the $\log(M_{\rm BH})$ and aperture velocity dispersion for our lens galaxy ($\sigma_{\rm v} = 260$ km s$^{-1}$; see Figure~\ref{fig:aperture_spec}):  
\begin{equation} \label{eq:bh}
    \log \left( \frac{M_{\rm BH}}{M_\odot} \right) = \alpha_{\rm BH} + \beta_{\rm BH} \log \left( \frac{\sigma_{\rm v}}{200 \text{ km s}^{-1}} \right).
\end{equation}
We use the zeropoint ($\alpha_{\rm BH} \sim 8.23 \pm 0.05$), slope ($\beta_{\rm BH} \sim 4.86 \pm 0.50$), and intrinsic root-mean squared scatter ($\Delta_{\log(M_{\rm BH})} = 0.34$) distributions of this linear relation from \citet{Graham2013}, derived from their sample of 28 massive elliptical galaxies.  And so, we include a prior of $\log(M_{\rm BH}/M_{\odot}) \sim \mathcal{N}(8.78, 0.35)$. 

\subsection{Internal mass-sheet parameter prior} \label{subsec:lambda_prior}
As shown in \sref{subsec:msd}, accounting for the internal mass-sheet parameter is paramount in recovering an unbiased measurement of cosmology.  Therefore, we set a physical upper bound for $\lambda_{\rm int}$ to ensure that the sampled $\kappa_{\rm gal}(\boldsymbol{\theta})$ realization must be monotonic.  
We find that when $\lambda_{\rm int} \ge 1.21$, there exists no reasonable sampling of the $\theta_{\rm E}$, $\gamma$, and $\theta_{\rm S}$ such that the true galaxy convergence maintains monotonicity.  
We also ensure that the total mass of the variable mass-sheet within a three-dimensional radius does not exceed the mass of the NFW profile in the same sphere \citep{Birrer_2020}.  
For our posteriors and range of scale radii, this effectively sets a lower limit of $\lambda_{\rm int} \ge 0.88$.  Previous analyses show that $\lambda_{\rm int}$ is close to unity and well within this range \citep{Shajib_2023,dinosI,dinosII,tdcosmo2025}.

In order to facilitate comparison with analyses that break the MSD by enforcing a power-law model without the parameter $\lambda_{\rm int}$ \citep{Wong2020,Millon_2020,q25} in Appendix~\ref{app:B}, we also run a set of dynamical modeling configurations where we fix $\lambda_{\rm int} = 1$.  This check highlights the importance of allowing it to be free for this system, and provides a lower bound on the precision attainable with this system alone using present data.

\subsection{Matter density parameter prior} \label{subsec:Om0_prior}

We include a prior on the matter density parameter of $\Omega_{\rm m, 0} = 0.3169 \pm 0.0065$ in the $\Lambda$CDM cosmology, from the DESI data release 2 (DESI-DR2), derived using baryon acoustic oscillation measurement \citep{desidr2}.  As an additional test for systematics, we also impose an alternative prior of $\Omega_{\rm m, 0} = 0.334 \pm 0.018$, from the Pantheon+ analysis of 1550 Type Ia supernovae from redshifts $0.001 \leq z \leq 2.26$ and also assuming a $\Lambda$CDM cosmology \citep{Brout_2022}.  Both priors are measurements from only using relative distances, and hence are independent of potential distance ladder biases.
{Although the time-delay distance $D_{\Delta t}$ for a relatively low-redshift lens system such as \j14 ($z_{\rm l} \approx 0.407$) is only weakly sensitive to $\Omega_{\rm m, 0}$ (with the effect suppressed by higher-order terms in the Taylor expansion of redshift), we explicitly marginalize over $\Omega_{\rm m, 0}$ rather than assuming a fixed value.  This ensures that any minor residual dependency is fully accounted for and its uncertainty is correctly propagated to our final $H_0$ error budget.  Additionally, comparing results across different independent priors for $\Omega_{\rm m, 0}$ (i.e., DESI-DR2 vs.\ Pantheon+) serves as a crucial sanity check to demonstrate the robustness and insensitivity of our $H_0$ measurement to the assumed matter density prior.}

We first run a given dynamical model with a uniform prior of $\Omega_{\rm m, 0} \sim \mathcal{U}(0.05, 0.5)$, then impose either the DESI-DR2 or Pantheon+ prior on $\Omega_{\rm m, 0}$ by importance sampling the converged chains.  This is effectively equivalent to rerunning the model with the normal prior established beforehand (assuming a sufficient number of samples).  

\begin{table} 
\begin{center}
 \caption{Our labeling scheme different dynamical model configurations.}   
 \label{tab:model_configs}
\begin{tabular}{cll}
    \hline
    \makecell{Dynamical\\model ID} & \makecell{Anisotropy\\model} & $\Omega_{\rm m, 0}$ prior \\ 
    \hline
    DyM1 & Cylindrical & DESI-DR2 \\
    DyM2 & Spherical & DESI-DR2 \\
    DyM3 & Cylindrical & Pantheon+ \\
    DyM4 & Spherical & Pantheon+ \\
    \hline
\end{tabular}
\end{center}
\end{table}

\section{Cosmological results} \label{sec:results}
For our primary results, we run four different dynamical model configurations, with different assumptions on cosmology and anisotropy.  See Table~\ref{tab:model_configs} for our labeling scheme moving forward.  We aggregate our models to generate a final set of posteriors by using the $\Delta$BIC (relative to the model with the lowest BIC) of each model to weight each individual distribution according to $w_{\rm BIC} \propto \exp(-\frac{1}{2}\Delta \text{BIC})$.  For each model, the BIC will be calculated using the median value for a given sampled parameter.  Hereafter, whenever we combine dynamical model results, it is assumed we weight each result by $w_{\rm BIC}$, which is given in Table~\ref{tab:model_final}.  

We present a table of our cosmological results in Table~\ref{tab:model_final_cosmo}.  We also show the full directly sampled posteriors for our dynamical models in Table~\ref{tab:model_final}.  

\begingroup
\renewcommand{\arraystretch}{1.5}
\begin{table*} 
\begin{center}
 \caption{Our final cosmological results for our dynamical models, aggregated by the anisotropy configurations (cylindrically and spherically aligned).  Left of the dashed line are cosmological parameters directly sampled from the dynamical modeling pipeline; right of the dashed line are informative parameters derived directly from the left parameters.}   
 \label{tab:model_final_cosmo}
\begin{tabular}{c|cccc:ccc}
    \hline
    \makecell{Dynamical\\model ID} & \makecell{${H}_0$\\{}[\ksmpc]} & \makecell{${\Omega}_{\rm m, 0}$\\{}} & \makecell{${\lambda}_{\rm int}$\\{}} & \makecell{$\kappa_{\rm ext}$\\{}} & \makecell{${\lambda}_{\rm total}$\\{}} & \makecell{${D}_{\Delta t}$\\{[Mpc]}} & \makecell{${D}_{\rm d}$\\{[Mpc]}} \\
    \hline
DyM1+2 (DESI-DR2) & 
$73.2^{+4.8}_{-4.7}$ & $0.30^{+0.01}_{-0.01}$ & $1.12^{+0.05}_{-0.06}$ & $-0.02^{+0.04}_{-0.04}$ & $1.14^{+0.05}_{-0.06}$ & $2032^{+141}_{-125}$ & $1072^{+74}_{-68}$ \\
DyM3+4 (Pantheon+) & 
$72.6^{+5.0}_{-4.7}$ & $0.33^{+0.02}_{-0.02}$ & $1.11^{+0.06}_{-0.06}$ & $-0.02^{+0.04}_{-0.03}$ & $1.13^{+0.05}_{-0.05}$ & $2046^{+142}_{-133}$ & $1069^{+75}_{-67}$ \\
\hline
\end{tabular}
\end{center}
\end{table*}
\endgroup 

    
    
    

\begin{figure*}
\begin{center}
 \includegraphics[width=0.85\linewidth]{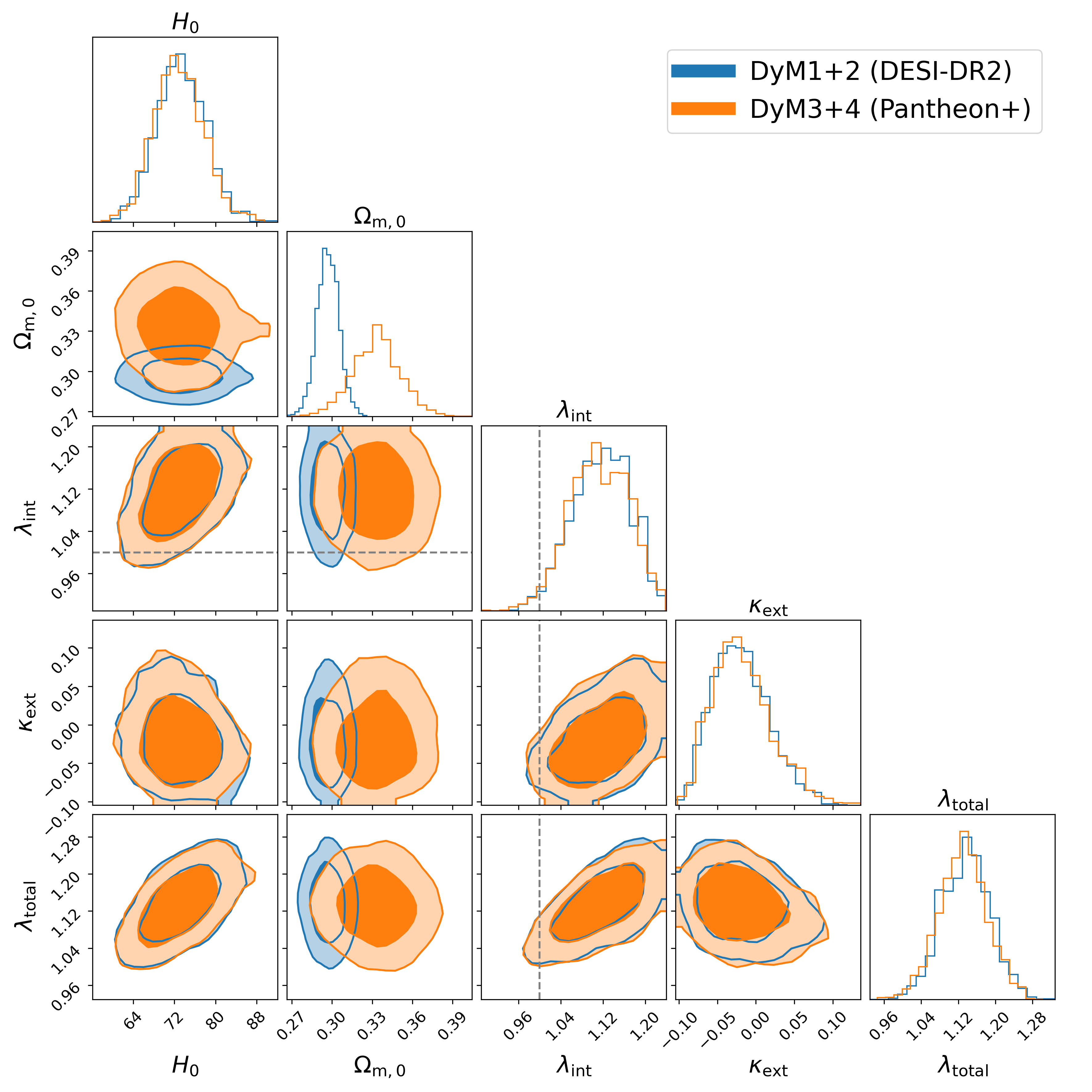}
 \caption{Our dynamical modeling results, illustrating our cosmological constraints.  See Table~\ref{tab:model_configs} for our labeling scheme, denoting different dynamical model configurations.  The dotted line indicates $\lambda_{\rm int} = 1$.  We find that our measurement of $H_0$ is robust against different assumptions of $\Omega_{\rm m, 0}$, and we report a significant measurement of $\lambda_{\rm int} > 1$.} 
 \label{fig:final_res_corner}
\end{center}
\end{figure*}

\subsection{Discussion}
\label{subsec:discussion}

In DyM1+2 (combining our anisotropy-alignment model, and assuming a DESI-DR2 prior on $\Omega_{\rm m, 0}$, which we treat as a baseline), we measure $H_0=73.2^{+4.8}_{-4.7}$, a 6.5\%-level precision.  Our result is most in line with late-universe measurements, such as those from the SH0ES team \citep[from the Cepheid-calibrated distance ladder;][]{Riess2022}.  However, given our uncertainty, we remain in approximate $1\sigma$ agreement with measurements from the CCHP team \citep[derived from the TRGB-calibrated distance ladder;][]{hoyt2025}, and even with early-universe measurements \citep[derived from the CMB power spectrum;][]{Planck2020}.

Our results highlight aspects we expect to see in a dynamical model such as this.  For example, we see that $H_0$ is positively correlated with $\lambda_{\rm int}$ and/or $\lambda_{\rm total}$, corresponding to the mass-sheet degeneracy.  Additionally, we find that our constraints on $H_0$ are robust against $\Omega_{\rm m, 0}$ (e.g., comparing DyM1 and 2 with DyM3 and 4), as expected, since time-delay cosmography for a single system should only be weakly dependent on cosmology outside of $H_0$.

We note some particularities that are more specific to \j14 compared to other analyses of time-delay lenses.  First, we find that cylindrically aligned anisotropic models are heavily preferred over spherically aligned ones for \j14 (comparing the $w_{\rm BIC}$ values of DyM1 and 3 versus DyM2 and 4 in Table~\ref{tab:model_final}).  We attribute this to \j14 being better described as a cylinder in the dynamical model, as evidenced by its significant rotation.  We do not believe the spherically aligned model priors on anisotropy are overly constraining and therefore causing this preference, because the $\beta_{\rm ani}$ constraints in Table~\ref{tab:model_final} are very similar between the alignment models.

Another feature to note is that we find that \j14 prefers an internal mass-sheet parameter greater than one ($\lambda_{\rm int} = 1.14^{+0.05}_{-0.06}$) at the $2\sigma$ level, implying the presence of a negative mass sheet.  This is not unrealistic, as some systems have been found to have $\lambda_{\rm int}>1$ using kinematics \citep{knabel2026slacsII}.  \citet{tdcosmo2025} measure a population mean of $\lambda_{\rm int} = 1.06^{+0.06}_{-0.06}$ (population $\sigma \approx 0.11$) from eight lensed quasars, the Sloan Lens ACS (SLACS), and the Strong Lensing Legacy Survey (SL2S) lens datasets, which is in agreement with our measurement for \j14.  

\subsection{Comparison with previous work}
\label{subsec:previous}

The most comparable analysis to the one presented here is that carried out by \citet{Shajib_2023} using KCWI data of the lens RXJ1131-1231 (henceforth RXJ1131).  While many elements of the analyses are similar, we highlight the differences here to gain insight into why our analysis achieves higher precision. The main differences are the following:

\begin{enumerate}
    \item We exploit the superior stellar velocity dispersion techniques introduced by \citet{Knabel2025}, developed after the \citet{Shajib_2023} analysis.  As a result, the covariance between stellar kinematics measurements is reduced from a few percent to sub-percent levels, improving overall accuracy. This is the main factor improving our ability to break the MSD.
    \item While the overall precision of the time delays is comparable between \j14 and RXJ1131, the latter is a cusp configuration with only one time delay measured to percent-level precision. In the lens analyzed here, the long delay is the most precise, but the other two are still measured with 6-7\% precision, providing some information to break degeneracies in the models.
    \item The clear detection of a kinematic axis aligned with the galaxy major axis allows us to constrain the intrinsic axis ratio of light, the inclination, and oblateness (versus prolateness) probability of \j14 much more tightly than RJX1131.  In the case of RXJ1131, \citet{Shajib_2023} find a probability of oblateness of 65\%, compared to a probability of oblateness of 99.97\% in \j14 (see \sref{subsec:kinematic_maps}).
    \item 
    Following \citet{Cappellari2026} we have imposed an informative prior on the anisotropy. Due to the clear rotation in \j14, we are able to test the systematics between two different anisotropy alignments (cylindrical and spherical), whereas a spherically aligned model was most suited for RXJ1131.  In \citet{Shajib_2023}, the authors use priors of $\sigma_\theta/\sigma_r \sim \mathcal{U}(0.78, 1.14)$ \citep{SAURON}, whereas we use updated constraints of $\sigma_r/\sigma_\theta \sim \sigma_\theta/\sigma_r \sim \mathcal{N}(1, 0.07)$ \citep{Cappellari2026}. 
    \item We have allowed for the presence of a supermassive black hole in the center of the deflector, which was not considered in the analysis of RXJ1131. 
\end{enumerate}

\section{Summary and conclusions} \label{sec:conclusion}
We present a blind time-delay cosmography measurement of the {Hubble-Lema\^{\i}tre} constant $H_0$ using the quadruply imaged quasar system SDSSJ1433+6007. To achieve this, we develop a comprehensive ``soup-to-nuts'' pipeline that combines data from multiple independent observations to reconstruct the lens and constrain cosmology.  Our analysis incorporates deep high-resolution \textit{HST} imaging, extended time-delay monitoring from the Wendelstein and Maidanak Observatories, spatially resolved stellar kinematics from KCWI/KCRM, and wide-field photometry from the DESI Legacy Survey DR10.

By combining these datasets, we construct a robust lens model and incorporate high signal-to-noise kinematic maps to break the mass-sheet degeneracy.  We lay out our reduction and analysis procedure for the IFU data, where we measure the lens galaxy's specific physical characteristics, identifying it as a rotating elliptical galaxy with significant inclination and oblate symmetry.  The inclusion of Maidanak Observatory data alongside the Wendelstein data yields a time-delay measurement uncertainty improved by a factor of 1.5 when compared to previously published results.  Furthermore, by characterizing the local environment and line-of-sight structures using wide-field photometry, we successfully constrain the external convergence to $\kappa_{\rm ext} = -0.041^{+0.047}_{-0.030}$.

We integrate these components into a highly flexible dynamical model to constrain cosmology while explicitly accounting for the mass-sheet transform.  Assuming a flat $\Lambda$CDM cosmology and utilizing an $\Omega_{\rm m,0}$ prior from DESI data release 2, our maximally flexible baseline models (which sample the internal mass-sheet parameter $\lambda_{\rm int}$) measure $H_0 = 73.2^{+4.8}_{-4.7}$ \ksmpc, with 6.5\% precision.

Extensive systematic testing prior to unblinding reveals several key physical insights about \j14.  We find that our cosmological constraints on $H_{0}$ are remarkably robust against the choice of matter density priors.  The system prefers a value greater than one for the mass-sheet parameter ($\lambda_{\rm int} = 1.14_{-0.06}^{+0.05}$ for our baseline model) at the $2\sigma$ level.  This implies the presence of a negative mass sheet, which is physically consistent with the galaxy being baryon-dominated at the Einstein radius.  Additionally, the kinematic data heavily prefer a cylindrically aligned anisotropic model over a spherically aligned one, which aligns with the clear rotational velocity profile observed in the lens.

\j14 is expected to receive further improvements, with cycle-5 observations by the \textit{James Webb Space Telescope} (\textit{JWST}) Near-Infrared Spectrograph (NIRSpec) to better assess the spatially resolved kinematics of the lens galaxy at higher angular resolution.  Furthermore, adaptive-optics-assisted IR imaging from the Keck Observatory OSIRIS instrument may further improve the resolution of arc features and hence better constrain the lens model.  

Ultimately, this work demonstrates the efficacy of our comprehensive, multi-probe pipeline in delivering a robust and independent measurement of the {Hubble-Lema\^{\i}tre} constant.  Explicitly accounting for the mass-sheet transform and breaking the mass-sheet degeneracy requires a highly integrated approach; our precise $H_0$ constraints rely equally on the synthesis of deep \textit{HST} imaging, extended time-delay monitoring, careful environmental characterization, and high-resolution spatially resolved kinematics.  Looking forward, by combining this fully integrated analysis of SDSSJ1433+6007 with the broader TDCOSMO dataset, we aim to drive down systematic uncertainties even further and achieve tighter constraints on $H_{0}$ to help resolve the current cosmological tension.   {Furthermore, while the absolute distance scale measured via time-delay cosmography is primarily sensitive to $H_0$ and is robust to the choice of the $\Omega_{\rm m, 0}$ prior, the latter parameter becomes highly significant when combining these results with other cosmological datasets.  Joint analyses of time-delay constraints with CMB anisotropy, galaxy clustering, and weak lensing measurements from forthcoming surveys (such as DESI, Euclid, Rubin, and Roman) will break the degeneracies between $H_0$ and $\Omega_{\rm m, 0}$, yielding tighter simultaneous constraints on both parameters and offering a powerful test of the standard $\Lambda$CDM framework and dark energy models.}
    

\begin{acknowledgements}
We thank all the friends of the TDCOSMO collaboration for useful feedback that improved this manuscript.
MM acknowledges support by the SNSF (Swiss National Science Foundation) through return CH grant P5R5PT\_225598 and Ambizione grant PZ00P2\_223738.
SB is supported by JWST-GO-07184 and acknowledges support by the Department of Physics and Astronomy, Stony Brook University.
KCW is supported by JSPS KAKENHI Grant Numbers JP24K07089, JP24H00221.
This research is based on observations made with the NASA/ESA Hubble Space Telescope obtained from the Space Telescope Science Institute, which is operated by the Association of Universities for Research in Astronomy, Inc., under NASA contract NAS 5–26555. These observations are associated with program(s) HST-GO-15320.
Some of the data presented herein were obtained at Keck Observatory, which is a private 501(c)3 non-profit organization operated as a scientific partnership among the California Institute of Technology, the University of California, and the National Aeronautics and Space Administration. The Observatory was made possible by the generous financial support of the W.~M.~Keck Foundation. 
The authors wish to recognize and acknowledge the very significant cultural role and reverence that the summit of Maunakea has always had within the Native Hawaiian community. We are most fortunate to have the opportunity to conduct observations from this mountain.
We acknowledge support by NSF through grants NSF-AST-2407277, and NSF-AST-1836016, and from the Moore Foundation through grant 8548.
The authors are grateful to the staff and observers of the Maidanak Observatory for supporting the observations. The Uzbekistan team acknowledged the support provided by the Departments of Galactic and Extragalactic Astronomy at the Astronomical Institute of the Academy of Sciences of Uzbekistan.  
The Wendelstein telescope project was funded by the Bavarian government and by the German Federal government. Part of the instrumentation was funded by the Excellence Cluster ORIGINS, funded by the Deutsche Forschungsgemeinschaft (DFG) under Germany’s Excellence Strategy (EXC-2094-390783311).
This work was supported by the Agencia Estatal de Investigación (AEI), Ministerio de Ciencia, Innovación y Universidades, Spain, under the project PID2024‑155455NB‑I00, within the 2024 Call for Knowledge Generation Projects.
{Lastly, we thank the anonymous referee for their insightful comments and discussion.}
\end{acknowledgements}

\appendix
\nolinenumbers
\section{Additional figures and data regarding the lens model}
\label{app:A}

\begin{figure*}
\begin{center}
 \includegraphics[width=1\linewidth]{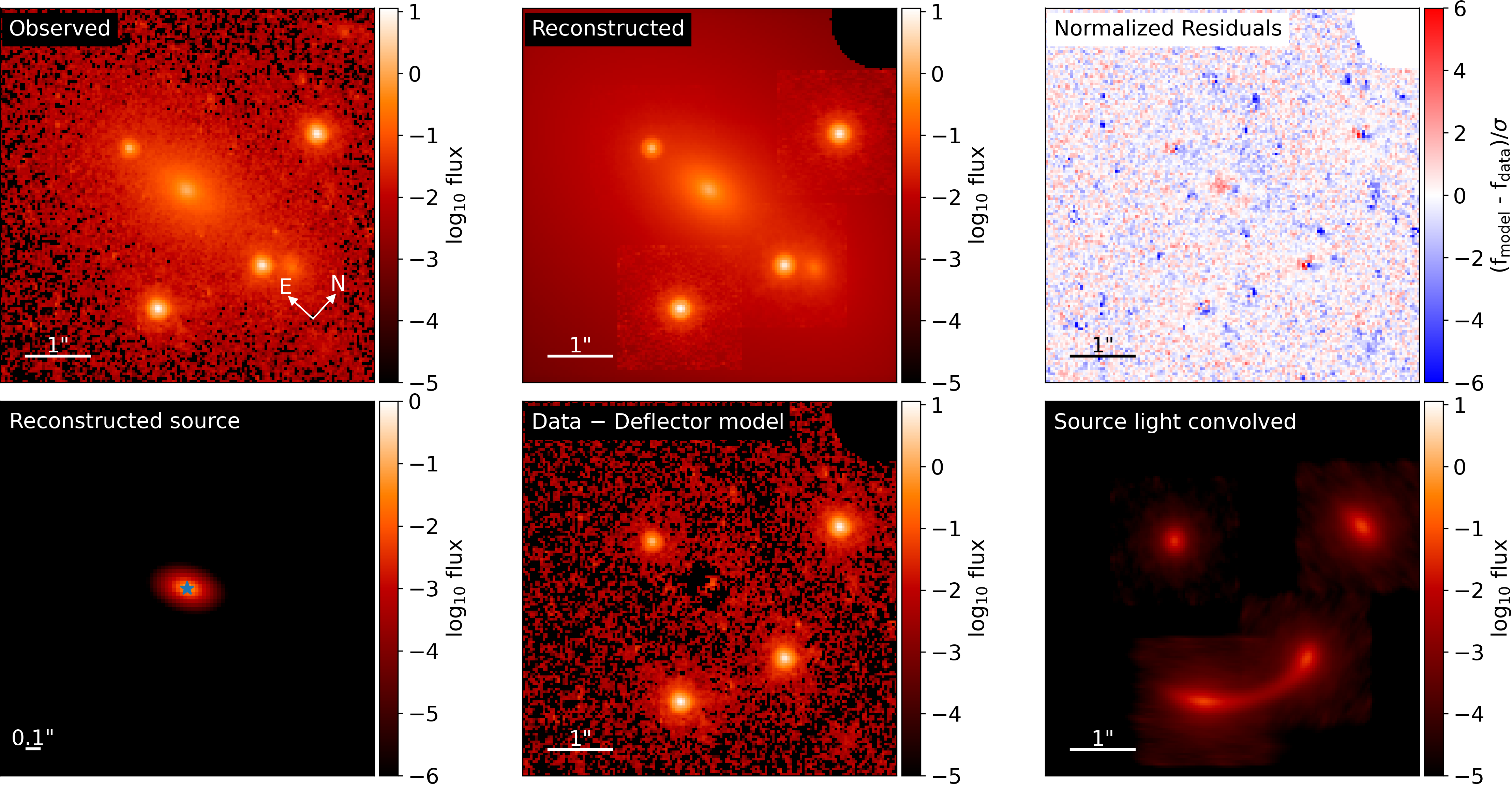}
 \caption{Representative plots illustrating the lens model of \j14, and its reconstruction of the \textit{HST} F814W filter data.  The layout of this figure (with the exception of the bottom right panel) is described in Figure~\ref{fig:f160w_model}.  The bottom right panel shows the convolved and lensed source light (excluding the point source) of our best-fit reconstruction.  We measure a $\chi^2_{\nu=21941}$ of 1.16 for the F814W data fit.}
 \label{fig:f814w_model}
\end{center}
\end{figure*}

\begin{figure*}
\begin{center}
 \includegraphics[width=1\linewidth]{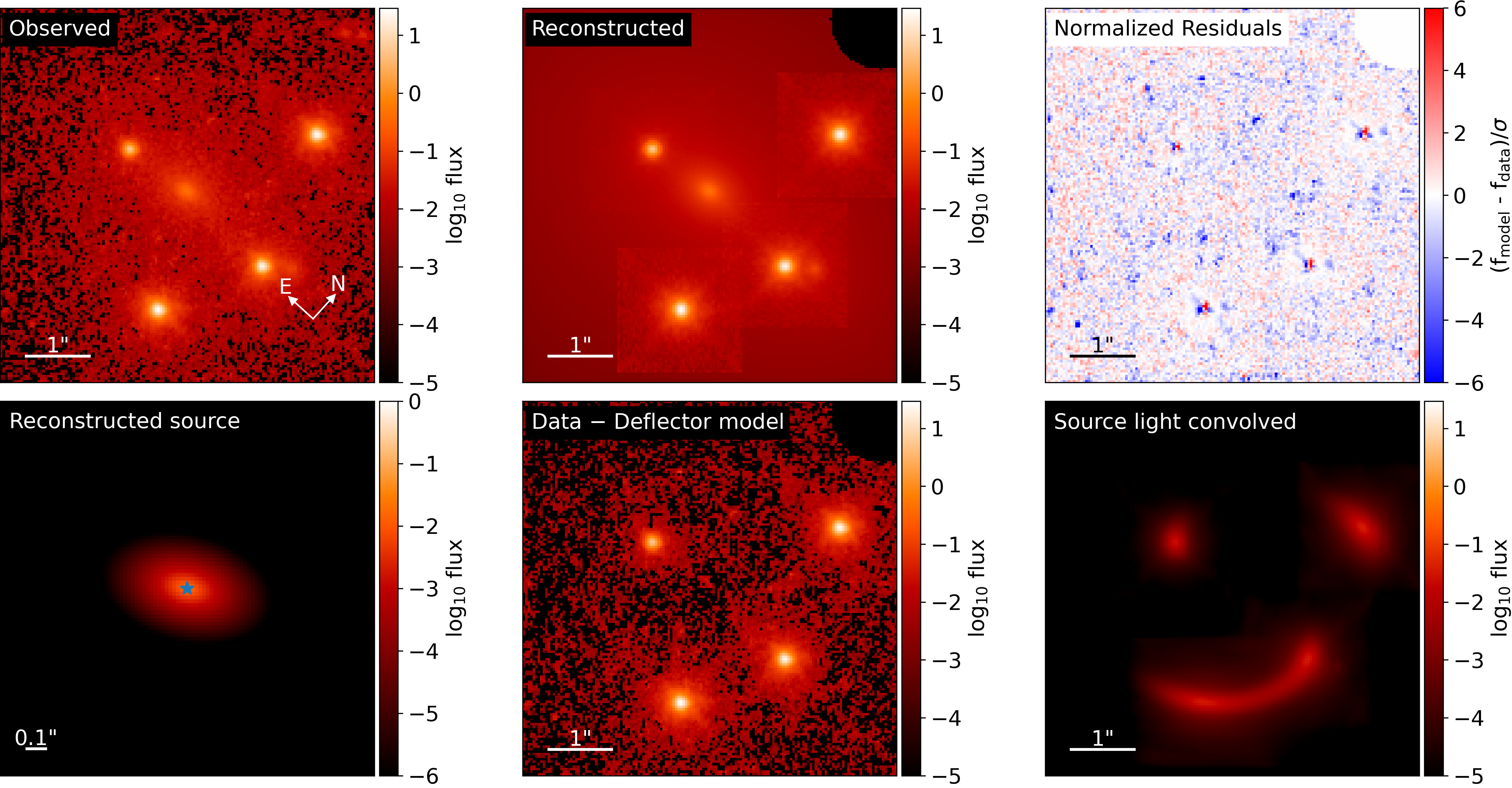}
 \caption{Representative plots illustrating the lens model of \j14, and its reconstruction of the \textit{HST} F475X filter data.  The layout of this figure (with the exception of the bottom right panel) is described in Figure~\ref{fig:f160w_model}.  The bottom right panel shows the convolved and lensed source light (excluding the point source) of our best-fit reconstruction.  We measure a $\chi^2_{\nu=21941}$ of 1.10 for the F475X data fit.}
 \label{fig:f475x_model}
\end{center}
\end{figure*}

Here we present the lens modeling results for the F814W and F475X filters (Figures~\ref{fig:f814w_model} and \ref{fig:f475x_model}, respectively).  We use the F475X fitted light profile of the main lens for our dynamical model in \sref{sec:dynamical_model}. 

We also present the constraints for all relevant lens model parameters, including the lensing parameters (Table~\ref{tab:full_lens_params}), the lens light parameters (Table~\ref{tab:full_lens_light_params}), and the source light parameters (Table~\ref{tab:full_source_params}).

\begin{table*} 
\begingroup
\renewcommand{\arraystretch}{1.25}
\begin{center}
 \caption{Full lensing parameter posteriors derived from our modeling procedures (see \sref{sec:lens_model}).  $\theta_{\rm E}$ represents the Einstein radius, $\gamma$ represents the logarithmic power-law profile slope, $q$ represents the minor-to-major axis ratio, $\phi$ represents the East-of-North position angle of the major axis, $\gamma_{\rm ext}$ represents the total external shear strength, and $\phi_{\rm ext}$ represents the position angle of the external shear.}   
 \label{tab:full_lens_params}
 \begin{tabular}{cc|ccccc}
  \hline
   \multicolumn{2}{c|}{Parameter} & Lens & Satellite & Aggregated perturber & Isolated perturber & External shear 
  \\
  \hline
$\theta_{\rm E}$ & [$''$] & $1.538^{+0.005}_{-0.007}$ & $0.159^{+0.003}_{-0.002}$ & $1.17^{+0.08}_{-0.07}$ & $0.64^{+0.08}_{-0.09}$ & - \\
$\gamma$ &  & $1.954^{+0.015}_{-0.015}$ & 2 & 2 & 2 & - \\
$q$ & & $0.754^{+0.006}_{-0.006}$ & $0.756^{+0.016}_{-0.015}$ & 1 & 1 & - \\
$\phi$ & [$^\circ$] & $-8.2^{+0.3}_{-0.3}$ & $18.5^{+2.0}_{-2.1}$ & - & - & - \\
$\gamma_{\rm ext}$ & & - & - & - & - & $0.053^{+0.004}_{-0.004}$\\
$\phi_{\rm ext}$ & [$^\circ$] & - & - & - & - & $-77.1^{+1.2}_{-1.2}$\\
\hline
\end{tabular}
\end{center}
\endgroup
\end{table*}
\begin{table*} 
\begingroup
\renewcommand{\arraystretch}{1.25}
\begin{center}
 \caption{Full lens light parameter posteriors derived from our modeling procedures (see \sref{sec:lens_model}).  For each filter, we give the distributions for the bulge (B) and disk (D) components of the lens, and the satellite.  $A$ represents the relative amplitude, $R_{\rm eff}$ represents the effective (half-light) radius, $n$ represents the Sérsic index, $q$ represents the minor-to-major axis ratio, and $\phi$ represents the East-of-North position angle of the major axis.}   
 \label{tab:full_lens_light_params}
 \begin{tabular}{cc|ccc|ccc|ccc}
  \hline
   & & \multicolumn{3}{c|}{F160W} & \multicolumn{3}{c|}{F814W} & \multicolumn{3}{c}{F475X}\\
   \multicolumn{2}{c|}{Parameter} & Lens (B) & Lens (D) & Satellite & Lens (B) & Lens (D) & Satellite & Lens (B) & Lens (D) & Satellite\\
  \hline


$A$ & & $112.7^{+7.5}_{-7.9}$ & $2.61^{+0.16}_{-0.16}$ & $46.5^{+1.2}_{-1.4}$ & $34.2^{+1.2}_{-1.1}$ & $1.92^{+0.08}_{-0.08}$ & $5.88^{+0.10}_{-0.10}$ & $9.54^{+0.71}_{-0.67}$ & $1.29^{+0.07}_{-0.08}$ & $2.03^{+0.03}_{-0.04}$\\
$R_{\rm eff}$ & [$''$] & $0.73^{+0.03}_{-0.03}$ & $4.92^{+0.24}_{-0.22}$ & $0.31^{+0.01}_{-0.01}$ & $0.48^{+0.01}_{-0.01}$ & $2.56^{+0.08}_{-0.08}$ & $0.42^{+0.01}_{-0.01}$ & $0.52^{+0.03}_{-0.03}$ & $3.36^{+0.19}_{-0.15}$ & $0.48^{+0.01}_{-0.01}$ \\
$n$ & & $4.48^{+0.19}_{-0.18}$ & 1 & 4 & 4 & 1 & 4 & 4 & 1 & 4 \\
$q$ & & $0.597^{+0.002}_{-0.002}$ & $0.37^{+0.02}_{-0.02}$ & $0.76^{+0.02}_{-0.02}$ & $0.628^{+0.004}_{-0.004}$ & $0.628^{+0.004}_{-0.004}$ & $0.76^{+0.02}_{-0.02}$ & $0.702^{+0.009}_{-0.009}$ & $0.702^{+0.009}_{-0.009}$ & $0.76^{+0.02}_{-0.02}$ \\
$\phi$ & [$^\circ$] & $-9.1^{+0.1}_{-0.2}$ & $1.5^{+0.8}_{-0.7}$ & $18.5^{+2.0}_{-2.1}$ & $-8.2^{+0.3}_{-0.3}$ & $-8.2^{+0.3}_{-0.3}$ & $18.5^{+2.0}_{-2.1}$ & $-8.2^{+0.3}_{-0.3}$ & $-8.2^{+0.3}_{-0.3}$ & $18.5^{+2.0}_{-2.1}$ \\
\hline
\end{tabular}
\end{center}
\endgroup
\end{table*}
\begin{table*} 
\begingroup
\renewcommand{\arraystretch}{1.25}
\begin{center}
 \caption{Sérsic source light parameter posteriors derived from our modeling procedures (see \sref{sec:lens_model}).  $A$ represents the relative amplitude, $R_{\rm eff}$ represents the effective (half-light) radius, $n$ represents the Sérsic index, $q$ represents the minor-to-major axis ratio, and $\phi$ represents the East-of-North position angle of the major axis.}  \label{tab:full_source_params}
 \begin{tabular}{cc|ccc}
  \hline
   \multicolumn{2}{c|}{Parameter} & F160W & F814W & F475X 
  \\
  \hline
$A$ & & $339^{+88}_{-99}$ & $21^{+11}_{-8}$ & $74^{+140}_{-62}$ \\
$R_{\rm eff}$ & [$''$] & $0.099^{+0.016}_{-0.017}$ & $0.038^{+0.006}_{-0.004}$ & $0.031^{+0.025}_{-0.008}$ \\
$n$ &  & 1 & 1 & 1 \\
$q$ & & $0.64^{+0.03}_{-0.03}$ & $0.64^{+0.03}_{-0.03}$ & $0.64^{+0.03}_{-0.03}$ \\
$\phi$ & [$^\circ$] & $-27^{+3}_{-3}$ & $-27^{+3}_{-3}$ & $-27^{+3}_{-3}$ \\
\hline
\end{tabular}
\end{center}
\endgroup
\end{table*}

\section{Analysis without accounting for the internal mass sheet}
\label{app:B}

To illustrate the importance of the MSD for this system, we run additional dynamical models where $\lambda_{\rm int}=1$ (i.e., removing the effect of an internal mass sheet). This is equivalent to enforcing a pure power-law model.  In Table~\ref{tab:model_final},  we present the full posteriors of our final dynamical models (for both our primary models DyM1$-$4, as well as for models where we fix $\lambda_{\rm int}=1$, denoted with a \mbox{``|$\lambda_{\rm int}=1$''} suffix in the dynamical model ID).  All parameters sampled are described in Table~\ref{tab:pop_names}.

We see that by removing the internal mass sheet altogether, we get a tighter constraint on $H_0$ (from 6.5\% to 5.2\%, while also shifting the median by $-7\%$).  This is to be expected, as $H_0$ is to first order linearly dependent on $\lambda_{\rm total}$. The increase in precision is appreciable but not dramatic, since other terms in the error budget (time delay, Fermat potential, $\kappa_{\rm ext}$) are comparable to the residual MSD and set an effective floor.

\begin{figure*}
\begin{center}
 \includegraphics[width=1\linewidth]{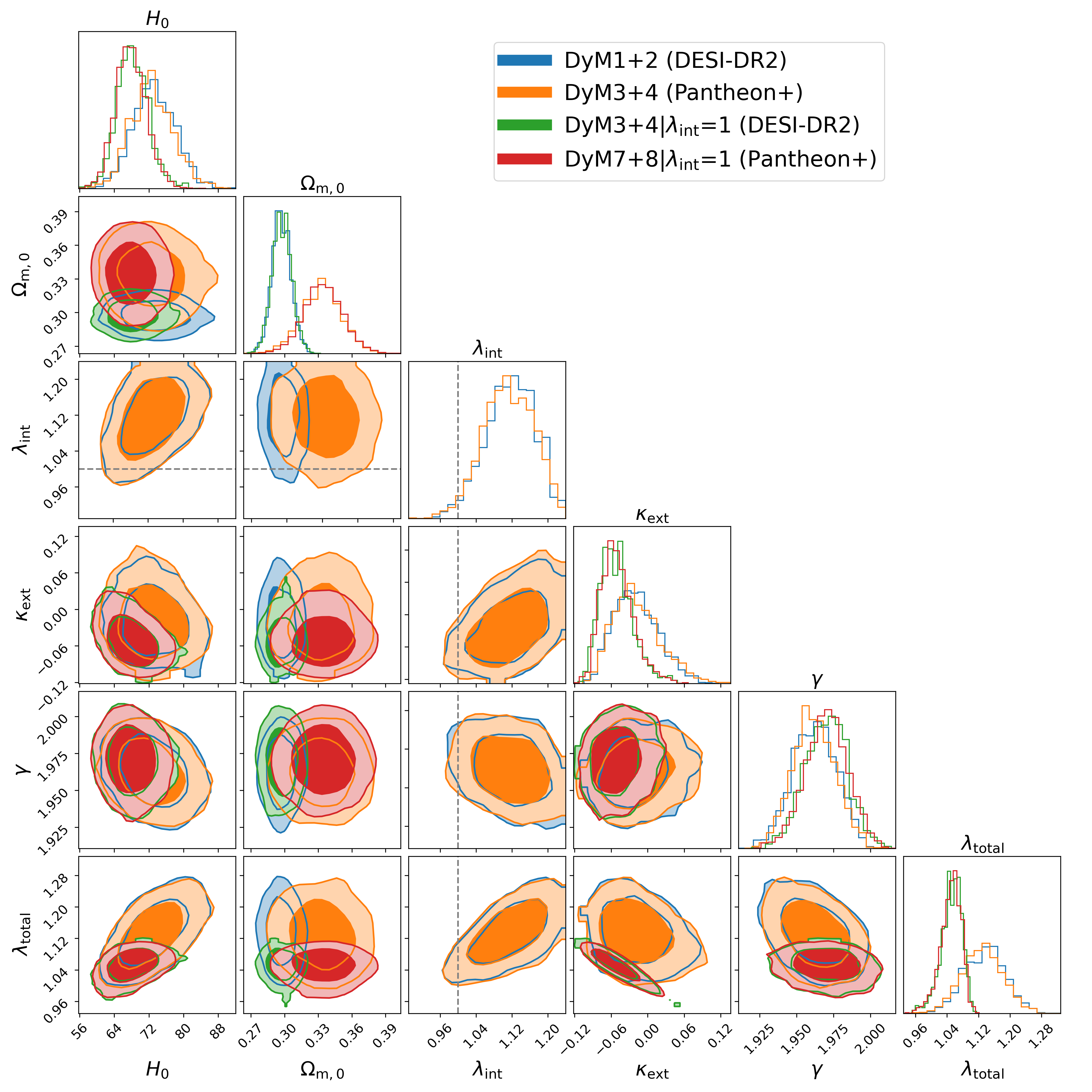}
 \caption{Our dynamical modeling results, with our primary and $\lambda_{\rm int} = 1$ model shown.  See Table~\ref{tab:model_configs} for our labeling scheme, denoting different dynamical model configurations.  The dotted line indicates $\lambda_{\rm int} = 1$.  By fixing $\lambda_{\rm int} = 1$, we observe a compensating effect on $\kappa_{\rm ext}$ (and $\gamma$ to a lesser extent), leading to a smaller discrepancy with $\lambda_{\rm total}$ and $H_0$.  All models have measurements of $H_0$ consistent with each other.} 
 \label{fig:final_res_corner_appendix}
\end{center}
\end{figure*}

Despite the tension between our models with free and fixed $\lambda_{\rm int}$, we observe a compensating effect with $\kappa_{\rm ext}$ (and to a lesser extent $\gamma$, which we address in \sref{subsec:gamma_robust}) in Figure~\ref{fig:final_res_corner_appendix}.  Without the freedom provided by an internal mass sheet, the model compensates by manifesting an external mass sheet such that the total mass sheet ($\lambda_{\rm total}$) is somewhat conserved.  This counterbalancing phenomenon decreases the effect of fixing $\lambda_{\rm int}$ on cosmology and $H_0$. 


\section{Robustness testing with the convergence slope} \label{subsec:gamma_robust}
Our results show that when the dynamical model is allowed additional freedom by not fixing $\lambda_{\rm int}$, it shifts the $\gamma$ distribution slightly higher (by $0.5\%$; see Figure~\ref{fig:final_res_corner} and Table~\ref{tab:model_final}).  This indicates that the kinematic data prefer a somewhat higher $\gamma$ than what is predicted by the lens model. To test whether this has an appreciable impact on $H_0$, we rerun DyM1+2 with double the uncertainty on $\gamma$ from the lens model posteriors.  This is done by multiplying the row and column corresponding to $\gamma$ in the lens model covariance matrix by two.  

We find that even with double the lensing uncertainty on $\gamma$, the resulting $\gamma$ dynamical posterior peak only shifts $1\%$ (from $\gamma = 1.96^{+0.02}_{-0.01}$ in DyM1+2, to $\gamma = 1.98^{+0.03}_{-0.02}$ in case 1).  The measured $H_0$ posterior peak only shifts by $-0.9\%$, a sub-percent level, much less than our estimated errors.


\begin{landscape}
\begingroup
\setlength{\tabcolsep}{3.1pt}
\renewcommand{\arraystretch}{2}
\begin{table} 
 \caption{The results from our final dynamical models, labeled by the first column, based on the configurations defined in Table~\ref{tab:model_configs}.  Excluding the columns that are directly being sampled (which are defined in Table~\ref{tab:pop_names}), we also add the column $w_{\rm BIC}$ which represents the associated BIC weight of a given model.  We see that our results for $H_0$ are especially robust against different model configurations.}   
 \label{tab:model_final}
\begin{tabular}{cc|ccccccccccccc}
    \hline
    \makecell{Dynamical\\model ID} & \makecell{$w_{\rm BIC}$\\{}} & \makecell{$\theta_{\rm E}$\\{}[$''$]} & \makecell{$\gamma$\\{}} & \makecell{$q_{\rm m}$\\{}} & \makecell{$R_{\rm eff}$\\{}[$''$]} & \makecell{$q_{\rm l}$\\{}} & \makecell{$i$\\{}[$^\circ$]} & \makecell{$\kappa_{\rm ext}$\\{}} & \makecell{$\theta_{\rm S}$\\{}[$''$]} & \makecell{{$\beta_{\rm ani}$}\\{}} & \makecell{$\log (M_{\rm BH})$\\{}[$\log M_{\odot}$]} & \makecell{${\lambda}_{\rm int}$\\{}} & \makecell{${\Omega}_{\rm m, 0}$\\{}} & \makecell{${H}_0$\\{}[\ksmpc]} \\
    \hline
DyM1 & 42.3 & 
$1.54^{+0.01}_{-0.01}$ & $1.96^{+0.02}_{-0.01}$ & $0.75^{+0.01}_{-0.01}$ & $3.05^{+0.13}_{-0.14}$ & $0.70^{+0.01}_{-0.01}$ & $89^{+22}_{-20}$ & $-0.02^{+0.04}_{-0.04}$ & $9.5^{+2.9}_{-1.4}$ & $0.11^{+0.10}_{-0.08}$ & $8.88^{+0.35}_{-0.38}$ & $1.12^{+0.05}_{-0.06}$ & $0.30^{+0.01}_{-0.01}$ & $73.1^{+4.9}_{-4.7}$ \\
DyM2 & 3.6 & 
$1.54^{+0.01}_{-0.01}$ & $1.96^{+0.01}_{-0.01}$ & $0.75^{+0.01}_{-0.01}$ & $3.05^{+0.14}_{-0.16}$ & $0.70^{+0.01}_{-0.01}$ & $88^{+20}_{-17}$ & $-0.03^{+0.04}_{-0.03}$ & $9.6^{+2.7}_{-1.5}$ & $0.10^{+0.10}_{-0.11}$ & $8.89^{+0.38}_{-0.38}$ & $1.13^{+0.05}_{-0.05}$ & $0.30^{+0.01}_{-0.01}$ & $74.4^{+4.8}_{-4.8}$ \\
DyM3 & 50.3 & 
$1.54^{+0.01}_{-0.01}$ & $1.96^{+0.02}_{-0.01}$ & $0.75^{+0.01}_{-0.01}$ & $3.04^{+0.16}_{-0.14}$ & $0.70^{+0.01}_{-0.01}$ & $90^{+21}_{-20}$ & $-0.02^{+0.05}_{-0.03}$ & $9.6^{+2.9}_{-1.4}$ & $0.11^{+0.09}_{-0.07}$ & $8.86^{+0.38}_{-0.36}$ & $1.11^{+0.06}_{-0.06}$ & $0.33^{+0.02}_{-0.02}$ & $72.5^{+5.1}_{-4.7}$ \\
DyM4 & 3.8 & 
$1.54^{+0.01}_{-0.01}$ & $1.96^{+0.01}_{-0.01}$ & $0.75^{+0.01}_{-0.01}$ & $3.04^{+0.16}_{-0.16}$ & $0.70^{+0.01}_{-0.01}$ & $91^{+19}_{-19}$ & $-0.03^{+0.04}_{-0.03}$ & $9.5^{+2.7}_{-1.3}$ & $0.09^{+0.10}_{-0.11}$ & $8.88^{+0.40}_{-0.39}$ & $1.13^{+0.05}_{-0.05}$ & $0.33^{+0.02}_{-0.02}$ & $74.5^{+4.7}_{-4.9}$ \\
\hdashline
DyM1|$\lambda_{\rm int}$=1 & 49.0 & 
$1.54^{+0.01}_{-0.01}$ & $1.97^{+0.01}_{-0.01}$ & $0.75^{+0.01}_{-0.01}$ & $3.04^{+0.17}_{-0.17}$ & $0.70^{+0.01}_{-0.01}$ & $90^{+20}_{-20}$ & $-0.05^{+0.03}_{-0.02}$ & $11.7^{+2.4}_{-2.7}$ & $0.16^{+0.09}_{-0.09}$ & $9.08^{+0.50}_{-0.47}$ & 1 & $0.30^{+0.01}_{-0.01}$ & $68.1^{+3.8}_{-3.3}$ \\
DyM2|$\lambda_{\rm int}$=1 & 2.4 & 
$1.54^{+0.01}_{-0.01}$ & $1.97^{+0.01}_{-0.01}$ & $0.75^{+0.01}_{-0.01}$ & $3.04^{+0.15}_{-0.15}$ & $0.70^{+0.01}_{-0.01}$ & $89^{+19}_{-17}$ & $-0.06^{+0.02}_{-0.02}$ & $11.6^{+2.8}_{-2.6}$ & $0.16^{+0.08}_{-0.10}$ & $9.11^{+0.47}_{-0.43}$ & 1 & $0.30^{+0.01}_{-0.01}$ & $68.4^{+3.5}_{-3.7}$ \\
DyM3|$\lambda_{\rm int}$=1 & 46.5 & 
$1.54^{+0.01}_{-0.01}$ & $1.97^{+0.01}_{-0.01}$ & $0.75^{+0.01}_{-0.01}$ & $3.04^{+0.15}_{-0.16}$ & $0.70^{+0.01}_{-0.01}$ & $91^{+20}_{-20}$ & $-0.05^{+0.03}_{-0.02}$ & $11.8^{+2.4}_{-3.0}$ & $0.15^{+0.09}_{-0.09}$ & $9.05^{+0.45}_{-0.46}$ & 1 & $0.33^{+0.02}_{-0.02}$ & $67.7^{+3.7}_{-3.3}$ \\
DyM4|$\lambda_{\rm int}$=1 & 2.1 & 
$1.54^{+0.01}_{-0.01}$ & $1.97^{+0.01}_{-0.01}$ & $0.75^{+0.01}_{-0.01}$ & $3.04^{+0.16}_{-0.14}$ & $0.70^{+0.01}_{-0.01}$ & $90^{+18}_{-18}$ & $-0.06^{+0.03}_{-0.02}$ & $11.6^{+2.6}_{-2.7}$ & $0.15^{+0.09}_{-0.10}$ & $9.07^{+0.46}_{-0.45}$ & 1 & $0.34^{+0.02}_{-0.02}$ & $68.3^{+3.7}_{-3.8}$ \\

    \hline
\end{tabular}
\end{table}
\endgroup 
\end{landscape}

\bibliographystyle{aa}
\bibliography{biblio}

@ARTICLE{B+T21,
       author = {{Birrer}, Simon and {Treu}, Tommaso},
        title = "{TDCOSMO. V. Strategies for precise and accurate measurements of the Hubble constant with strong lensing}",
      journal = {\aap},
         year = 2021,
        month = may,
       volume = {649},
          eid = {A61},
        pages = {A61},
          doi = {10.1051/0004-6361/202039179},
archivePrefix = {arXiv},
       eprint = {2008.06157},
 primaryClass = {astro-ph.CO},
       adsurl = {https://ui.adsabs.harvard.edu/abs/2021A&A...649A..61B}
}

@ARTICLE{FKM2026,
       author = {{For{\'e}s-Toribio}, R. and {Kochanek}, C.~S. and {Mu{\~n}oz}, J.~A.},
        title = "{Dynamical Systematics on Time Delay Lenses and the Impact on the Hubble Constant}",
      journal = {arXiv e-prints},
         year = 2026,
        month = feb,
          eid = {arXiv:2602.03934},
        pages = {arXiv:2602.03934},
          doi = {10.48550/arXiv.2602.03934},
archivePrefix = {arXiv},
       eprint = {2602.03934},
 primaryClass = {astro-ph.GA},
       adsurl = {https://ui.adsabs.harvard.edu/abs/2026arXiv260203934F}
}

@ARTICLE{Cappellari2016,
       author = {{Cappellari}, Michele},
        title = "{Structure and Kinematics of Early-Type Galaxies from Integral Field Spectroscopy}",
      journal = {\araa},
         year = 2016,
        month = sep,
       volume = {54},
        pages = {597-665},
          doi = {10.1146/annurev-astro-082214-122432},
archivePrefix = {arXiv},
       eprint = {1602.04267},
 primaryClass = {astro-ph.GA},
       adsurl = {https://ui.adsabs.harvard.edu/abs/2016ARA&A..54..597C}
}

@ARTICLE{VM2026,
       author = {{Verma}, Vishal and {Minor}, Quinn},
        title = "{The stellar velocity anisotropy of strong lensing massive elliptical galaxies and its role in the inference of the Hubble parameter $H_0$ using spatially resolved kinematics}",
      journal = {arXiv e-prints},
         year = 2026,
        month = feb,
          eid = {arXiv:2602.07159},
        pages = {arXiv:2602.07159},
          doi = {10.48550/arXiv.2602.07159},
archivePrefix = {arXiv},
       eprint = {2602.07159},
 primaryClass = {astro-ph.GA},
       adsurl = {https://ui.adsabs.harvard.edu/abs/2026arXiv260207159V}
}

@ARTICLE{hoyt2025,
       author = {{Hoyt}, Taylor J. and {Jang}, In Sung and {Freedman}, Wendy L. and {Madore}, Barry F. and {Owens}, Kayla A. and {Lee}, Abigail J.},
        title = "{The Chicago Carnegie Hubble Program: Improving the Calibration of SNe Ia with JWST Measurements of the Tip of the Red Giant Branch}",
      journal = {arXiv e-prints},
         year = 2025,
        month = mar,
          eid = {arXiv:2503.11769},
        pages = {arXiv:2503.11769},
          doi = {10.48550/arXiv.2503.11769},
archivePrefix = {arXiv},
       eprint = {2503.11769},
 primaryClass = {astro-ph.GA},
       adsurl = {https://ui.adsabs.harvard.edu/abs/2025arXiv250311769H}
}

@ARTICLE{schwarzschild79,
       author = {{Schwarzschild}, M.},
        title = "{A numerical model for a triaxial stellar system in dynamical equilibrium.}",
      journal = {\apj},
         year = 1979,
        month = aug,
       volume = {232},
        pages = {236-247},
          doi = {10.1086/157282},
       adsurl = {https://ui.adsabs.harvard.edu/abs/1979ApJ...232..236S}
}

@ARTICLE{Ospikov,
       author = {{Osipkov}, L.~P.},
        title = "{Spherical systems of gravitating bodies with an ellipsoidal velocity distribution}",
      journal = {Soviet Astronomy Letters},
         year = 1979,
        month = jan,
       volume = {5},
        pages = {42-44},
       adsurl = {https://ui.adsabs.harvard.edu/abs/1979SvAL....5...42O}
}

@ARTICLE{Merritt1985,
       author = {{Merritt}, D.},
        title = "{Spherical stellar systems with spheroidal velocity distributions}",
      journal = {\aj},
         year = 1985,
        month = jun,
       volume = {90},
        pages = {1027-1037},
          doi = {10.1086/113810},
       adsurl = {https://ui.adsabs.harvard.edu/abs/1985AJ.....90.1027M}
}

@ARTICLE{Paic2026,
       author = {{Paic}, Eric and {Courbin}, Fr{\'e}d{\'e}ric and {Fassnacht}, Christopher D. and {Galan}, Aymeric and {Millon}, Martin and {Sluse}, Dominique and {Williams}, Devon M. and {Birrer}, Simon and {Buckley-Geer}, Elizabeth J. and {Cappellari}, Michele and {Dux}, Fr{\'e}d{\'e}ric and {Huang}, Xiang-Yu and {Knabel}, Shawn and {Lemon}, Cameron and {Shajib}, Anowar J. and {Suyu}, Sherry H. and {Treu}, Tommaso and {Wong}, Kenneth C. and {Christensen}, Lise and {Motta}, Veronica and {Sonnenfeld}, Alessandro},
        title = "{TDCOSMO: XXIII. Measurement of the Hubble constant from the doubly lensed quasar HE 1104{\ensuremath{-}}1805}",
      journal = {\aap},
         year = 2026,
        month = feb,
       volume = {706},
          eid = {A270},
        pages = {A270},
          doi = {10.1051/0004-6361/202556411},
archivePrefix = {arXiv},
       eprint = {2512.03178},
 primaryClass = {astro-ph.GA},
       adsurl = {https://ui.adsabs.harvard.edu/abs/2026A&A...706A.270P}
}

@ARTICLE{Suyu2010,
       author = {{Suyu}, S.~H. and {Marshall}, P.~J. and {Auger}, M.~W. and {Hilbert}, S. and {Blandford}, R.~D. and {Koopmans}, L.~V.~E. and {Fassnacht}, C.~D. and {Treu}, T.},
        title = "{Dissecting the Gravitational lens B1608+656. II. Precision Measurements of the Hubble Constant, Spatial Curvature, and the Dark Energy Equation of State}",
      journal = {\apj},
         year = 2010,
        month = mar,
       volume = {711},
       number = {1},
        pages = {201-221},
          doi = {10.1088/0004-637X/711/1/201},
archivePrefix = {arXiv},
       eprint = {0910.2773},
 primaryClass = {astro-ph.CO},
       adsurl = {https://ui.adsabs.harvard.edu/abs/2010ApJ...711..201S}
}

@ARTICLE{Kelly2015,
       author = {{Kelly}, Patrick L. and {Rodney}, Steven A. and {Treu}, Tommaso and {Foley}, Ryan J. and {Brammer}, Gabriel and {Schmidt}, Kasper B. and {Zitrin}, Adi and {Sonnenfeld}, Alessandro and {Strolger}, Louis-Gregory and {Graur}, Or and {Filippenko}, Alexei V. and {Jha}, Saurabh W. and {Riess}, Adam G. and {Bradac}, Marusa and {Weiner}, Benjamin J. and {Scolnic}, Daniel and {Malkan}, Matthew A. and {von der Linden}, Anja and {Trenti}, Michele and {Hjorth}, Jens and {Gavazzi}, Raphael and {Fontana}, Adriano and {Merten}, Julian C. and {McCully}, Curtis and {Jones}, Tucker and {Postman}, Marc and {Dressler}, Alan and {Patel}, Brandon and {Cenko}, S. Bradley and {Graham}, Melissa L. and {Tucker}, Bradley E.},
        title = "{Multiple images of a highly magnified supernova formed by an early-type cluster galaxy lens}",
      journal = {Science},
         year = 2015,
        month = mar,
       volume = {347},
       number = {6226},
        pages = {1123-1126},
          doi = {10.1126/science.aaa3350},
archivePrefix = {arXiv},
       eprint = {1411.6009},
 primaryClass = {astro-ph.CO},
       adsurl = {https://ui.adsabs.harvard.edu/abs/2015Sci...347.1123K}
}

@ARTICLE{Kelly2023,
       author = {{Kelly}, Patrick L. and {Rodney}, Steven and {Treu}, Tommaso and {Oguri}, Masamune and {Chen}, Wenlei and {Zitrin}, Adi and {Birrer}, Simon and {Bonvin}, Vivien and {Dessart}, Luc and {Diego}, Jose M. and {Filippenko}, Alexei V. and {Foley}, Ryan J. and {Gilman}, Daniel and {Hjorth}, Jens and {Jauzac}, Mathilde and {Mandel}, Kaisey and {Millon}, Martin and {Pierel}, Justin and {Sharon}, Keren and {Thorp}, Stephen and {Williams}, Liliya and {Broadhurst}, Tom and {Dressler}, Alan and {Graur}, Or and {Jha}, Saurabh and {McCully}, Curtis and {Postman}, Marc and {Schmidt}, Kasper Borello and {Tucker}, Brad E. and {von der Linden}, Anja},
        title = "{Constraints on the Hubble constant from supernova Refsdal's reappearance}",
      journal = {Science},
         year = 2023,
        month = jun,
       volume = {380},
       number = {6649},
          eid = {abh1322},
        pages = {abh1322},
          doi = {10.1126/science.abh1322},
archivePrefix = {arXiv},
       eprint = {2305.06367},
 primaryClass = {astro-ph.CO},
       adsurl = {https://ui.adsabs.harvard.edu/abs/2023Sci...380.1322K}
}

@ARTICLE{Jensen2025,
       author = {{Jensen}, Joseph B. and {Blakeslee}, John P. and {Cantiello}, Michele and {Cowles}, Mikaela and {Anand}, Gagandeep S. and {Tully}, R. Brent and {Kourkchi}, Ehsan and {Raimondo}, Gabriella},
        title = "{The TRGB{\ensuremath{-}}SBF Project. III. Refining the HST Surface Brightness Fluctuation Distance Scale Calibration with JWST}",
      journal = {\apj},
         year = 2025,
        month = jul,
       volume = {987},
       number = {1},
          eid = {87},
        pages = {87},
          doi = {10.3847/1538-4357/addfd6},
archivePrefix = {arXiv},
       eprint = {2502.15935},
 primaryClass = {astro-ph.CO},
       adsurl = {https://ui.adsabs.harvard.edu/abs/2025ApJ...987...87J}
}

@ARTICLE{Pesce2020,
       author = {{Pesce}, D.~W. and {Braatz}, J.~A. and {Reid}, M.~J. and {Riess}, A.~G. and {Scolnic}, D. and {Condon}, J.~J. and {Gao}, F. and {Henkel}, C. and {Impellizzeri}, C.~M.~V. and {Kuo}, C.~Y. and {Lo}, K.~Y.},
        title = "{The Megamaser Cosmology Project. XIII. Combined Hubble Constant Constraints}",
      journal = {\apjl},
         year = 2020,
        month = mar,
       volume = {891},
       number = {1},
          eid = {L1},
        pages = {L1},
          doi = {10.3847/2041-8213/ab75f0},
archivePrefix = {arXiv},
       eprint = {2001.09213},
 primaryClass = {astro-ph.CO},
       adsurl = {https://ui.adsabs.harvard.edu/abs/2020ApJ...891L...1P}
}

@ARTICLE{TreuMarshall2016,
       author = {{Treu}, Tommaso and {Marshall}, Philip J.},
        title = "{Time delay cosmography}",
      journal = {\aapr},
         year = 2016,
        month = jul,
       volume = {24},
       number = {1},
          eid = {11},
        pages = {11},
          doi = {10.1007/s00159-016-0096-8},
archivePrefix = {arXiv},
       eprint = {1605.05333},
 primaryClass = {astro-ph.CO},
       adsurl = {https://ui.adsabs.harvard.edu/abs/2016A&ARv..24...11T}
}

@INPROCEEDINGS{kcrm,
       author = {{McGurk}, Rosalie C. and {Matuszewski}, Mateusz and {Neill}, James D. and {Martin}, Chris and {Bertz}, Robert and {Rockosi}, Constance and {Kassis}, Marc F.},
        title = "{The Keck Cosmic Reionization Mapper project: adding red spectroscopy to the Keck Cosmic Web Imager Integral Field Spectrograph}",
    booktitle = {Ground-based and Airborne Instrumentation for Astronomy X},
         year = 2024,
       editor = {{Bryant}, Julia J. and {Motohara}, Kentaro and {Vernet}, Jo{\"e}l. R.~D.},
       series = {Society of Photo-Optical Instrumentation Engineers (SPIE) Conference Series},
       volume = {13096},
        month = jul,
          eid = {1309647},
        pages = {1309647},
          doi = {10.1117/12.3020646},
       adsurl = {https://ui.adsabs.harvard.edu/abs/2024SPIE13096E..47M}
}

@article{Dux_2024,
   title={lightcurver: A Python Pipeline for Precise Photometry
of Multiple-Epoch Wide-Field Images},
   volume={9},
   ISSN={2475-9066},
   url={http://dx.doi.org/10.21105/joss.06775},
   DOI={10.21105/joss.06775},
   number={102},
   journal={Journal of Open Source Software},
   publisher={The Open Journal},
   author={Dux, Frédéric},
   year={2024},
   month=oct, pages={6775} }

@article{Millon_2020,
   title={COSMOGRAIL: XIX. Time delays in 18 strongly lensed quasars from 15 years of optical monitoring},
   volume={640},
   ISSN={1432-0746},
   url={http://dx.doi.org/10.1051/0004-6361/202037740},
   DOI={10.1051/0004-6361/202037740},
   journal={Astronomy \& Astrophysics},
   publisher={EDP Sciences},
   author={Millon, M. and Courbin, F. and Bonvin, V. and Paic, E. and Meylan, G. and Tewes, M. and Sluse, D. and Magain, P. and Chan, J. H. H. and Galan, A. and Joseph, R. and Lemon, C. and Tihhonova, O. and Anderson, R. I. and Marmier, M. and Chazelas, B. and Lendl, M. and Triaud, A. H. M. J. and Wyttenbach, A.},
   year={2020},
   month=aug, pages={A105} }

@article{Dux_2025,
   title={TDCOSMO: XVII. New time delays in 22 lensed quasars from optical monitoring with the ESO-VST 2.6m and MPG 2.2m telescopes},
   volume={697},
   ISSN={1432-0746},
   url={http://dx.doi.org/10.1051/0004-6361/202553807},
   DOI={10.1051/0004-6361/202553807},
   journal={Astronomy \& Astrophysics},
   publisher={EDP Sciences},
   author={Dux, F. and Millon, M. and Galan, A. and Paic, E. and Lemon, C. and Courbin, F. and Bonvin, V. and Anguita, T. and Auger, M. and Birrer, S. and Buckley-Geer, E. and Fassnacht, C. D. and Frieman, J. and McMahon, R. G. and Marshall, P. J. and Melo, A. and Motta, V. and Neira, F. and Sluse, D. and Suyu, S. H. and Treu, T. and Agnello, A. and Ávila, F. and Chan, J. and Chijani, M. and Rojas, K. and Hempel, A. and Hempel, M. and Kim, S. and Eigenthaler, P. and Lachaume, R. and Rabus, M.},
   year={2025},
   month=may, pages={A139} }

@ARTICLE{Moffat1969,
       author = {{Moffat}, A.~F.~J.},
        title = "{A Theoretical Investigation of Focal Stellar Images in the Photographic Emulsion and Application to Photographic Photometry}",
      journal = {\aap},
         year = 1969,
        month = dec,
       volume = {3},
        pages = {455},
       adsurl = {https://ui.adsabs.harvard.edu/abs/1969A&A.....3..455M}
}

@article{Freedman_2025,
doi = {10.3847/1538-4357/adce78},
url = {https://doi.org/10.3847/1538-4357/adce78},
year = {2025},
month = {may},
publisher = {The American Astronomical Society},
volume = {985},
number = {2},
pages = {203},
author = {Freedman, Wendy L. and Madore, Barry F. and Hoyt, Taylor J. and Jang, In Sung and Lee, Abigail J. and Owens, Kayla A.},
title = {Status Report on the Chicago-Carnegie Hubble Program (CCHP): Measurement of the Hubble Constant Using the Hubble and James Webb Space Telescopes},
journal = {The Astrophysical Journal}
}

@ARTICLE{desidr2,
       author = {{DESI Collaboration} and {Abdul Karim}, M. and {Aguilar}, J. and {Ahlen}, S. and {Alam}, S. and {Allen}, L. and {Allende Prieto}, C. and {Alves}, O. and {Anand}, A. and {Andrade}, U. and {Armengaud}, E. and {Aviles}, A. and {Bailey}, S. and {Baltay}, C. and {Bansal}, P. and {Bault}, A. and {Behera}, J. and {BenZvi}, S. and {Bianchi}, D. and {Blake}, C. and {Brieden}, S. and {Brodzeller}, A. and {Brooks}, D. and {Buckley-Geer}, E. and {Burtin}, E. and {Calderon}, R. and {Canning}, R. and {Rosell}, A. Carnero and {Carrilho}, P. and {Casas}, L. and {Castander}, F.~J. and {Charles}, M. and {Chaussidon}, E. and {Chaves-Montero}, J. and {Chebat}, D. and {Chen}, X. and {Claybaugh}, T. and {Cole}, S. and {Cooper}, A.~P. and {Cuceu}, A. and {Dawson}, K.~S. and {de la Macorra}, A. and {de Mattia}, A. and {Deiosso}, N. and {Della Costa}, J. and {Demina}, R. and {Dey}, A. and {Dey}, B. and {Ding}, Z. and {Doel}, P. and {Edelstein}, J. and {Eisenstein}, D.~J. and {Elbers}, W. and {Fagrelius}, P. and {Fanning}, K. and {Fern{\'a}ndez-Garc{\'\i}a}, E. and {Ferraro}, S. and {Font-Ribera}, A. and {Forero-Romero}, J.~E. and {Frenk}, C.~S. and {Garcia-Quintero}, C. and {Garrison}, L.~H. and {Gazta{\~n}aga}, E. and {Gil-Mar{\'\i}n}, H. and {Gontcho A Gontcho}, S. and {Gonzalez}, D. and {Gonzalez-Morales}, A.~X. and {Gordon}, C. and {Green}, D. and {Gutierrez}, G. and {Guy}, J. and {Hadzhiyska}, B. and {Hahn}, C. and {He}, S. and {Herbold}, M. and {Herrera-Alcantar}, H.~K. and {Ho}, M.-F. and {Honscheid}, K. and {Howlett}, C. and {Huterer}, D. and {Ishak}, M. and {Juneau}, S. and {Kamble}, N.~V. and {Kara{\c{c}}ayl{\i}}, N.~G. and {Kehoe}, R. and {Kent}, S. and {Kim}, A.~G. and {Kirkby}, D. and {Kisner}, T. and {Koposov}, S.~E. and {Kremin}, A. and {Krolewski}, A. and {Lahav}, O. and {Lamman}, C. and {Landriau}, M. and {Lang}, D. and {Lasker}, J. and {Le Goff}, J.~M. and {Le Guillou}, L. and {Leauthaud}, A. and {Levi}, M.~E. and {Li}, Q. and {Li}, T.~S. and {Lodha}, K. and {Lokken}, M. and {Lozano-Rodr{\'\i}guez}, F. and {Magneville}, C. and {Manera}, M. and {Martini}, P. and {Matthewson}, W.~L. and {Meisner}, A. and {Mena-Fern{\'a}ndez}, J. and {Menegas}, A. and {Mergulh{\~a}o}, T. and {Miquel}, R. and {Moustakas}, J. and {Mu{\~n}oz-Guti{\'e}rrez}, A. and {Mu{\~n}oz-Santos}, D. and {Myers}, A.~D. and {Nadathur}, S. and {Naidoo}, K. and {Napolitano}, L. and {Newman}, J.~A. and {Niz}, G. and {Noriega}, H.~E. and {Paillas}, E. and {Palanque-Delabrouille}, N. and {Pan}, J. and {Peacock}, J.~A. and {Pellejero Ibanez}, M. and {Percival}, W.~J. and {P{\'e}rez-Fern{\'a}ndez}, A. and {P{\'e}rez-R{\`a}fols}, I. and {Pieri}, M.~M. and {Poppett}, C. and {Prada}, F. and {Rabinowitz}, D. and {Raichoor}, A. and {Ram{\'\i}rez-P{\'e}rez}, C. and {Rashkovetskyi}, M. and {Ravoux}, C. and {Rich}, J. and {Rocher}, A. and {Rockosi}, C. and {Rohlf}, J. and {Rom{\'a}n-Herrera}, J.~O. and {Ross}, A.~J. and {Rossi}, G. and {Ruggeri}, R. and {Ruhlmann-Kleider}, V. and {Samushia}, L. and {Sanchez}, E. and {Sanders}, N. and {Schlegel}, D. and {Schubnell}, M. and {Seo}, H. and {Shafieloo}, A. and {Sharples}, R. and {Silber}, J. and {Sinigaglia}, F. and {Sprayberry}, D. and {Tan}, T. and {Tarl{\'e}}, G. and {Taylor}, P. and {Turner}, W. and {Ure{\~n}a-L{\'o}pez}, L.~A. and {Vaisakh}, R. and {Valdes}, F. and {Valogiannis}, G. and {Vargas-Maga{\~n}a}, M. and {Verde}, L. and {Walther}, M. and {Weaver}, B.~A. and {Weinberg}, D.~H. and {White}, M. and {Wolfson}, M. and {Y{\`e}che}, C. and {Yu}, J. and {Zaborowski}, E.~A. and {Zarrouk}, P. and {Zhai}, Z. and {Zhang}, H. and {Zhao}, C. and {Zhao}, G.~B. and {Zhou}, R. and {Zou}, H. and {DESI Collaboration}},
        title = "{DESI DR2 results. II. Measurements of baryon acoustic oscillations and cosmological constraints}",
      journal = {\prd},
         year = 2025,
        month = oct,
       volume = {112},
       number = {8},
          eid = {083515},
        pages = {083515},
          doi = {10.1103/tr6y-kpc6},
archivePrefix = {arXiv},
       eprint = {2503.14738},
 primaryClass = {astro-ph.CO},
       adsurl = {https://ui.adsabs.harvard.edu/abs/2025PhRvD.112h3515A}
}

@article{Sheuc_2024,
   title={The Carousel Lens: A Well-modeled Strong Lens with Multiple Sources Spectroscopically Confirmed by VLT/MUSE},
   volume={973},
   ISSN={1538-4357},
   url={http://dx.doi.org/10.3847/1538-4357/ad65d3},
   DOI={10.3847/1538-4357/ad65d3},
   number={1},
   journal={The Astrophysical Journal},
   publisher={American Astronomical Society},
   author={Sheu, William and Cikota, Aleksandar and Huang, Xiaosheng and Glazebrook, Karl and Storfer, Christopher and Agarwal, Shrihan and Schlegel, David J. and Suzuki, Nao and Barone, Tania M. and Bian, Fuyan and Jeltema, Tesla and Jones, Tucker and Kacprzak, Glenn G. and O’Donnell, Jackson H. and G. C., Keerthi Vasan},
   year={2024},
   month=sep, pages={3} }

@ARTICLE{COSMOVERSE25,
       author = {Eleonora Di Valentino and {Said}, Jackson Levi and {Riess}, Adam and {Pollo}, Agnieszka and {Poulin}, Vivian and {G{\'o}mez-Valent}, Adri{\`a} and {Weltman}, Amanda and {Palmese}, Antonella and {Huang}, Caroline D. and {Bruck}, Carsten van de and {Saraf}, Chandra Shekhar and {Kuo}, Cheng-Yu and {Uhlemann}, Cora and {Grand{\'o}n}, Daniela and {Paz}, Dante and {Eckert}, Dominique and {Teixeira}, Elsa M. and {Saridakis}, Emmanuel N. and {Colg{\'a}in}, Eoin {\'O}. and {Beutler}, Florian and {Niedermann}, Florian and {Bajardi}, Francesco and {Barenboim}, Gabriela and {Gubitosi}, Giulia and {Musella}, Ilaria and {Banik}, Indranil and {Szapudi}, Istvan and {Singal}, Jack and {Cases}, Jaume Haro and {Chluba}, Jens and {Torrado}, Jes{\'u}s and {Mifsud}, Jurgen and {Jedamzik}, Karsten and {Said}, Khaled and {Dialektopoulos}, Konstantinos and {Herold}, Laura and {Perivolaropoulos}, Leandros and {Zu}, Lei and {Galbany}, Llu{\'\i}s and {Breuval}, Louise and {Visinelli}, Luca and {Escamilla}, Luis A. and {Anchordoqui}, Luis A. and {Sheikh-Jabbari}, M.~M. and {Lembo}, Margherita and {Dainotti}, Maria Giovanna and {Vincenzi}, Maria and {Asgari}, Marika and {Gerbino}, Martina and {Forconi}, Matteo and {Cantiello}, Michele and {Moresco}, Michele and {Benetti}, Micol and {Sch{\"o}neberg}, Nils and {Akarsu}, {\"O}zg{\"u}r and {Nunes}, Rafael C. and {Bernardo}, Reginald Christian and {Ch{\'a}vez}, Ricardo and {Anderson}, Richard I. and {Watkins}, Richard and {Capozziello}, Salvatore and {Li}, Siyang and {Vagnozzi}, Sunny and {Pan}, Supriya and {Treu}, Tommaso and {Irsic}, Vid and {Handley}, Will and {Giar{\`e}}, William and {Murakami}, Yukei and {Banihashemi}, Abdolali and {Poudou}, Ad{\`e}le and {Heavens}, Alan and {Kogut}, Alan and {Domi}, Alba and {Lenart}, Aleksander {\L}ukasz and {Melchiorri}, Alessandro and {Vadal{\`a}}, Alessandro and {Amon}, Alexandra and {Rivera}, Alexander Bonilla and {Reeves}, Alexander and {Zhuk}, Alexander and {Bonanno}, Alfio and {{\"O}vg{\"u}n}, Ali and {Pisani}, Alice and {Talebian}, Alireza and {Abebe}, Amare and {Aboubrahim}, Amin and {Mor{\'a}n}, Ana Luisa Gonz{\'a}lez and {Kov{\'a}cs}, Andr{\'a}s and {Lymperis}, Andreas and {Papatriantafyllou}, Andreas and {Liddle}, Andrew R. and {Paliathanasis}, Andronikos and {Borowiec}, Andrzej and {Yadav}, Anil Kumar and {Yadav}, Anita and {Sen}, Anjan Ananda and {William}, Anjitha John and {Davis}, Anne Christine and {Shajib}, Anowar J. and {Walters}, Anthony and {Lonappan}, Anto Idicherian and {Chudaykin}, Anton and {Capodagli}, Antonio and {Silva}, Antonio da and {Felice}, Antonio De and {Racioppi}, Antonio and {Oficial}, Araceli Soler and {Montiel}, Ariadna and {Favale}, Arianna and {Bernui}, Armando and {Velasco}, Arrianne Crystal and {Heinesen}, Asta and {Bakopoulos}, Athanasios and {Chatzistavrakidis}, Athanasios and {Khanpour}, Bahman and {Sathyaprakash}, Bangalore S. and {Zgirski}, Bartek and {L'Huillier}, Benjamin and {Famaey}, Benoit and {Jain}, Bhuvnesh and {Zhang}, Bing and {Karmakar}, Biswajit and {Dragovich}, Branko and {Thomas}, Brooks and {Correa}, Carlos and {Boiza}, Carlos G. and {Marques}, Catarina and {Escamilla-Rivera}, Celia and {Tzerefos}, Charalampos and {Zhang}, Chi and {Leo}, Chiara De and {Pfeifer}, Christian and {Lee}, Christine and {Venter}, Christo and {Gomes}, Cl{\'a}udio and {bom}, Clecio Roque De and {Moreno-Pulido}, Cristian and {Iosifidis}, Damianos and {Grin}, Dan and {Blixt}, Daniel and {Scolnic}, Dan and {Oriti}, Daniele and {Dobrycheva}, Daria and {Bettoni}, Dario and {Benisty}, David and {Fern{\'a}ndez-Arenas}, David and {Wiltshire}, David L. and {Cid}, David Sanchez and {Tamayo}, David and {Valls-Gabaud}, David and {Pedrotti}, Davide and {Wang}, Deng and {Staicova}, Denitsa and {Totolou}, Despoina and {Rubiera-Garcia}, Diego and {Milakovi{\'c}}, Dinko and {Pesce}, Dominic W. and {Sluse}, Dominique and {Borka}, Du{\v{s}}ko and {Yusofi}, Ebrahim and {Giusarma}, Elena and {Terlevich}, Elena and {Tomasetti}, Elena and {Vagenas}, Elias C. and {Fazzari}, Elisa and {Ferreira}, Elisa G.~M. and {Barakovic}, Elvis and {Dimastrogiovanni}, Emanuela and {Holm}, Emil Brinch and {Mottola}, Emil and {{\"O}z{\"u}lker}, Emre and {Specogna}, Enrico and {Brocato}, Enzo and {Jensko}, Erik and {Enriquez}, Erika Antonette and {Bhatia}, Esha and {Bresolin}, Fabio and {Avila}, Felipe and {Bouch{\`e}}, Filippo and {Bombacigno}, Flavio and {Anagnostopoulos}, Fotios K. and {Pace}, Francesco and {Sorrenti}, Francesco and {Lobo}, Francisco S.~N. and {Courbin}, Fr{\'e}d{\'e}ric and {Hansen}, Frode K. and {Sloan}, Greg and {Farrugia}, Gabriel and {Lynch}, Gabriel and {Garcia-Arroyo}, Gabriela and {Raimondo}, Gabriella and {Lambiase}, Gaetano and {Anand}, Gagandeep S. and {Poulot}, Gaspard and {Leon}, Genly and {Kouniatalis}, Gerasimos and {Nardini}, Germano and {Cs{\"o}rnyei}, G{\'e}za and {Galloni}, Giacomo},
        title = "{The CosmoVerse White Paper: Addressing observational tensions in cosmology with systematics and fundamental physics}",
      journal = {Physics of the Dark Universe},
         year = 2025,
        month = sep,
       volume = {49},
          eid = {101965},
        pages = {101965},
          doi = {10.1016/j.dark.2025.101965},
archivePrefix = {arXiv},
       eprint = {2504.01669},
 primaryClass = {astro-ph.CO},
       adsurl = {https://ui.adsabs.harvard.edu/abs/2025PDU....4901965D}
}

@Article{Krajnovic2006,
    title = {Kinemetry: a generalization of photometry to the higher moments
        of the line-of-sight velocity distribution},
    author = {{Krajnovi{\'c}}, D. and {Cappellari}, M. and {de Zeeuw}, P.~T.
        and {Copin}, Y.},
    journal = {MNRAS},
    eprint = {arXiv:astro-ph/0512200},
    year = {2006},
    pages = {787-802},
    volume = {366},
    doi = {10.1111/j.1365-2966.2005.09902.x}
}

@article{Brout_2022,
   title={The Pantheon+ Analysis: Cosmological Constraints},
   volume={938},
   ISSN={1538-4357},
   url={http://dx.doi.org/10.3847/1538-4357/ac8e04},
   DOI={10.3847/1538-4357/ac8e04},
   number={2},
   journal={The Astrophysical Journal},
   publisher={American Astronomical Society},
   author={Brout, Dillon and Scolnic, Dan and Popovic, Brodie and Riess, Adam G. and Carr, Anthony and Zuntz, Joe and Kessler, Rick and Davis, Tamara M. and Hinton, Samuel and Jones, David and Kenworthy, W. D’Arcy and Peterson, Erik R. and Said, Khaled and Taylor, Georgie and Ali, Noor and Armstrong, Patrick and Charvu, Pranav and Dwomoh, Arianna and Meldorf, Cole and Palmese, Antonella and Qu, Helen and Rose, Benjamin M. and Sanchez, Bruno and Stubbs, Christopher W. and Vincenzi, Maria and Wood, Charlotte M. and Brown, Peter J. and Chen, Rebecca and Chambers, Ken and Coulter, David A. and Dai, Mi and Dimitriadis, Georgios and Filippenko, Alexei V. and Foley, Ryan J. and Jha, Saurabh W. and Kelsey, Lisa and Kirshner, Robert P. and Möller, Anais and Muir, Jessie and Nadathur, Seshadri and Pan, Yen-Chen and Rest, Armin and Rojas-Bravo, Cesar and Sako, Masao and Siebert, Matthew R. and Smith, Mat and Stahl, Benjamin E. and Wiseman, Phil},
   year={2022},
   month=oct, pages={110} }

@article{Birrer_2020,
   title={TDCOSMO: IV. Hierarchical time-delay cosmography – joint inference of the Hubble constant and galaxy density profiles},
   volume={643},
   ISSN={1432-0746},
   url={http://dx.doi.org/10.1051/0004-6361/202038861},
   DOI={10.1051/0004-6361/202038861},
   journal={Astronomy \& Astrophysics},
   publisher={EDP Sciences},
   author={Birrer, S. and Shajib, A. J. and Galan, A. and Millon, M. and Treu, T. and Agnello, A. and Auger, M. and Chen, G. C.-F. and Christensen, L. and Collett, T. and Courbin, F. and Fassnacht, C. D. and Koopmans, L. V. E. and Marshall, P. J. and Park, J.-W. and Rusu, C. E. and Sluse, D. and Spiniello, C. and Suyu, S. H. and Wagner-Carena, S. and Wong, K. C. and Barnabè, M. and Bolton, A. S. and Czoske, O. and Ding, X. and Frieman, J. A. and Van de Vyvere, L.},
   year={2020},
   month=nov, pages={A165} }

@Article{Cappellari2002,
    author = {Cappellari, Michele},
    title = {Efficient multi-Gaussian expansion of galaxies},
    journal = {MNRAS},
    eprint = {arXiv:astro-ph/0201430},
    year = {2002},
    volume = {333},
    pages = {400-410},
    doi = {10.1046/j.1365-8711.2002.05412.x}
}

@ARTICLE{blum2020,
       author = {{Blum}, Kfir and {Castorina}, Emanuele and {Simonovi{\'c}}, Marko},
        title = "{Could Quasar Lensing Time Delays Hint to a Core Component in Halos, Instead of H$_{0}$ Tension?}",
      journal = {\apjl},
         year = 2020,
        month = apr,
       volume = {892},
       number = {2},
          eid = {L27},
        pages = {L27},
          doi = {10.3847/2041-8213/ab8012},
archivePrefix = {arXiv},
       eprint = {2001.07182},
 primaryClass = {astro-ph.CO},
       adsurl = {https://ui.adsabs.harvard.edu/abs/2020ApJ...892L..27B}
}

@article{Li_2018,
   title={SDSS-IV MaNGA: The Intrinsic Shape of Slow Rotator Early-type Galaxies},
   volume={863},
   ISSN={2041-8213},
   url={http://dx.doi.org/10.3847/2041-8213/aad54b},
   DOI={10.3847/2041-8213/aad54b},
   number={2},
   journal={The Astrophysical Journal Letters},
   publisher={American Astronomical Society},
   author={Li, Hongyu and Mao, Shude and Cappellari, Michele and Graham, Mark T. and Emsellem, Eric and Long, R. J.},
   year={2018},
   month=aug, pages={L19} }

@ARTICLE{Springel2005,
       author = {{Springel}, Volker and {White}, Simon D.~M. and {Jenkins}, Adrian and {Frenk}, Carlos S. and {Yoshida}, Naoki and {Gao}, Liang and {Navarro}, Julio and {Thacker}, Robert and {Croton}, Darren and {Helly}, John and {Peacock}, John A. and {Cole}, Shaun and {Thomas}, Peter and {Couchman}, Hugh and {Evrard}, August and {Colberg}, J{\"o}rg and {Pearce}, Frazer},
        title = "{Simulations of the formation, evolution and clustering of galaxies and quasars}",
      journal = {\nat},
         year = 2005,
        month = jun,
       volume = {435},
       number = {7042},
        pages = {629-636},
          doi = {10.1038/nature03597},
archivePrefix = {arXiv},
       eprint = {astro-ph/0504097},
 primaryClass = {astro-ph},
       adsurl = {https://ui.adsabs.harvard.edu/abs/2005Natur.435..629S}
}

@article{Poulin_2019,
   title={Early Dark Energy can Resolve the Hubble Tension},
   volume={122},
   ISSN={1079-7114},
   url={http://dx.doi.org/10.1103/PhysRevLett.122.221301},
   DOI={10.1103/physrevlett.122.221301},
   number={22},
   journal={Physical Review Letters},
   publisher={American Physical Society (APS)},
   author={Poulin, Vivian and Smith, Tristan L. and Karwal, Tanvi and Kamionkowski, Marc},
   year={2019},
   month=jun }

@ARTICLE{dinosII,
       author = {{Sheu}, William and {Shajib}, Anowar J. and {Treu}, Tommaso and {Sonnenfeld}, Alessandro and {Birrer}, Simon and {Cappellari}, Michele and {Oldham}, Lindsay J. and {Tan}, Chin Yi},
        title = "{Project Dinos II: redshift evolution of dark and luminous matter density profiles in strong-lensing elliptical galaxies across 0.1 < z < 0.9}",
      journal = {\mnras},
         year = 2025,
        month = jul,
       volume = {541},
       number = {1},
        pages = {1-27},
          doi = {10.1093/mnras/staf976},
archivePrefix = {arXiv},
       eprint = {2408.10316},
 primaryClass = {astro-ph.GA},
       adsurl = {https://ui.adsabs.harvard.edu/abs/2025MNRAS.541....1S}
}

@ARTICLE{dinosI,
       author = {{Tan}, Chin Yi and {Shajib}, Anowar J. and {Birrer}, Simon and {Sonnenfeld}, Alessandro and {Treu}, Tommaso and {Wells}, Patrick and {Williams}, Devon and {Buckley-Geer}, Elizabeth J. and {Drlica-Wagner}, Alex and {Frieman}, Joshua},
        title = "{Project Dinos I: A joint lensing-dynamics constraint on the deviation from the power law in the mass profile of massive ellipticals}",
      journal = {\mnras},
         year = 2024,
        month = may,
       volume = {530},
       number = {2},
        pages = {1474-1505},
          doi = {10.1093/mnras/stae884},
archivePrefix = {arXiv},
       eprint = {2311.09307},
 primaryClass = {astro-ph.GA},
       adsurl = {https://ui.adsabs.harvard.edu/abs/2024MNRAS.530.1474T}
}

@ARTICLE{Graham2013,
       author = {{Graham}, Alister W. and {Scott}, Nicholas},
        title = "{The M $_{BH}$-L $_{spheroid}$ Relation at High and Low Masses, the Quadratic Growth of Black Holes, and Intermediate-mass Black Hole Candidates}",
      journal = {\apj},
         year = 2013,
        month = feb,
       volume = {764},
       number = {2},
          eid = {151},
        pages = {151},
          doi = {10.1088/0004-637X/764/2/151},
archivePrefix = {arXiv},
       eprint = {1211.3199},
 primaryClass = {astro-ph.CO},
       adsurl = {https://ui.adsabs.harvard.edu/abs/2013ApJ...764..151G}
}

@INPROCEEDINGS{Cappellari2026,
       author = {{Cappellari}, Michele},
        title = "{Early-type galaxies: Elliptical and S0 galaxies, or fast and slow rotators}",
    booktitle = {Encyclopedia of Astrophysics, Volume 4},
         year = 2026,
       volume = {4},
        month = jan,
        pages = {122-152},
          doi = {10.1016/B978-0-443-21439-4.00109-7},
archivePrefix = {arXiv},
       eprint = {2503.02746},
 primaryClass = {astro-ph.GA},
       adsurl = {https://ui.adsabs.harvard.edu/abs/2026enap....4..122C}
}

@article{Knox_2020,
   title={Hubble constant hunter’s guide},
   volume={101},
   ISSN={2470-0029},
   url={http://dx.doi.org/10.1103/PhysRevD.101.043533},
   DOI={10.1103/physrevd.101.043533},
   number={4},
   journal={Physical Review D},
   publisher={American Physical Society (APS)},
   author={Knox, L. and Millea, M.},
   year={2020},
   month=feb }

@ARTICLE{Cappellari2003,
    author = {{Cappellari}, M. and {Copin}, Y.},
    title = "{Adaptive spatial binning of integral-field spectroscopic
        data using Voronoi tessellations}",
    journal = {MNRAS},
    eprint = {astro-ph/0302262},
    year = 2003,
    volume = 342,
    pages = {345-354},
    doi = {10.1046/j.1365-8711.2003.06541.x}
}

@INPROCEEDINGS{Kennedy1995,
  author={Kennedy, J. and Eberhart, R.},
  booktitle={Proceedings of ICNN'95 - International Conference on Neural Networks}, 
  title={Particle swarm optimization}, 
  year={1995},
  volume={4},
  number={},
  pages={1942-1948 vol.4},
  doi={10.1109/ICNN.1995.488968}}

@article{emcee,
   title={<tt>emcee</tt>: The MCMC Hammer},
   volume={125},
   ISSN={1538-3873},
   url={http://dx.doi.org/10.1086/670067},
   DOI={10.1086/670067},
   number={925},
   journal={Publications of the Astronomical Society of the Pacific},
   publisher={IOP Publishing},
   author={Foreman-Mackey, Daniel and Hogg, David W. and Lang, Dustin and Goodman, Jonathan},
   year={2013},
   month=mar, pages={306–312} }

@ARTICLE{Cappellari2008,
    author = {Cappellari, Michele},
    title = "{Measuring the inclination and mass-to-light ratio of axisymmetric
        galaxies via anisotropic Jeans models of stellar kinematics}",
    journal = {MNRAS},
    eprint = {0806.0042},
    year = 2008,
    volume = 390,
    pages = {71-86},
    doi = {10.1111/j.1365-2966.2008.13754.x}
}

@ARTICLE{Cappellari2020,
    author = {{Cappellari}, Michele},
    title = "{Efficient solution of the anisotropic spherically-aligned axisymmetric
        Jeans equations of stellar hydrodynamics for galactic dynamics}",
    journal = {MNRAS},
    eprint = {1907.09894},
    year = 2020,
    volume = 494,
    pages = {4819-4837},
    doi = {10.1093/mnras/staa959}
}

@article{McCully_2017,
   title={Quantifying Environmental and Line-of-sight Effects in Models of Strong Gravitational Lens Systems},
   volume={836},
   ISSN={1538-4357},
   url={http://dx.doi.org/10.3847/1538-4357/836/1/141},
   DOI={10.3847/1538-4357/836/1/141},
   number={1},
   journal={The Astrophysical Journal},
   publisher={American Astronomical Society},
   author={McCully, Curtis and Keeton, Charles R. and Wong, Kenneth C. and Zabludoff, Ann I.},
   year={2017},
   month=feb, pages={141} }

@Article{Cappellari2017,
  author        = {Cappellari, M.},
  journal       = {\mnras},
  title         = {{Improving the full spectrum fitting method: accurate convolution with Gauss-Hermite functions}},
  year          = {2017},
  month         = apr,
  pages         = {798--811},
  volume        = {466},
  adsurl        = {https://ui.adsabs.harvard.edu/abs/2017MNRAS.466..798C},
  archiveprefix = {arXiv},
  doi           = {10.1093/mnras/stw3020},
  eprint        = {1607.08538},
}

@ARTICLE{Cappellari2023,
    author = {{Cappellari}, M.},
    title = "{Full spectrum fitting with photometry in PPXF: stellar population
        versus dynamical masses, non-parametric star formation history and
        metallicity for 3200 LEGA-C galaxies at redshift $z\approx0.8$}",
    journal = {MNRAS},
    eprint = {2208.14974},
    year = 2023,
    volume = 526,
    pages = {3273-3300},
    doi = {10.1093/mnras/stad2597}
}

@article{miles11,
   title={An updated MILES stellar library and stellar population models},
   volume={532},
   ISSN={1432-0746},
   url={http://dx.doi.org/10.1051/0004-6361/201116842},
   DOI={10.1051/0004-6361/201116842},
   journal={Astronomy \& Astrophysics},
   publisher={EDP Sciences},
   author={Falcón-Barroso, J. and Sánchez-Blázquez, P. and Vazdekis, A. and Ricciardelli, E. and Cardiel, N. and Cenarro, A. J. and Gorgas, J. and Peletier, R. F.},
   year={2011},
   month=jul, pages={A95} }

@ARTICLE{indous04,
       author = {{Valdes}, Francisco and {Gupta}, Ranjan and {Rose}, James A. and {Singh}, Harinder P. and {Bell}, David J.},
        title = "{The Indo-US Library of Coud{\'e} Feed Stellar Spectra}",
      journal = {\apjs},
         year = 2004,
        month = jun,
       volume = {152},
       number = {2},
        pages = {251-259},
          doi = {10.1086/386343},
archivePrefix = {arXiv},
       eprint = {astro-ph/0402435},
 primaryClass = {astro-ph},
       adsurl = {https://ui.adsabs.harvard.edu/abs/2004ApJS..152..251V}
}

@article{xshooter22,
   title={The X-shooter Spectral Library (XSL): Data Release 3},
   volume={660},
   ISSN={1432-0746},
   url={http://dx.doi.org/10.1051/0004-6361/202142388},
   DOI={10.1051/0004-6361/202142388},
   journal={Astronomy \& Astrophysics},
   publisher={EDP Sciences},
   author={Verro, K. and Trager, S. C. and Peletier, R. F. and Lançon, A. and Gonneau, A. and Vazdekis, A. and Prugniel, P. and Chen, Y.-P. and Coelho, P. R. T. and Sánchez-Blázquez, P. and Martins, L. and Arentsen, A. and Lyubenova, M. and Falcón-Barroso, J. and Dries, M.},
   year={2022},
   month=apr, pages={A34} }

@Article{birrer2018,
  author    = {Simon Birrer and Adam Amara},
  journal   = {Physics of the Dark Universe},
  title     = {lenstronomy: Multi-purpose gravitational lens modelling software package},
  year      = {2018},
  month     = {dec},
  pages     = {189--201},
  volume    = {22},
  doi       = {10.1016/j.dark.2018.11.002},
  publisher = {Elsevier {BV}},
  url       = {https://doi.org/10.1016%2Fj.dark.2018.11.002},
}

@ARTICLE{lenstronomyII,
       author = {{Birrer}, Simon and {Shajib}, Anowar and {Gilman}, Daniel and {Galan}, Aymeric and {Aalbers}, Jelle and {Millon}, Martin and {Morgan}, Robert and {Pagano}, Giulia and {Park}, Ji and {Teodori}, Luca and {Tessore}, Nicolas and {Ueland}, Madison and {Van de Vyvere}, Lyne and {Wagner-Carena}, Sebastian and {Wempe}, Ewoud and {Yang}, Lilan and {Ding}, Xuheng and {Schmidt}, Thomas and {Sluse}, Dominique and {Zhang}, Ming and {Amara}, Adam},
        title = "{lenstronomy II: A gravitational lensing software ecosystem}",
      journal = {The Journal of Open Source Software},
         year = 2021,
        month = jun,
       volume = {6},
       number = {62},
          eid = {3283},
        pages = {3283},
          doi = {10.21105/joss.03283},
archivePrefix = {arXiv},
       eprint = {2106.05976},
 primaryClass = {astro-ph.CO},
       adsurl = {https://ui.adsabs.harvard.edu/abs/2021JOSS....6.3283B}
}

@article{shapelets1,
   title={Shapelets -- I. A method for image analysis},
   volume={338},
   ISSN={1365-2966},
   url={http://dx.doi.org/10.1046/j.1365-8711.2003.05901.x},
   DOI={10.1046/j.1365-8711.2003.05901.x},
   number={1},
   journal={Monthly Notices of the Royal Astronomical Society},
   publisher={Oxford University Press (OUP)},
   author={Refregier, A.},
   year={2003},
   month=jan, pages={35–47} }

@article{shapelets2,
   title={Shapelets -- II. A method for weak lensing measurements},
   volume={338},
   ISSN={1365-2966},
   url={http://dx.doi.org/10.1046/j.1365-8711.2003.05902.x},
   DOI={10.1046/j.1365-8711.2003.05902.x},
   number={1},
   journal={Monthly Notices of the Royal Astronomical Society},
   publisher={Oxford University Press (OUP)},
   author={Refregier, A. and Bacon, D.},
   year={2003},
   month=jan, pages={48–56} }

@ARTICLE{vaucouleurs1948,
       author = {{de Vaucouleurs}, Gerard},
        title = "{Recherches sur les Nebuleuses Extragalactiques}",
      journal = {Annales d'Astrophysique},
         year = 1948,
        month = jan,
       volume = {11},
        pages = {247},
       adsurl = {https://ui.adsabs.harvard.edu/abs/1948AnAp...11..247D}
}

@ARTICLE{sersic,
       author = {{S{\'e}rsic}, J.~L.},
        title = "{Influence of the atmospheric and instrumental dispersion on the brightness distribution in a galaxy}",
      journal = {Boletin de la Asociacion Argentina de Astronomia La Plata Argentina},
         year = 1963,
        month = feb,
       volume = {6},
        pages = {41-43},
       adsurl = {https://ui.adsabs.harvard.edu/abs/1963BAAA....6...41S}
}

@article{Tessore_2015,
   title={The elliptical power law profile lens},
   volume={580},
   ISSN={1432-0746},
   url={http://dx.doi.org/10.1051/0004-6361/201526773},
   DOI={10.1051/0004-6361/201526773},
   journal={Astronomy \& Astrophysics},
   publisher={EDP Sciences},
   author={Tessore, Nicolas and Benton Metcalf, R.},
   year={2015},
   month=aug, pages={A79} }

@INPROCEEDINGS{Silva2016,
       author = {{Silva}, David R. and {Blum}, Robert D. and {Allen}, Lori and {Dey}, Arjun and {Schlegel}, David J. and {Lang}, Dustin and {Moustakas}, John and {Meisner}, Aaron M. and {Valdes}, Francisco and {Patej}, Anna and {Myers}, Adam D. and {Sprayberry}, David and {Saha}, Abi and {Olsen}, Knut A. and {Gaines}, Sasha and {Yang}, Qian and {Burleigh}, Kaylan J. and {MzLS Team}},
        title = "{The Mayall z-band Legacy Survey}",
    booktitle = {American Astronomical Society Meeting Abstracts \#228},
         year = 2016,
       series = {American Astronomical Society Meeting Abstracts},
       volume = {228},
        month = jun,
          eid = {317.02},
        pages = {317.02},
       adsurl = {https://ui.adsabs.harvard.edu/abs/2016AAS...22831702S}
}

@article{gaia2018,
   title={Gaia Data Release 2: Summary of the contents and survey properties},
   volume={616},
   ISSN={1432-0746},
   url={http://dx.doi.org/10.1051/0004-6361/201833051},
   DOI={10.1051/0004-6361/201833051},
   journal={Astronomy \& Astrophysics},
   publisher={EDP Sciences},
   author={Brown, A. G. A. and Vallenari, A. and Prusti, T. and de Bruijne, J. H. J. and Babusiaux, C. and Bailer-Jones, C. A. L. and Biermann, M. and Evans, D. W. and Eyer, L. and Jansen, F. and Jordi, C. and Klioner, S. A. and Lammers, U. and Lindegren, L. and Luri, X. and Mignard, F. and Panem, C. and Pourbaix, D. and Randich, S. and Sartoretti, P. and Siddiqui, H. I. and Soubiran, C. and van Leeuwen, F. and Walton, N. A. and Arenou, F. and Bastian, U. and Cropper, M. and Drimmel, R. and Katz, D. and Lattanzi, M. G. and Bakker, J. and Cacciari, C. and Castañeda, J. and Chaoul, L. and Cheek, N. and De Angeli, F. and Fabricius, C. and Guerra, R. and Holl, B. and Masana, E. and Messineo, R. and Mowlavi, N. and Nienartowicz, K. and Panuzzo, P. and Portell, J. and Riello, M. and Seabroke, G. M. and Tanga, P. and Thévenin, F. and Gracia-Abril, G. and Comoretto, G. and Garcia-Reinaldos, M. and Teyssier, D. and Altmann, M. and Andrae, R. and Audard, M. and Bellas-Velidis, I. and Benson, K. and Berthier, J. and Blomme, R. and Burgess, P. and Busso, G. and Carry, B. and Cellino, A. and Clementini, G. and Clotet, M. and Creevey, O. and Davidson, M. and De Ridder, J. and Delchambre, L. and Dell’Oro, A. and Ducourant, C. and Fernández-Hernández, J. and Fouesneau, M. and Frémat, Y. and Galluccio, L. and García-Torres, M. and González-Núñez, J. and González-Vidal, J. J. and Gosset, E. and Guy, L. P. and Halbwachs, J.-L. and Hambly, N. C. and Harrison, D. L. and Hernández, J. and Hestroffer, D. and Hodgkin, S. T. and Hutton, A. and Jasniewicz, G. and Jean-Antoine-Piccolo, A. and Jordan, S. and Korn, A. J. and Krone-Martins, A. and Lanzafame, A. C. and Lebzelter, T. and Löffler, W. and Manteiga, M. and Marrese, P. M. and Martín-Fleitas, J. M. and Moitinho, A. and Mora, A. and Muinonen, K. and Osinde, J. and Pancino, E. and Pauwels, T. and Petit, J.-M. and Recio-Blanco, A. and Richards, P. J. and Rimoldini, L. and Robin, A. C. and Sarro, L. M. and Siopis, C. and Smith, M. and Sozzetti, A. and Süveges, M. and Torra, J. and van Reeven, W. and Abbas, U. and Abreu Aramburu, A. and Accart, S. and Aerts, C. and Altavilla, G. and Álvarez, M. A. and Alvarez, R. and Alves, J. and Anderson, R. I. and Andrei, A. H. and Anglada Varela, E. and Antiche, E. and Antoja, T. and Arcay, B. and Astraatmadja, T. L. and Bach, N. and Baker, S. G. and Balaguer-Núñez, L. and Balm, P. and Barache, C. and Barata, C. and Barbato, D. and Barblan, F. and Barklem, P. S. and Barrado, D. and Barros, M. and Barstow, M. A. and Bartholomé Muñoz, S. and Bassilana, J.-L. and Becciani, U. and Bellazzini, M. and Berihuete, A. and Bertone, S. and Bianchi, L. and Bienaymé, O. and Blanco-Cuaresma, S. and Boch, T. and Boeche, C. and Bombrun, A. and Borrachero, R. and Bossini, D. and Bouquillon, S. and Bourda, G. and Bragaglia, A. and Bramante, L. and Breddels, M. A. and Bressan, A. and Brouillet, N. and Brüsemeister, T. and Brugaletta, E. and Bucciarelli, B. and Burlacu, A. and Busonero, D. and Butkevich, A. G. and Buzzi, R. and Caffau, E. and Cancelliere, R. and Cannizzaro, G. and Cantat-Gaudin, T. and Carballo, R. and Carlucci, T. and Carrasco, J. M. and Casamiquela, L. and Castellani, M. and Castro-Ginard, A. and Charlot, P. and Chemin, L. and Chiavassa, A. and Cocozza, G. and Costigan, G. and Cowell, S. and Crifo, F. and Crosta, M. and Crowley, C. and Cuypers†, J. and Dafonte, C. and Damerdji, Y. and Dapergolas, A. and David, P. and David, M. and de Laverny, P. and De Luise, F. and De March, R. and de Martino, D. and de Souza, R. and de Torres, A. and Debosscher, J. and del Pozo, E. and Delbo, M. and Delgado, A. and Delgado, H. E. and Di Matteo, P. and Diakite, S. and Diener, C. and Distefano, E. and Dolding, C. and Drazinos, P. and Durán, J. and Edvardsson, B. and Enke, H. and Eriksson, K. and Esquej, P. and Eynard Bontemps, G. and Fabre, C. and Fabrizio, M. and Faigler, S. and Falcão, A. J. and Farràs Casas, M. and Federici, L. and Fedorets, G. and Fernique, P. and Figueras, F. and Filippi, F. and Findeisen, K. and Fonti, A. and Fraile, E. and Fraser, M. and Frézouls, B. and Gai, M. and Galleti, S. and Garabato, D. and García-Sedano, F. and Garofalo, A. and Garralda, N. and Gavel, A. and Gavras, P. and Gerssen, J. and Geyer, R. and Giacobbe, P. and Gilmore, G. and Girona, S. and Giuffrida, G. and Glass, F. and Gomes, M. and Granvik, M. and Gueguen, A. and Guerrier, A. and Guiraud, J. and Gutiérrez-Sánchez, R. and Haigron, R. and Hatzidimitriou, D. and Hauser, M. and Haywood, M. and Heiter, U. and Helmi, A. and Heu, J. and Hilger, T. and Hobbs, D. and Hofmann, W. and Holland, G. and Huckle, H. E. and Hypki, A. and Icardi, V. and Janßen, K. and Jevardat de Fombelle, G. and Jonker, P. G. and Juhász, Á. L. and Julbe, F. and Karampelas, A. and Kewley, A. and Klar, J. and Kochoska, A. and Kohley, R. and Kolenberg, K. and Kontizas, M. and Kontizas, E. and Koposov, S. E. and Kordopatis, G. and Kostrzewa-Rutkowska, Z. and Koubsky, P. and Lambert, S. and Lanza, A. F. and Lasne, Y. and Lavigne, J.-B. and Le Fustec, Y. and Le Poncin-Lafitte, C. and Lebreton, Y. and Leccia, S. and Leclerc, N. and Lecoeur-Taibi, I. and Lenhardt, H. and Leroux, F. and Liao, S. and Licata, E. and Lindstrøm, H. E. P. and Lister, T. A. and Livanou, E. and Lobel, A. and López, M. and Managau, S. and Mann, R. G. and Mantelet, G. and Marchal, O. and Marchant, J. M. and Marconi, M. and Marinoni, S. and Marschalkó, G. and Marshall, D. J. and Martino, M. and Marton, G. and Mary, N. and Massari, D. and Matijevič, G. and Mazeh, T. and McMillan, P. J. and Messina, S. and Michalik, D. and Millar, N. R. and Molina, D. and Molinaro, R. and Molnár, L. and Montegriffo, P. and Mor, R. and Morbidelli, R. and Morel, T. and Morris, D. and Mulone, A. F. and Muraveva, T. and Musella, I. and Nelemans, G. and Nicastro, L. and Noval, L. and O’Mullane, W. and Ordénovic, C. and Ordóñez-Blanco, D. and Osborne, P. and Pagani, C. and Pagano, I. and Pailler, F. and Palacin, H. and Palaversa, L. and Panahi, A. and Pawlak, M. and Piersimoni, A. M. and Pineau, F.-X. and Plachy, E. and Plum, G. and Poggio, E. and Poujoulet, E. and Prša, A. and Pulone, L. and Racero, E. and Ragaini, S. and Rambaux, N. and Ramos-Lerate, M. and Regibo, S. and Reylé, C. and Riclet, F. and Ripepi, V. and Riva, A. and Rivard, A. and Rixon, G. and Roegiers, T. and Roelens, M. and Romero-Gómez, M. and Rowell, N. and Royer, F. and Ruiz-Dern, L. and Sadowski, G. and Sagristà Sellés, T. and Sahlmann, J. and Salgado, J. and Salguero, E. and Sanna, N. and Santana-Ros, T. and Sarasso, M. and Savietto, H. and Schultheis, M. and Sciacca, E. and Segol, M. and Segovia, J. C. and Ségransan, D. and Shih, I-C. and Siltala, L. and Silva, A. F. and Smart, R. L. and Smith, K. W. and Solano, E. and Solitro, F. and Sordo, R. and Soria Nieto, S. and Souchay, J. and Spagna, A. and Spoto, F. and Stampa, U. and Steele, I. A. and Steidelmüller, H. and Stephenson, C. A. and Stoev, H. and Suess, F. F. and Surdej, J. and Szabados, L. and Szegedi-Elek, E. and Tapiador, D. and Taris, F. and Tauran, G. and Taylor, M. B. and Teixeira, R. and Terrett, D. and Teyssandier, P. and Thuillot, W. and Titarenko, A. and Torra Clotet, F. and Turon, C. and Ulla, A. and Utrilla, E. and Uzzi, S. and Vaillant, M. and Valentini, G. and Valette, V. and van Elteren, A. and Van Hemelryck, E. and van Leeuwen, M. and Vaschetto, M. and Vecchiato, A. and Veljanoski, J. and Viala, Y. and Vicente, D. and Vogt, S. and von Essen, C. and Voss, H. and Votruba, V. and Voutsinas, S. and Walmsley, G. and Weiler, M. and Wertz, O. and Wevers, T. and Wyrzykowski, Ł. and Yoldas, A. and Žerjal, M. and Ziaeepour, H. and Zorec, J. and Zschocke, S. and Zucker, S. and Zurbach, C. and Zwitter, T.},
   year={2018},
   month=aug, pages={A1} }

@article{Dawson_2012,
   title={THE BARYON OSCILLATION SPECTROSCOPIC SURVEY OF SDSS-III},
   volume={145},
   ISSN={1538-3881},
   url={http://dx.doi.org/10.1088/0004-6256/145/1/10},
   DOI={10.1088/0004-6256/145/1/10},
   number={1},
   journal={The Astronomical Journal},
   publisher={American Astronomical Society},
   author={Dawson, Kyle S. and Schlegel, David J. and Ahn, Christopher P. and Anderson, Scott F. and Aubourg, Éric and Bailey, Stephen and Barkhouser, Robert H. and Bautista, Julian E. and Beifiori, Alessandra and Berlind, Andreas A. and Bhardwaj, Vaishali and Bizyaev, Dmitry and Blake, Cullen H. and Blanton, Michael R. and Blomqvist, Michael and Bolton, Adam S. and Borde, Arnaud and Bovy, Jo and Brandt, W. N. and Brewington, Howard and Brinkmann, Jon and Brown, Peter J. and Brownstein, Joel R. and Bundy, Kevin and Busca, N. G. and Carithers, William and Carnero, Aurelio R. and Carr, Michael A. and Chen, Yanmei and Comparat, Johan and Connolly, Natalia and Cope, Frances and Croft, Rupert A. C. and Cuesta, Antonio J. and da Costa, Luiz N. and Davenport, James R. A. and Delubac, Timothée and de Putter, Roland and Dhital, Saurav and Ealet, Anne and Ebelke, Garrett L. and Eisenstein, Daniel J. and Escoffier, S. and Fan, Xiaohui and Filiz Ak, N. and Finley, Hayley and Font-Ribera, Andreu and Génova-Santos, R. and Gunn, James E. and Guo, Hong and Haggard, Daryl and Hall, Patrick B. and Hamilton, Jean-Christophe and Harris, Ben and Harris, David W. and Ho, Shirley and Hogg, David W. and Holder, Diana and Honscheid, Klaus and Huehnerhoff, Joe and Jordan, Beatrice and Jordan, Wendell P. and Kauffmann, Guinevere and Kazin, Eyal A. and Kirkby, David and Klaene, Mark A. and Kneib, Jean-Paul and Le Goff, Jean-Marc and Lee, Khee-Gan and Long, Daniel C. and Loomis, Craig P. and Lundgren, Britt and Lupton, Robert H. and Maia, Marcio A. G. and Makler, Martin and Malanushenko, Elena and Malanushenko, Viktor and Mandelbaum, Rachel and Manera, Marc and Maraston, Claudia and Margala, Daniel and Masters, Karen L. and McBride, Cameron K. and McDonald, Patrick and McGreer, Ian D. and McMahon, Richard G. and Mena, Olga and Miralda-Escudé, Jordi and Montero-Dorta, Antonio D. and Montesano, Francesco and Muna, Demitri and Myers, Adam D. and Naugle, Tracy and Nichol, Robert C. and Noterdaeme, Pasquier and Nuza, Sebastián E. and Olmstead, Matthew D. and Oravetz, Audrey and Oravetz, Daniel J. and Owen, Russell and Padmanabhan, Nikhil and Palanque-Delabrouille, Nathalie and Pan, Kaike and Parejko, John K. and Pâris, Isabelle and Percival, Will J. and Pérez-Fournon, Ismael and Pérez-Ràfols, Ignasi and Petitjean, Patrick and Pfaffenberger, Robert and Pforr, Janine and Pieri, Matthew M. and Prada, Francisco and Price-Whelan, Adrian M. and Raddick, M. Jordan and Rebolo, Rafael and Rich, James and Richards, Gordon T. and Rockosi, Constance M. and Roe, Natalie A. and Ross, Ashley J. and Ross, Nicholas P. and Rossi, Graziano and Rubiño-Martin, J. A. and Samushia, Lado and Sánchez, Ariel G. and Sayres, Conor and Schmidt, Sarah J. and Schneider, Donald P. and Scóccola, C. G. and Seo, Hee-Jong and Shelden, Alaina and Sheldon, Erin and Shen, Yue and Shu, Yiping and Slosar, Anže and Smee, Stephen A. and Snedden, Stephanie A. and Stauffer, Fritz and Steele, Oliver and Strauss, Michael A. and Streblyanska, Alina and Suzuki, Nao and Swanson, Molly E. C. and Tal, Tomer and Tanaka, Masayuki and Thomas, Daniel and Tinker, Jeremy L. and Tojeiro, Rita and Tremonti, Christy A. and Vargas Magaña, M. and Verde, Licia and Viel, Matteo and Wake, David A. and Watson, Mike and Weaver, Benjamin A. and Weinberg, David H. and Weiner, Benjamin J. and West, Andrew A. and White, Martin and Wood-Vasey, W. M. and Yeche, Christophe and Zehavi, Idit and Zhao, Gong-Bo and Zheng, Zheng},
   year={2012},
   month=dec, pages={10} }

@article{KCWI2018,
   title={The Keck Cosmic Web Imager Integral Field Spectrograph},
   volume={864},
   ISSN={1538-4357},
   url={http://dx.doi.org/10.3847/1538-4357/aad597},
   DOI={10.3847/1538-4357/aad597},
   number={1},
   journal={The Astrophysical Journal},
   publisher={American Astronomical Society},
   author={Morrissey, Patrick and Matuszewski, Matuesz and Martin, D. Christopher and Neill, James D. and Epps, Harland and Fucik, Jason and Weber, Bob and Darvish, Behnam and Adkins, Sean and Allen, Steve and Bartos, Randy and Belicki, Justin and Cabak, Jerry and Callahan, Shawn and Cowley, Dave and Crabill, Marty and Deich, Willian and Delecroix, Alex and Doppman, Greg and Hilyard, David and James, Ean and Kaye, Steve and Kokorowski, Michael and Kwok, Shui and Lanclos, Kyle and Milner, Steve and Moore, Anna and O’Sullivan, Donal and Parihar, Prachi and Park, Sam and Phillips, Andrew and Rizzi, Luca and Rockosi, Constance and Rodriguez, Hector and Salaun, Yves and Seaman, Kirk and Sheikh, David and Weiss, Jason and Zarzaca, Ray},
   year={2018},
   month=sep, pages={93} }

@ARTICLE{pascale2025,
       author = {{Pascale}, Massimo and {Frye}, Brenda L. and {Pierel}, Justin D.~R. and {Chen}, Wenlei and {Kelly}, Patrick L. and {Cohen}, Seth H. and {Windhorst}, Rogier A. and {Riess}, Adam G. and {Kamieneski}, Patrick S. and {Diego}, Jos{\'e} M. and {Meena}, Ashish K. and {Cha}, Sangjun and {Oguri}, Masamune and {Zitrin}, Adi and {Jee}, M. James and {Foo}, Nicholas and {Leimbach}, Reagen and {Koekemoer}, Anton M. and {Conselice}, C.~J. and {Dai}, Liang and {Goobar}, Ariel and {Siebert}, Matthew R. and {Strolger}, Lou and {Willner}, S.~P.},
        title = "{SN H0pe: The First Measurement of H$_{0}$ from a Multiply Imaged Type Ia Supernova, Discovered by JWST}",
      journal = {\apj},
         year = 2025,
        month = jan,
       volume = {979},
       number = {1},
          eid = {13},
        pages = {13},
          doi = {10.3847/1538-4357/ad9928},
archivePrefix = {arXiv},
       eprint = {2403.18902},
 primaryClass = {astro-ph.CO},
       adsurl = {https://ui.adsabs.harvard.edu/abs/2025ApJ...979...13P}
}

@article{lemon2018,
   title={Gravitationally lensed quasars in Gaia – II. Discovery of 24 lensed quasars},
   volume={479},
   ISSN={1365-2966},
   url={http://dx.doi.org/10.1093/mnras/sty911},
   DOI={10.1093/mnras/sty911},
   number={4},
   journal={Monthly Notices of the Royal Astronomical Society},
   publisher={Oxford University Press (OUP)},
   author={Lemon, Cameron A and Auger, Matthew W and McMahon, Richard G and Ostrovski, Fernanda},
   year={2018},
   month=apr, pages={5060–5074} }

@article{quimby2014,
   title={Detection of the Gravitational Lens Magnifying a Type Ia Supernova},
   volume={344},
   ISSN={1095-9203},
   url={http://dx.doi.org/10.1126/science.1250903},
   DOI={10.1126/science.1250903},
   number={6182},
   journal={Science},
   publisher={American Association for the Advancement of Science (AAAS)},
   author={Quimby, Robert M. and Oguri, Masamune and More, Anupreeta and More, Surhud and Moriya, Takashi J. and Werner, Marcus C. and Tanaka, Masayuki and Folatelli, Gaston and Bersten, Melina C. and Maeda, Keiichi and Nomoto, Ken’ichi},
   year={2014},
   month=apr, pages={396–399} }

@article{lemon2022,
   title={Gravitationally lensed quasars in Gaia – IV. 150 new lenses, quasar pairs, and projected quasars},
   volume={520},
   ISSN={1365-2966},
   url={http://dx.doi.org/10.1093/mnras/stac3721},
   DOI={10.1093/mnras/stac3721},
   number={3},
   journal={Monthly Notices of the Royal Astronomical Society},
   publisher={Oxford University Press (OUP)},
   author={Lemon, C and Anguita, T and Auger-Williams, M W and Courbin, F and Galan, A and McMahon, R and Neira, F and Oguri, M and Schechter, P and Shajib, A and Treu, T and Agnello, A and Spiniello, C},
   year={2022},
   month=dec, pages={3305–3328} }

@article{sheu2024,
   title={A Targeted Search for Variable Gravitationally Lensed Quasars},
   volume={973},
   ISSN={1538-4357},
   url={http://dx.doi.org/10.3847/1538-4357/ad5dad},
   DOI={10.3847/1538-4357/ad5dad},
   number={1},
   journal={The Astrophysical Journal},
   publisher={American Astronomical Society},
   author={Sheu, William and Huang, Xiaosheng and Cikota, Aleksandar and Suzuki, Nao and Palmese, Antonella and Schlegel, David J. and Storfer, Christopher},
   year={2024},
   month=sep, pages={24} }

@ARTICLE{S+S13,
       author = {{Schneider}, Peter and {Sluse}, Dominique},
        title = "{Mass-sheet degeneracy, power-law models and external convergence: Impact on the determination of the Hubble constant from gravitational lensing}",
      journal = {\aap},
         year = 2013,
        month = nov,
       volume = {559},
          eid = {A37},
        pages = {A37},
          doi = {10.1051/0004-6361/201321882},
archivePrefix = {arXiv},
       eprint = {1306.0901},
 primaryClass = {astro-ph.CO},
       adsurl = {https://ui.adsabs.harvard.edu/abs/2013A&A...559A..37S}
}

@article{sheu2023,
   title={Retrospective Search for Strongly Lensed Supernovae in the DESI Legacy Imaging Surveys},
   volume={952},
   ISSN={1538-4357},
   url={http://dx.doi.org/10.3847/1538-4357/acd1e4},
   DOI={10.3847/1538-4357/acd1e4},
   number={1},
   journal={The Astrophysical Journal},
   publisher={American Astronomical Society},
   author={Sheu, William and Huang, Xiaosheng and Cikota, Aleksandar and Suzuki, Nao and Schlegel, David J. and Storfer, Christopher},
   year={2023},
   month=jul, pages={10} }

@ARTICLE{pycs30,
       author = {{Tewes}, M. and {Courbin}, F. and {Meylan}, G.},
        title = "{COSMOGRAIL: the COSmological MOnitoring of GRAvItational Lenses. XI. Techniques for time delay measurement in presence of microlensing}",
      journal = {\aap},
         year = 2013,
        month = may,
       volume = {553},
          eid = {A120},
        pages = {A120},
          doi = {10.1051/0004-6361/201220123},
archivePrefix = {arXiv},
       eprint = {1208.5598},
 primaryClass = {astro-ph.CO},
       adsurl = {https://ui.adsabs.harvard.edu/abs/2013A&A...553A.120T}
}

@article{pycs31,
  author       = {Martin Millon and
                  Malte Tewes and
                  Vivien Bonvin and
                  Bastian Lengen and
                  Frederic Courbin},
  title        = {PyCS3: A Python toolbox for time-delay
                   measurements in lensed quasars
                  },
  month        = sep,
  year         = 2020,
  journal    = {Journal of Open Source Software},
  doi          = {10.21105/joss.02654},
  url          = {https://doi.org/10.21105/joss.02654},
}

@ARTICLE{han2025,
       author = {{Wang}, Han and {Suyu}, Sherry H. and {Galan}, Aymeric and {Halkola}, Aleksi and {Cappellari}, Michele and {Shajib}, Anowar J. and {Cernetic}, Miha},
        title = "{GPU-Accelerated Gravitational Lensing and Dynamical (GLaD) modeling for cosmology and galaxies}",
      journal = {\aap},
         year = 2025,
        month = sep,
       volume = {701},
          eid = {A280},
        pages = {A280},
          doi = {10.1051/0004-6361/202554861},
archivePrefix = {arXiv},
       eprint = {2504.01302},
 primaryClass = {astro-ph.CO},
       adsurl = {https://ui.adsabs.harvard.edu/abs/2025A&A...701A.280W}
}

@ARTICLE{SAURON,
       author = {{Cappellari}, Michele and {Emsellem}, Eric and {Bacon}, R. and {Bureau}, M. and {Davies}, Roger L. and {de Zeeuw}, P.~T. and {Falc{\'o}n-Barroso}, Jes{\'u}s and {Krajnovi{\'c}}, Davor and {Kuntschner}, Harald and {McDermid}, Richard M. and {Peletier}, Reynier F. and {Sarzi}, Marc and {van den Bosch}, Remco C.~E. and {van de Ven}, Glenn},
        title = "{The SAURON project - X. The orbital anisotropy of elliptical and lenticular galaxies: revisiting the (V/{\ensuremath{\sigma}}, ɛ) diagram with integral-field stellar kinematics}",
      journal = {\mnras},
         year = 2007,
        month = aug,
       volume = {379},
       number = {2},
        pages = {418-444},
          doi = {10.1111/j.1365-2966.2007.11963.x},
archivePrefix = {arXiv},
       eprint = {astro-ph/0703533},
 primaryClass = {astro-ph},
       adsurl = {https://ui.adsabs.harvard.edu/abs/2007MNRAS.379..418C}
}

@article{Schmidt_2022,
   title={STRIDES: automated uniform models for 30 quadruply imaged quasars},
   volume={518},
   ISSN={1365-2966},
   url={http://dx.doi.org/10.1093/mnras/stac2235},
   DOI={10.1093/mnras/stac2235},
   number={1},
   journal={Monthly Notices of the Royal Astronomical Society},
   publisher={Oxford University Press (OUP)},
   author={Schmidt, T and Treu, T and Birrer, S and Shajib, A J and Lemon, C and Millon, M and Sluse, D and Agnello, A and Anguita, T and Auger-Williams, M W and McMahon, R G and Motta, V and Schechter, P and Spiniello, C and Kayo, I and Courbin, F and Ertl, S and Fassnacht, C D and Frieman, J A and More, A and Schuldt, S and Suyu, S H and Aguena, M and Andrade-Oliveira, F and Annis, J and Bacon, D and Bertin, E and Brooks, D and Burke, D L and Carnero Rosell, A and Carrasco Kind, M and Carretero, J and Conselice, C and Costanzi, M and da Costa, L N and Pereira, M E S and De Vicente, J and Desai, S and Doel, P and Everett, S and Ferrero, I and Friedel, D and García-Bellido, J and Gaztanaga, E and Gruen, D and Gruendl, R A and Gschwend, J and Gutierrez, G and Hinton, S R and Hollowood, D L and Honscheid, K and James, D J and Kuehn, K and Lahav, O and Menanteau, F and Miquel, R and Palmese, A and Paz-Chinchón, F and Pieres, A and Plazas Malagón, A A and Prat, J and Rodriguez-Monroy, M and Romer, A K and Sanchez, E and Scarpine, V and Sevilla-Noarbe, I and Smith, M and Suchyta, E and Tarle, G and To, C and Varga, T N},
   year={2022},
   month=nov, pages={1260–1300} }

@article{Ertl_2023,
   title={TDCOSMO: X. Automated modeling of nine strongly lensed quasars and comparison between lens-modeling software},
   volume={672},
   ISSN={1432-0746},
   url={http://dx.doi.org/10.1051/0004-6361/202244909},
   DOI={10.1051/0004-6361/202244909},
   journal={Astronomy \& Astrophysics},
   publisher={EDP Sciences},
   author={Ertl, S. and Schuldt, S. and Suyu, S. H. and Schmidt, T. and Treu, T. and Birrer, S. and Shajib, A. J. and Sluse, D.},
   year={2023},
   month=mar, pages={A2} }

@article{Rusu2017,
   title={H0LiCOW – III. Quantifying the effect of mass along the line of sight to the gravitational lens HE 0435−1223 through weighted galaxy counts★},
   volume={467},
   ISSN={1365-2966},
   url={http://dx.doi.org/10.1093/mnras/stx285},
   DOI={10.1093/mnras/stx285},
   number={4},
   journal={Monthly Notices of the Royal Astronomical Society},
   publisher={Oxford University Press (OUP)},
   author={Rusu, Cristian E. and Fassnacht, Christopher D. and Sluse, Dominique and Hilbert, Stefan and Wong, Kenneth C. and Huang, Kuang-Han and Suyu, Sherry H. and Collett, Thomas E. and Marshall, Philip J. and Treu, Tommaso and Koopmans, Leon V. E.},
   year={2017},
   month=feb, pages={4220–4242} }

@article{Shajib_2022,
   title={TDCOSMO: IX. Systematic comparison between lens modelling software programs: Time-delay prediction for WGD 2038−4008},
   volume={667},
   ISSN={1432-0746},
   url={http://dx.doi.org/10.1051/0004-6361/202243401},
   DOI={10.1051/0004-6361/202243401},
   journal={Astronomy \& Astrophysics},
   publisher={EDP Sciences},
   author={Shajib, A. J. and Wong, K. C. and Birrer, S. and Suyu, S. H. and Treu, T. and Buckley-Geer, E. J. and Lin, H. and Rusu, C. E. and Poh, J. and Palmese, A. and Agnello, A. and Auger-Williams, M. W. and Galan, A. and Schuldt, S. and Sluse, D. and Courbin, F. and Frieman, J. and Millon, M.},
   year={2022},
   month=nov, pages={A123} }

@article{Emsellem_2007,
   title={The SAURON project - IX. A kinematic classification for early-type galaxies},
   volume={379},
   ISSN={1365-2966},
   url={http://dx.doi.org/10.1111/j.1365-2966.2007.11752.x},
   DOI={10.1111/j.1365-2966.2007.11752.x},
   number={2},
   journal={Monthly Notices of the Royal Astronomical Society},
   publisher={Oxford University Press (OUP)},
   author={Emsellem, E. and Cappellari, M. and Krajnovi , D. and Van De Ven, G. and Bacon, R. and Bureau, M. and Davies, R. L. and De Zeeuw, P. T. and Falcon-Barroso, J. and Kuntschner, H. and McDermid, R. and Peletier, R. F. and Sarzi, M.},
   year={2007},
   month=aug, pages={401–417} }

@ARTICLE{knabelslac,
       author = {{Knabel}, Shawn and {Treu}, Tommaso and {Cappellari}, Michele and {Shajib}, Anowar J. and {Chen}, Chih-Fan and {Birrer}, Simon and {Bennert}, Vardha N.},
        title = "{Spatially Resolved Kinematics of SLACS Lens Galaxies. I. Data and Kinematic Classification}",
      journal = {\apj},
         year = 2025,
        month = sep,
       volume = {990},
       number = {1},
          eid = {51},
        pages = {51},
          doi = {10.3847/1538-4357/adea94},
archivePrefix = {arXiv},
       eprint = {2409.10631},
 primaryClass = {astro-ph.CO},
       adsurl = {https://ui.adsabs.harvard.edu/abs/2025ApJ...990...51K}
}

@ARTICLE{starred1,
       author = {{Millon}, Martin and {Michalewicz}, Kevin and {Dux}, Fr{\'e}d{\'e}ric and {Courbin}, Fr{\'e}d{\'e}ric and {Marshall}, Philip J.},
        title = "{Image Deconvolution and Point-spread Function Reconstruction with STARRED: A Wavelet-based Two-channel Method Optimized for Light-curve Extraction}",
      journal = {\aj},
         year = 2024,
        month = aug,
       volume = {168},
       number = {2},
          eid = {55},
        pages = {55},
          doi = {10.3847/1538-3881/ad4da7},
archivePrefix = {arXiv},
       eprint = {2402.08725},
 primaryClass = {astro-ph.IM},
       adsurl = {https://ui.adsabs.harvard.edu/abs/2024AJ....168...55M}
}

@ARTICLE{starred0,
       author = {{Michalewicz}, Kevin and {Millon}, Martin and {Dux}, Fr{\'e}d{\'e}ric and {Courbin}, Fr{\'e}d{\'e}ric},
        title = "{STARRED: a two-channel deconvolution method with Starlet regularization}",
      journal = {The Journal of Open Source Software},
         year = 2023,
        month = may,
       volume = {8},
       number = {85},
          eid = {5340},
        pages = {5340},
          doi = {10.21105/joss.05340},
archivePrefix = {arXiv},
       eprint = {2305.18526},
 primaryClass = {astro-ph.IM},
       adsurl = {https://ui.adsabs.harvard.edu/abs/2023JOSS....8.5340M}
}

@misc{treu2023,
      title={Strong Lensing and $H_0$}, 
      author={Tommaso Treu and Anowar J. Shajib},
      year={2023},
      eprint={2307.05714},
      archivePrefix={arXiv},
      primaryClass={id='astro-ph.CO' full_name='Cosmology and Nongalactic Astrophysics' is_active=True alt_name=None in_archive='astro-ph' is_general=False description='Phenomenology of early universe, cosmic microwave background, cosmological parameters, primordial element abundances, extragalactic distance scale, large-scale structure of the universe. Groups, superclusters, voids, intergalactic medium. Particle astrophysics: dark energy, dark matter, baryogenesis, leptogenesis, inflationary models, reheating, monopoles, WIMPs, cosmic strings, primordial black holes, cosmological gravitational radiation'}
}

@Article{Burkhonov_etal,

    title = "{Time-delay and lens galaxy redshift in the doubly imaged quasar PS J2305+3714}",

    author = {{Burkhonov}, O.~A. and {Shalyapin}, V.~N. and {Sergeyev}, A.~V. and {Nurmamatov}, Sh E. and {Ehgamberdiev}, Sh A. and {Akhunov}, T.~A. and {Dux}, F. and {Courbin}, F. and {Muminov}, M.~M.},

    journal = {MNRAS},

    year = 2026,

    month = feb,

    volume = {546},

    number = {2},

        pages = {stag023},

    doi = {10.1093/mnras/stag023},

        eprint = {arXiv:2601.01230}, 

}

@Article{Ehgamberdiev_2018,

    title = "{Modern astronomy at the Maidanak observatory in Uzbekistan}",

    author = {{Ehgamberdiev}, Shuhrat},

    journal = {Nature Astronomy},

    year = 2018,

    month = may,

    volume = {2},

    pages = {349-351},

    doi = {10.1038/s41550-018-0459-3},

}

@misc{birrer2024,
      title={Time-Delay Cosmography: Measuring the Hubble Constant and other cosmological parameters with strong gravitational lensing}, 
      author={S. Birrer and M. Millon and D. Sluse and A. J. Shajib and F. Courbin and L. V. E. Koopmans and S. H. Suyu and T. Treu},
      year={2024},
      eprint={2210.10833},
      archivePrefix={arXiv},
      primaryClass={id='astro-ph.CO' full_name='Cosmology and Nongalactic Astrophysics' is_active=True alt_name=None in_archive='astro-ph' is_general=False description='Phenomenology of early universe, cosmic microwave background, cosmological parameters, primordial element abundances, extragalactic distance scale, large-scale structure of the universe. Groups, superclusters, voids, intergalactic medium. Particle astrophysics: dark energy, dark matter, baryogenesis, leptogenesis, inflationary models, reheating, monopoles, WIMPs, cosmic strings, primordial black holes, cosmological gravitational radiation'}
}

@article{tdcosmo2025,
   title={TDCOSMO 2025: Cosmological constraints from strong lensing time delays},
   volume={704},
   ISSN={1432-0746},
   url={http://dx.doi.org/10.1051/0004-6361/202555801},
   DOI={10.1051/0004-6361/202555801},
   journal={Astronomy \& Astrophysics},
   publisher={EDP Sciences},
   author={{TDCOSMO Collaboration} and Birrer, Simon and Buckley-Geer, Elizabeth J. and Cappellari, Michele and Courbin, Frédéric and Dux, Frédéric and Fassnacht, Christopher D. and Frieman, Joshua A. and Galan, Aymeric and Gilman, Daniel and Huang, Xiang-Yu and Knabel, Shawn and Langeroodi, Danial and Lin, Huan and Millon, Martin and Morishita, Takahiro and Motta, Veronica and Mozumdar, Pritom and Paic, Eric and Shajib, Anowar J. and Sheu, William and Sluse, Dominique and Sonnenfeld, Alessandro and Spiniello, Chiara and Stiavelli, Massimo and Suyu, Sherry H. and Tan, Chin Yi and Treu, Tommaso and Van de Vyvere, Lyne and Wang, Han and Wells, Patrick and Williams, Devon M. and Wong, Kenneth C.},
   year={2025},
   month=dec, pages={A63} }

@article{valentino2021,
	doi = {10.1088/1361-6382/ac086d},
  
	url = {https://doi.org/10.1088%2F1361-6382%2Fac086d},
  
	year = 2021,
	month = {jul},
  
	publisher = {{IOP} Publishing},
  
	volume = {38},
  
	number = {15},
  
	pages = {153001},
  
	author = {Eleonora Di Valentino and Olga Mena and Supriya Pan and Luca Visinelli and Weiqiang Yang and Alessandro Melchiorri and David F Mota and Adam G Riess and Joseph Silk},
  
	title = {In the realm of the Hubble tension{\textemdash}a review of solutions
								            $\less$sup$\greater${\ast}$\less$/sup$\greater$},
  
	journal = {Classical and Quantum Gravity}
}

@article{Ding2021,
   title={Time delay lens modelling challenge},
   volume={503},
   ISSN={1365-2966},
   url={http://dx.doi.org/10.1093/mnras/stab484},
   DOI={10.1093/mnras/stab484},
   number={1},
   journal={Monthly Notices of the Royal Astronomical Society},
   publisher={Oxford University Press (OUP)},
   author={Ding, X and Treu, T and Birrer, S and Chen, G C-F and Coles, J and Denzel, P and Frigo, M and Galan, A and Marshall, P J and Millon, M and More, A and Shajib, A J and Sluse, D and Tak, H and Xu, D and Auger, M W and Bonvin, V and Chand, H and Courbin, F and Despali, G and Fassnacht, C D and Gilman, D and Hilbert, S and Kumar, S R and Lin, J Y-Y and Park, J W and Saha, P and Vegetti, S and Van de Vyvere, L and Williams, L L R},
   year={2021},
   month=feb, pages={1096–1123} }

@article{Knabel2025,
   title={TDCOSMO: XIX. Measuring stellar velocity dispersion with sub-percent accuracy for cosmography},
   volume={703},
   ISSN={1432-0746},
   url={http://dx.doi.org/10.1051/0004-6361/202554229},
   DOI={10.1051/0004-6361/202554229},
   journal={Astronomy \& Astrophysics},
   publisher={EDP Sciences},
   author={Knabel, Shawn and Mozumdar, Pritom and Shajib, Anowar J. and Treu, Tommaso and Cappellari, Michele and Spiniello, Chiara and Birrer, Simon},
   year={2025},
   month=nov, pages={A117} }

@ARTICLE{Falco1985,
       author = {{Falco}, E.~E. and {Gorenstein}, M.~V. and {Shapiro}, I.~I.},
        title = "{On model-dependent bounds on H 0 from gravitational images : application to Q 0957+561 A, B.}",
      journal = {\apjl},
         year = 1985,
        month = feb,
       volume = {289},
        pages = {L1-L4},
          doi = {10.1086/184422},
       adsurl = {https://ui.adsabs.harvard.edu/abs/1985ApJ...289L...1F}
}

@article{Suyu2017,
   title={H0LiCOW – I. H0 Lenses in COSMOGRAIL’s Wellspring: program overview},
   volume={468},
   ISSN={1365-2966},
   url={http://dx.doi.org/10.1093/mnras/stx483},
   DOI={10.1093/mnras/stx483},
   number={3},
   journal={Monthly Notices of the Royal Astronomical Society},
   publisher={Oxford University Press (OUP)},
   author={Suyu, S. H. and Bonvin, V. and Courbin, F. and Fassnacht, C. D. and Rusu, C. E. and Sluse, D. and Treu, T. and Wong, K. C. and Auger, M. W. and Ding, X. and Hilbert, S. and Marshall, P. J. and Rumbaugh, N. and Sonnenfeld, A. and Tewes, M. and Tihhonova, O. and Agnello, A. and Blandford, R. D. and Chen, G. C.-F. and Collett, T. and Koopmans, L. V. E. and Liao, K. and Meylan, G. and Spiniello, C.},
   year={2017},
   month=feb, pages={2590–2604} }

@ARTICLE{Wong2020,
       author = {{Wong}, Kenneth C. and {Suyu}, Sherry H. and {Chen}, Geoff C.-F. and {Rusu}, Cristian E. and {Millon}, Martin and {Sluse}, Dominique and {Bonvin}, Vivien and {Fassnacht}, Christopher D. and {Taubenberger}, Stefan and {Auger}, Matthew W. and {Birrer}, Simon and {Chan}, James H.~H. and {Courbin}, Frederic and {Hilbert}, Stefan and {Tihhonova}, Olga and {Treu}, Tommaso and {Agnello}, Adriano and {Ding}, Xuheng and {Jee}, Inh and {Komatsu}, Eiichiro and {Shajib}, Anowar J. and {Sonnenfeld}, Alessandro and {Blandford}, Roger D. and {Koopmans}, L{\'e}on V.~E. and {Marshall}, Philip J. and {Meylan}, Georges},
        title = "{H0LiCOW - XIII. A 2.4 per cent measurement of H$_{0}$ from lensed quasars: 5.3{\ensuremath{\sigma}} tension between early- and late-Universe probes}",
      journal = {\mnras},
         year = 2020,
        month = oct,
       volume = {498},
       number = {1},
        pages = {1420-1439},
          doi = {10.1093/mnras/stz3094},
archivePrefix = {arXiv},
       eprint = {1907.04869},
 primaryClass = {astro-ph.CO},
       adsurl = {https://ui.adsabs.harvard.edu/abs/2020MNRAS.498.1420W}
}

@ARTICLE{Refsdal1964,
       author = {{Refsdal}, S.},
        title = "{On the possibility of determining Hubble's parameter and the masses of galaxies from the gravitational lens effect}",
      journal = {\mnras},
         year = 1964,
        month = jan,
       volume = {128},
        pages = {307},
          doi = {10.1093/mnras/128.4.307},
       adsurl = {https://ui.adsabs.harvard.edu/abs/1964MNRAS.128..307R}
}

@ARTICLE{Freedman2021,
       author = {{Freedman}, Wendy L.},
        title = "{Measurements of the Hubble Constant: Tensions in Perspective}",
      journal = {\apj},
         year = 2021,
        month = sep,
       volume = {919},
       number = {1},
          eid = {16},
        pages = {16},
          doi = {10.3847/1538-4357/ac0e95},
archivePrefix = {arXiv},
       eprint = {2106.15656},
 primaryClass = {astro-ph.CO},
       adsurl = {https://ui.adsabs.harvard.edu/abs/2021ApJ...919...16F}
}

@article{Riess2022,
   title={A Comprehensive Measurement of the Local Value of the Hubble Constant with 1 km s−1 Mpc−1 Uncertainty from the Hubble Space Telescope and the SH0ES Team},
   volume={934},
   ISSN={2041-8213},
   url={http://dx.doi.org/10.3847/2041-8213/ac5c5b},
   DOI={10.3847/2041-8213/ac5c5b},
   number={1},
   journal={The Astrophysical Journal Letters},
   publisher={American Astronomical Society},
   author={Riess, Adam G. and Yuan, Wenlong and Macri, Lucas M. and Scolnic, Dan and Brout, Dillon and Casertano, Stefano and Jones, David O. and Murakami, Yukei and Anand, Gagandeep S. and Breuval, Louise and Brink, Thomas G. and Filippenko, Alexei V. and Hoffmann, Samantha and Jha, Saurabh W. and D’arcy Kenworthy, W. and Mackenty, John and Stahl, Benjamin E. and Zheng, WeiKang},
   year={2022},
   month=jul, pages={L7} }

@article{Planck2020,
   title={Planck 2018 results: VI. Cosmological parameters},
   volume={641},
   ISSN={1432-0746},
   url={http://dx.doi.org/10.1051/0004-6361/201833910},
   DOI={10.1051/0004-6361/201833910},
   journal={Astronomy \& Astrophysics},
   publisher={EDP Sciences},
   author={{Planck Collaboration} and Aghanim, N. and Akrami, Y. and Ashdown, M. and Aumont, J. and Baccigalupi, C. and Ballardini, M. and Banday, A. J. and Barreiro, R. B. and Bartolo, N. and Basak, S. and Battye, R. and Benabed, K. and Bernard, J.-P. and Bersanelli, M. and Bielewicz, P. and Bock, J. J. and Bond, J. R. and Borrill, J. and Bouchet, F. R. and Boulanger, F. and Bucher, M. and Burigana, C. and Butler, R. C. and Calabrese, E. and Cardoso, J.-F. and Carron, J. and Challinor, A. and Chiang, H. C. and Chluba, J. and Colombo, L. P. L. and Combet, C. and Contreras, D. and Crill, B. P. and Cuttaia, F. and de Bernardis, P. and de Zotti, G. and Delabrouille, J. and Delouis, J.-M. and Di Valentino, E. and Diego, J. M. and Doré, O. and Douspis, M. and Ducout, A. and Dupac, X. and Dusini, S. and Efstathiou, G. and Elsner, F. and Enßlin, T. A. and Eriksen, H. K. and Fantaye, Y. and Farhang, M. and Fergusson, J. and Fernandez-Cobos, R. and Finelli, F. and Forastieri, F. and Frailis, M. and Fraisse, A. A. and Franceschi, E. and Frolov, A. and Galeotta, S. and Galli, S. and Ganga, K. and Génova-Santos, R. T. and Gerbino, M. and Ghosh, T. and González-Nuevo, J. and Górski, K. M. and Gratton, S. and Gruppuso, A. and Gudmundsson, J. E. and Hamann, J. and Handley, W. and Hansen, F. K. and Herranz, D. and Hildebrandt, S. R. and Hivon, E. and Huang, Z. and Jaffe, A. H. and Jones, W. C. and Karakci, A. and Keihänen, E. and Keskitalo, R. and Kiiveri, K. and Kim, J. and Kisner, T. S. and Knox, L. and Krachmalnicoff, N. and Kunz, M. and Kurki-Suonio, H. and Lagache, G. and Lamarre, J.-M. and Lasenby, A. and Lattanzi, M. and Lawrence, C. R. and Le Jeune, M. and Lemos, P. and Lesgourgues, J. and Levrier, F. and Lewis, A. and Liguori, M. and Lilje, P. B. and Lilley, M. and Lindholm, V. and López-Caniego, M. and Lubin, P. M. and Ma, Y.-Z. and Macías-Pérez, J. F. and Maggio, G. and Maino, D. and Mandolesi, N. and Mangilli, A. and Marcos-Caballero, A. and Maris, M. and Martin, P. G. and Martinelli, M. and Martínez-González, E. and Matarrese, S. and Mauri, N. and McEwen, J. D. and Meinhold, P. R. and Melchiorri, A. and Mennella, A. and Migliaccio, M. and Millea, M. and Mitra, S. and Miville-Deschênes, M.-A. and Molinari, D. and Montier, L. and Morgante, G. and Moss, A. and Natoli, P. and Nørgaard-Nielsen, H. U. and Pagano, L. and Paoletti, D. and Partridge, B. and Patanchon, G. and Peiris, H. V. and Perrotta, F. and Pettorino, V. and Piacentini, F. and Polastri, L. and Polenta, G. and Puget, J.-L. and Rachen, J. P. and Reinecke, M. and Remazeilles, M. and Renzi, A. and Rocha, G. and Rosset, C. and Roudier, G. and Rubiño-Martín, J. A. and Ruiz-Granados, B. and Salvati, L. and Sandri, M. and Savelainen, M. and Scott, D. and Shellard, E. P. S. and Sirignano, C. and Sirri, G. and Spencer, L. D. and Sunyaev, R. and Suur-Uski, A.-S. and Tauber, J. A. and Tavagnacco, D. and Tenti, M. and Toffolatti, L. and Tomasi, M. and Trombetti, T. and Valenziano, L. and Valiviita, J. and Van Tent, B. and Vibert, L. and Vielva, P. and Villa, F. and Vittorio, N. and Wandelt, B. D. and Wehus, I. K. and White, M. and White, S. D. M. and Zacchei, A. and Zonca, A.},
   year={2020},
   month=sep, pages={A6} }

@ARTICLE{Krolewski2025,
       author = {{Krolewski}, Alex and {Percival}, Will J. and {Woodfinden}, Alex},
        title = "{New Method to Determine the Hubble Parameter from Cosmological Energy-Density Measurements}",
      journal = {\prl},
         year = 2025,
        month = mar,
       volume = {134},
       number = {10},
          eid = {101002},
        pages = {101002},
          doi = {10.1103/PhysRevLett.134.101002},
archivePrefix = {arXiv},
       eprint = {2403.19227},
 primaryClass = {astro-ph.CO},
       adsurl = {https://ui.adsabs.harvard.edu/abs/2025PhRvL.134j1002K}
}

@ARTICLE{knabel2026slacsII,
       author = {{Knabel}, Shawn and {Treu}, Tommaso and {Cappellari}, Michele and {Birrer}, Simon and {Huang}, Xiang-Yu and {Shajib}, Anowar J. and {Sheu}, William},
        title = "{Spatially Resolved Kinematics of SLACS Lens Galaxies. II: Breaking Degeneracies with Lensing and Dynamical Models}",
      journal = {arXiv e-prints},
         year = 2026,
        month = apr,
          eid = {arXiv:2604.12155},
        pages = {arXiv:2604.12155},
archivePrefix = {arXiv},
       eprint = {2604.12155},
 primaryClass = {astro-ph.GA},
       adsurl = {https://ui.adsabs.harvard.edu/abs/2026arXiv260412155K}
}

@article{Dey_2019,
   title={Overview of the DESI Legacy Imaging Surveys},
   volume={157},
   ISSN={1538-3881},
   url={http://dx.doi.org/10.3847/1538-3881/ab089d},
   DOI={10.3847/1538-3881/ab089d},
   number={5},
   journal={The Astronomical Journal},
   publisher={American Astronomical Society},
   author={Dey, Arjun and Schlegel, David J. and Lang, Dustin and Blum, Robert and Burleigh, Kaylan and Fan, Xiaohui and Findlay, Joseph R. and Finkbeiner, Doug and Herrera, David and Juneau, Stéphanie and Landriau, Martin and Levi, Michael and McGreer, Ian and Meisner, Aaron and Myers, Adam D. and Moustakas, John and Nugent, Peter and Patej, Anna and Schlafly, Edward F. and Walker, Alistair R. and Valdes, Francisco and Weaver, Benjamin A. and Yèche, Christophe and Zou, Hu and Zhou, Xu and Abareshi, Behzad and Abbott, T. M. C. and Abolfathi, Bela and Aguilera, C. and Alam, Shadab and Allen, Lori and Alvarez, A. and Annis, James and Ansarinejad, Behzad and Aubert, Marie and Beechert, Jacqueline and Bell, Eric F. and BenZvi, Segev Y. and Beutler, Florian and Bielby, Richard M. and Bolton, Adam S. and Briceño, César and Buckley-Geer, Elizabeth J. and Butler, Karen and Calamida, Annalisa and Carlberg, Raymond G. and Carter, Paul and Casas, Ricard and Castander, Francisco J. and Choi, Yumi and Comparat, Johan and Cukanovaite, Elena and Delubac, Timothée and DeVries, Kaitlin and Dey, Sharmila and Dhungana, Govinda and Dickinson, Mark and Ding, Zhejie and Donaldson, John B. and Duan, Yutong and Duckworth, Christopher J. and Eftekharzadeh, Sarah and Eisenstein, Daniel J. and Etourneau, Thomas and Fagrelius, Parker A. and Farihi, Jay and Fitzpatrick, Mike and Font-Ribera, Andreu and Fulmer, Leah and Gänsicke, Boris T. and Gaztanaga, Enrique and George, Koshy and Gerdes, David W. and A Gontcho, Satya Gontcho and Gorgoni, Claudio and Green, Gregory and Guy, Julien and Harmer, Diane and Hernandez, M. and Honscheid, Klaus and Huang, Lijuan (Wendy) and James, David J. and Jannuzi, Buell T. and Jiang, Linhua and Joyce, Richard and Karcher, Armin and Karkar, Sonia and Kehoe, Robert and Kneib, Jean-Paul and Kueter-Young, Andrea and Lan, Ting-Wen and Lauer, Tod R. and Guillou, Laurent Le and Van Suu, Auguste Le and Lee, Jae Hyeon and Lesser, Michael and Levasseur, Laurence Perreault and Li, Ting S. and Mann, Justin L. and Marshall, Robert and Martínez-Vázquez, C. E. and Martini, Paul and du Mas des Bourboux, Hélion and McManus, Sean and Meier, Tobias Gabriel and Ménard, Brice and Metcalfe, Nigel and Muñoz-Gutiérrez, Andrea and Najita, Joan and Napier, Kevin and Narayan, Gautham and Newman, Jeffrey A. and Nie, Jundan and Nord, Brian and Norman, Dara J. and Olsen, Knut A. G. and Paat, Anthony and Palanque-Delabrouille, Nathalie and Peng, Xiyan and Poppett, Claire L. and Poremba, Megan R. and Prakash, Abhishek and Rabinowitz, David and Raichoor, Anand and Rezaie, Mehdi and Robertson, A. N. and Roe, Natalie A. and Ross, Ashley J. and Ross, Nicholas P. and Rudnick, Gregory and Gaines, Sasha and Saha, Abhijit and Sánchez, F. Javier and Savary, Elodie and Schweiker, Heidi and Scott, Adam and Seo, Hee-Jong and Shan, Huanyuan and Silva, David R. and Slepian, Zachary and Soto, Christian and Sprayberry, David and Staten, Ryan and Stillman, Coley M. and Stupak, Robert J. and Summers, David L. and Tie, Suk Sien and Tirado, H. and Vargas-Magaña, Mariana and Vivas, A. Katherina and Wechsler, Risa H. and Williams, Doug and Yang, Jinyi and Yang, Qian and Yapici, Tolga and Zaritsky, Dennis and Zenteno, A. and Zhang, Kai and Zhang, Tianmeng and Zhou, Rongpu and Zhou, Zhimin},
   year={2019},
   month=apr, pages={168} }

@article{Wells_2023,
   title={TDCOSMO: XIV. Practical techniques for estimating external convergence of strong gravitational lens systems and applications to the SDSS J0924+0219 system},
   volume={676},
   ISSN={1432-0746},
   url={http://dx.doi.org/10.1051/0004-6361/202346093},
   DOI={10.1051/0004-6361/202346093},
   journal={Astronomy \& Astrophysics},
   publisher={EDP Sciences},
   author={Wells, Patrick and Fassnacht, Christopher D. and Rusu, C. E.},
   year={2023},
   month=aug, pages={A95} }

@ARTICLE{zhou2023,
       author = {{Zhou}, Rongpu and {Ferraro}, Simone and {White}, Martin and {DeRose}, Joseph and {Sailer}, Noah and {Aguilar}, Jessica and {Ahlen}, Steven and {Bailey}, Stephen and {Brooks}, David and {Claybaugh}, Todd and {Dawson}, Kyle and {de la Macorra}, Axel and {Dey}, Biprateep and {Doel}, Peter and {Font-Ribera}, Andreu and {Forero-Romero}, Jaime E. and {Gontcho A Gontcho}, Satya and {Guy}, Julien and {Kremin}, Anthony and {Lambert}, Andrew and {Le Guillou}, Laurent and {Levi}, Michael and {Magneville}, Christophe and {Manera}, Marc and {Meisner}, Aaron and {Miquel}, Ramon and {Moustakas}, John and {Myers}, Adam D. and {Newman}, Jeffrey A. and {Nie}, Jundan and {Percival}, Will and {Rezaie}, Mehdi and {Rossi}, Graziano and {Sanchez}, Eusebio and {Schlegel}, David and {Schubnell}, Michael and {Seo}, Hee-Jong and {Tarl{\'e}}, Gregory and {Zhou}, Zhimin},
        title = "{DESI luminous red galaxy samples for cross-correlations}",
      journal = {\jcap},
         year = 2023,
        month = nov,
       volume = {2023},
       number = {11},
          eid = {097},
        pages = {097},
          doi = {10.1088/1475-7516/2023/11/097},
archivePrefix = {arXiv},
       eprint = {2309.06443},
 primaryClass = {astro-ph.CO},
       adsurl = {https://ui.adsabs.harvard.edu/abs/2023JCAP...11..097Z}
}

@article{zahid2016,
   title={THE SCALING OF STELLAR MASS AND CENTRAL STELLAR VELOCITY DISPERSION FOR QUIESCENT GALAXIES AT z<0.7},
   volume={832},
   ISSN={1538-4357},
   url={http://dx.doi.org/10.3847/0004-637X/832/2/203},
   DOI={10.3847/0004-637x/832/2/203},
   number={2},
   journal={The Astrophysical Journal},
   publisher={American Astronomical Society},
   author={Zahid, H. Jabran and Geller, Margaret J. and Fabricant, Daniel G. and Hwang, Ho Seong},
   year={2016},
   month=dec, pages={203} }

@article{Abazajian_2009,
   title={THE SEVENTH DATA RELEASE OF THE SLOAN DIGITAL SKY SURVEY},
   volume={182},
   ISSN={1538-4365},
   url={http://dx.doi.org/10.1088/0067-0049/182/2/543},
   DOI={10.1088/0067-0049/182/2/543},
   number={2},
   journal={The Astrophysical Journal Supplement Series},
   publisher={American Astronomical Society},
   author={Abazajian, Kevork N. and Adelman-McCarthy, Jennifer K. and Agüeros, Marcel A. and Allam, Sahar S. and Prieto, Carlos Allende and An, Deokkeun and Anderson, Kurt S. J. and Anderson, Scott F. and Annis, James and Bahcall, Neta A. and Bailer-Jones, C. A. L. and Barentine, J. C. and Bassett, Bruce A. and Becker, Andrew C. and Beers, Timothy C. and Bell, Eric F. and Belokurov, Vasily and Berlind, Andreas A. and Berman, Eileen F. and Bernardi, Mariangela and Bickerton, Steven J. and Bizyaev, Dmitry and Blakeslee, John P. and Blanton, Michael R. and Bochanski, John J. and Boroski, William N. and Brewington, Howard J. and Brinchmann, Jarle and Brinkmann, J. and Brunner, Robert J. and Budavári, Tamás and Carey, Larry N. and Carliles, Samuel and Carr, Michael A. and Castander, Francisco J. and Cinabro, David and Connolly, A. J. and Csabai, István and Cunha, Carlos E. and Czarapata, Paul C. and Davenport, James R. A. and de Haas, Ernst and Dilday, Ben and Doi, Mamoru and Eisenstein, Daniel J. and Evans, Michael L. and Evans, N. W. and Fan, Xiaohui and Friedman, Scott D. and Frieman, Joshua A. and Fukugita, Masataka and Gänsicke, Boris T. and Gates, Evalyn and Gillespie, Bruce and Gilmore, G. and Gonzalez, Belinda and Gonzalez, Carlos F. and Grebel, Eva K. and Gunn, James E. and Györy, Zsuzsanna and Hall, Patrick B. and Harding, Paul and Harris, Frederick H. and Harvanek, Michael and Hawley, Suzanne L. and Hayes, Jeffrey J. E. and Heckman, Timothy M. and Hendry, John S. and Hennessy, Gregory S. and Hindsley, Robert B. and Hoblitt, J. and Hogan, Craig J. and Hogg, David W. and Holtzman, Jon A. and Hyde, Joseph B. and Ichikawa, Shin-ichi and Ichikawa, Takashi and Im, Myungshin and Ivezić, Željko and Jester, Sebastian and Jiang, Linhua and Johnson, Jennifer A. and Jorgensen, Anders M. and Jurić, Mario and Kent, Stephen M. and Kessler, R. and Kleinman, S. J. and Knapp, G. R. and Konishi, Kohki and Kron, Richard G. and Krzesinski, Jurek and Kuropatkin, Nikolay and Lampeitl, Hubert and Lebedeva, Svetlana and Lee, Myung Gyoon and Lee, Young Sun and Leger, R. French and Lépine, Sébastien and Li, Nolan and Lima, Marcos and Lin, Huan and Long, Daniel C. and Loomis, Craig P. and Loveday, Jon and Lupton, Robert H. and Magnier, Eugene and Malanushenko, Olena and Malanushenko, Viktor and Mandelbaum, Rachel and Margon, Bruce and Marriner, John P. and Martínez-Delgado, David and Matsubara, Takahiko and McGehee, Peregrine M. and McKay, Timothy A. and Meiksin, Avery and Morrison, Heather L. and Mullally, Fergal and Munn, Jeffrey A. and Murphy, Tara and Nash, Thomas and Nebot, Ada and Neilsen, Eric H. and Newberg, Heidi Jo and Newman, Peter R. and Nichol, Robert C. and Nicinski, Tom and Nieto-Santisteban, Maria and Nitta, Atsuko and Okamura, Sadanori and Oravetz, Daniel J. and Ostriker, Jeremiah P. and Owen, Russell and Padmanabhan, Nikhil and Pan, Kaike and Park, Changbom and Pauls, George and Peoples, John and Percival, Will J. and Pier, Jeffrey R. and Pope, Adrian C. and Pourbaix, Dimitri and Price, Paul A. and Purger, Norbert and Quinn, Thomas and Raddick, M. Jordan and Fiorentin, Paola Re and Richards, Gordon T. and Richmond, Michael W. and Riess, Adam G. and Rix, Hans-Walter and Rockosi, Constance M. and Sako, Masao and Schlegel, David J. and Schneider, Donald P. and Scholz, Ralf-Dieter and Schreiber, Matthias R. and Schwope, Axel D. and Seljak, Uroš and Sesar, Branimir and Sheldon, Erin and Shimasaku, Kazu and Sibley, Valena C. and Simmons, A. E. and Sivarani, Thirupathi and Smith, J. Allyn and Smith, Martin C. and Smolčić, Vernesa and Snedden, Stephanie A. and Stebbins, Albert and Steinmetz, Matthias and Stoughton, Chris and Strauss, Michael A. and SubbaRao, Mark and Suto, Yasushi and Szalay, Alexander S. and Szapudi, István and Szkody, Paula and Tanaka, Masayuki and Tegmark, Max and Teodoro, Luis F. A. and Thakar, Aniruddha R. and Tremonti, Christy A. and Tucker, Douglas L. and Uomoto, Alan and Vanden Berk, Daniel E. and Vandenberg, Jan and Vidrih, S. and Vogeley, Michael S. and Voges, Wolfgang and Vogt, Nicole P. and Wadadekar, Yogesh and Watters, Shannon and Weinberg, David H. and West, Andrew A. and White, Simon D. M. and Wilhite, Brian C. and Wonders, Alainna C. and Yanny, Brian and Yocum, D. R. and York, Donald G. and Zehavi, Idit and Zibetti, Stefano and Zucker, Daniel B.},
   year={2009},
   month=may, pages={543–558} }

@article{Agnello_2017,
   title={Discovery and first models of the quadruply lensed quasar SDSS J1433+6007},
   volume={474},
   ISSN={1365-2966},
   url={http://dx.doi.org/10.1093/mnras/stx2950},
   DOI={10.1093/mnras/stx2950},
   number={3},
   journal={Monthly Notices of the Royal Astronomical Society},
   publisher={Oxford University Press (OUP)},
   author={Agnello, Adriano and Grillo, Claudio and Jones, Tucker and Treu, Tommaso and Bonamigo, Mario and Suyu, Sherry H},
   year={2017},
   month=nov, pages={3391–3396} }

@article{q25,
   title={Time-delay cosmography: analysis of quadruply lensed QSO SDSSJ1433 from Wendelstein Observatory},
   volume={542},
   ISSN={1365-2966},
   url={http://dx.doi.org/10.1093/mnras/staf1186},
   DOI={10.1093/mnras/staf1186},
   number={1},
   journal={Monthly Notices of the Royal Astronomical Society},
   publisher={Oxford University Press (OUP)},
   author={Queirolo, G and Seitz, S and Riffeser, A and Kluge, M and Ecker, L R and Bender, R and Gössl, C and Hopp, U and Ries, C and Schmidt, M and Zöller, R},
   year={2025},
   month=jul, pages={170–202} }

@article{Shajib_2023,
   title={TDCOSMO: XII. Improved Hubble constant measurement from lensing time delays using spatially resolved stellar kinematics of the lens galaxy},
   volume={673},
   ISSN={1432-0746},
   url={http://dx.doi.org/10.1051/0004-6361/202345878},
   DOI={10.1051/0004-6361/202345878},
   journal={Astronomy \& Astrophysics},
   publisher={EDP Sciences},
   author={Shajib, Anowar J. and Mozumdar, Pritom and Chen, Geoff C.-F. and Treu, Tommaso and Cappellari, Michele and Knabel, Shawn and Suyu, Sherry H. and Bennert, Vardha N. and Frieman, Joshua A. and Sluse, Dominique and Birrer, Simon and Courbin, Frederic and Fassnacht, Christopher D. and Villafaña, Lizvette and Williams, Peter R.},
   year={2023},
   month=apr, pages={A9} }

@article{Mozumdar_2023,
   title={TDCOSMO: XI. New lensing galaxy redshift and velocity dispersion measurements from Keck spectroscopy of eight lensed quasar systems},
   volume={672},
   ISSN={1432-0746},
   url={http://dx.doi.org/10.1051/0004-6361/202245082},
   DOI={10.1051/0004-6361/202245082},
   journal={Astronomy \& Astrophysics},
   publisher={EDP Sciences},
   author={Mozumdar, P. and Fassnacht, C. D. and Treu, T. and Spiniello, C. and Shajib, A. J.},
   year={2023},
   month=mar, pages={A20} }

\end{document}